\documentclass[12pt]{article}
\usepackage{times}
\usepackage{here}
\usepackage[english]{babel}

\usepackage{tikz}
\usepackage{amsmath,amssymb}
\usepackage{here}
\usepackage{bm}
\usepackage{steinmetz} 

\usepackage[hidelinks]{hyperref}
\usepackage{xurl}   
\usepackage{multirow} 

\usepackage[font=small,labelfont=bf]{caption}

\usepackage{makecell}

\title{Morphogenetic mechanical metamaterials: \\ \Large \textit{Emerging tensor properties from self-organized structures}}

\author{\bfseries Thomas Froment\`eze,$^{1,2,*}$ Philippe Michaud,$^{3}$\\  \bfseries Ali Hassny,$^{4}$ and Vincent Pateloup$^{3}$\\
	\normalsize{$^{1}$ University of Limoges, CNRS, XLIM, UMR 7252, F-87000 Limoges, France}\\
	\normalsize{$^{2}$ Institut universitaire de France (IUF)}\\
	\normalsize{$^{3}$ University of Limoges, CNRS, IRCER, UMR 7315, F-87000 Limoges, France} \\
	\normalsize{$^{4}$ University of Limoges, IUT du Limousin, F-87000 Limoges, France} \\
	\normalsize{$^{*}$  Corresponding author: thomas.fromenteze@unilim.fr}}

\date{}

\definecolor{blue}{RGB}{0,0,0} 
\definecolor{revtwo}{RGB}{0,0,0} 
\definecolor{reveditor}{RGB}{0,0,0} 

\usepackage{booktabs}

\begin{document}

\maketitle

\bfseries

Understanding how living organisms spontaneously develop complex functional structures motivates new strategies in engineering design. Here, we introduce a decentralized generative model based on morphogenesis to autonomously grow mechanical structures with controlled tensorial properties. By adapting Turing's reaction-diffusion concept through anisotropic diffusion, our approach enables the local emergence of microstructures exhibiting tailored stiffness and anisotropy\textcolor{blue}{,} achieving target orthotropic tensors without adjoint or topology optimization loops. The synthesis of these structures relies on a database linking morphogenetic parameters to effective elastic tensors obtained through homogenization techniques. We experimentally demonstrate this concept through a mechanical cloaking example, validating our method's capability to independently control local anisotropy and rigidity, and effectively conceal structural defects from mechanical fields. \textcolor{blue}{This approach circumvents the iterative global solves required by topology optimization, while preserving local control over anisotropy and stiffness across large design domains.}

\normalfont
\newpage

\section*{Introduction}

How do living organisms manage to give rise to highly functional and complex structures, even before the emergence of a central nervous system capable of orchestrating decision-making? Despite rapid technological advances in the digital era, current design methodologies remain largely limited when compared to the richness of complex systems observed in nature.

Among the effective design strategies known today, shaped by a long process of evolutionary optimization, nature appears to have found a way to decentralize its structuring mechanisms through local interactions and seemingly simple rules. The emergence of form in living systems was precisely the focus of Alan Turing's final publication, in which he proposed a mechanism based on the reaction and diffusion of chemical species~\cite{turing1952chemical}. \textcolor{blue}{This process, known as morphogenesis, governs how tissues and organs self-organize into complex functional shapes without central direction.} This foundational work led to the development of numerous generative models, capable of producing spatial patterns with essential features such as self-duplication, self-organization, and adaptation to arbitrarily shaped boundary conditions, all while relying on a compact formalism and purely local interactions~\cite{gray1983autocatalytic,lee1994experimental}.

\textcolor{blue}{In this work, we propose to harness these biological principles to generate mechanical metamaterials. Specifically, we adapt reaction-diffusion systems to synthesize cellular structures through a fully decentralized growth process.} The design of mechanical metamaterials represents a major challenge in engineering, paving the way toward the free manipulation of stress and strain fields~\cite{bertoldi2017flexible,jiao2023mechanical}. This concept originates from electromagnetism, where the homogenization of sub-wavelength patterns has enabled the emergence of unexpected properties such as negative refractive indices~\cite{smith2004metamaterials}. When combined with spatial transformation methods, these approaches have proven particularly effective in manipulating wave propagation and have led to early demonstrations of invisibility cloaks~\cite{pendry2006controlling}. Since then, these concepts have been transposed to various areas of physics~\cite{milton2006cloaking,farhat2009cloaking, norris2011elastic, kadic2013metamaterials}, leveraging the generality and flexibility of fundamental homogenization and multiscale structuring principles. Extending this logic to solid mechanics, mechanical metamaterials enable effective properties that are counterintuitive at the continuum scale. Within this landscape, auxetic mechanical metamaterials, exhibiting a negative Poisson's ratio, illustrate how geometry can program large, reversible responses, with recent advances spanning rational design and cataloging, data-driven exploration, and highly stretchable implementations~\cite{kolken2017auxetic,reid2018auxetic,wilt2020accelerating,zhang2024stretchable}.

\textcolor{blue}{We first focus here on the concept of mechanical cloaking: a structured material distribution that guides mechanical stress waves around an object (such as a cavity or inclusion) so that the displacement and stress fields outside the cloaking region remain identical to those of a homogeneous medium, effectively rendering the object invisible to mechanical probing.} \textcolor{blue}{The methods initially developed for electromagnetic cloaking rely on the general invariance of Maxwell's equations under spatial transformations~\cite{pendry2006controlling,smith2004metamaterials}. This condition does not hold for the classical equations of elasticity in continuum mechanics, which fundamentally limits the applicability of direct transformation-based cloaking approaches in solid mechanics~\cite{milton2006cloaking}. This difference stems from the fact that, in isotropic dielectric materials, the defining parameters are independent scalars, despite the vectorial nature of electromagnetic equations. In contrast, solid mechanics requires the definition of two coupled quantities to characterize an isotropic material. The bulk modulus $B$ and the shear modulus $G$ thus respectively quantify the resistance to volumetric and shape deformations~\cite{buckmann2014elasto}. Transformation-based methods typically induce anisotropic distortions, which in turn couple volumetric and shear deformation modes. As a result, the material properties required for an ideal mechanical cloak involve extreme ratios between these elastic moduli, which are difficult to achieve with conventional or isotropic materials~\cite{kadic2012practicability, buckmann2014elasto, norris2011elastic}.}

\begin{figure}[h!]
	\centering
	\includegraphics[width = 1\textwidth]{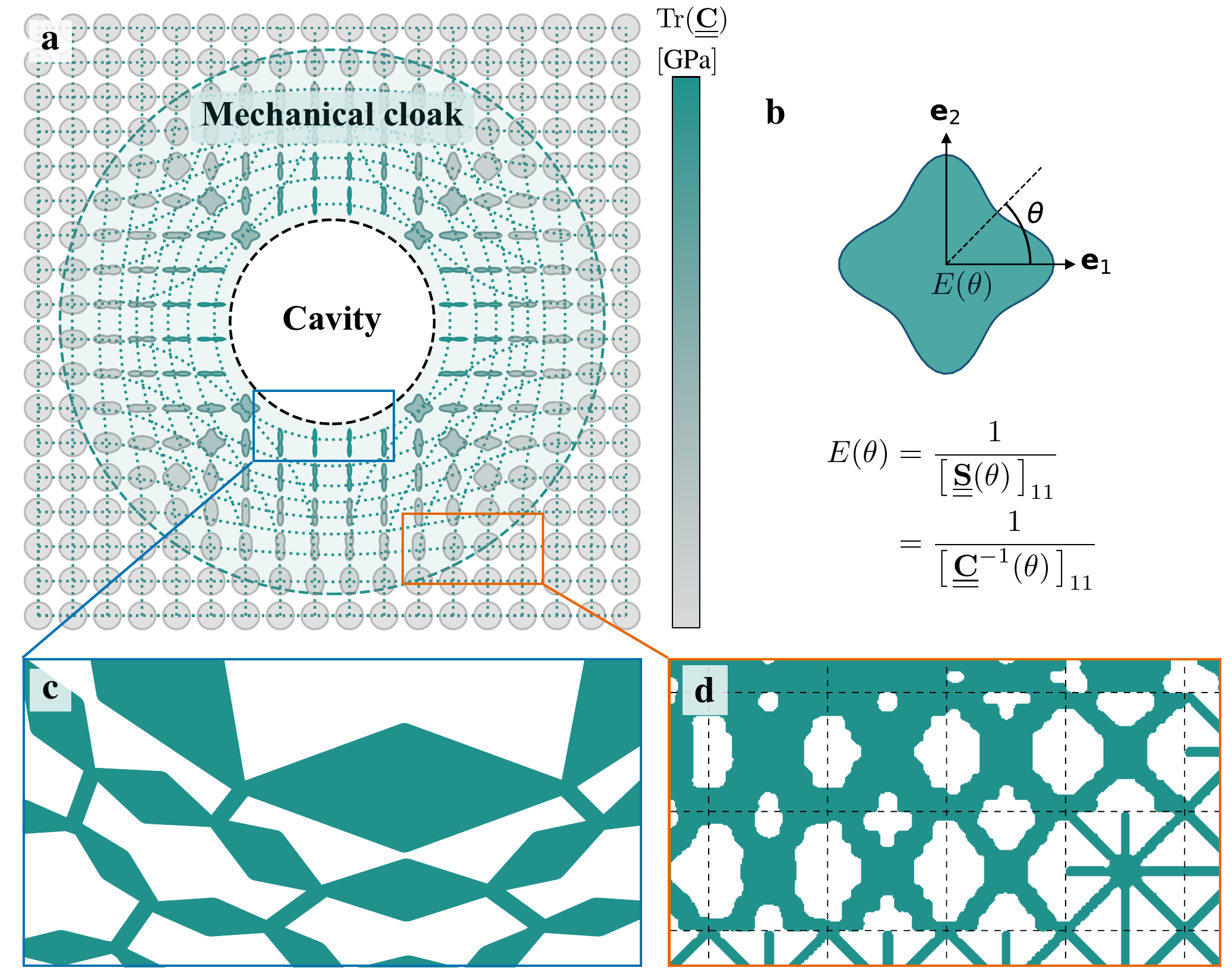}
	\caption{\color{revtwo} Mechanical cloaking synthesis strategies. The objective of a mechanical cloak is to ensure that the displacement and stress fields outside the cloaking region remain identical to those of a homogeneous medium, effectively rendering the enclosed region invisible to mechanical probing. Achieving this theoretical design (a) requires synthesizing spatially distributed and locally controlled elasticity tensors $\mathbf{\underline{\underline{C}}}$. \textcolor{reveditor}{In (a), the colormap encodes the stiffness trace $\mathrm{Tr}(\mathbf{\underline{\underline{C}}})$: dark tones near the cavity correspond to high overall stiffness (amplified by the spatial transformation); light tones at the outer boundary approach the reference stiffness of the homogeneous isotropic host material.} Designing the corresponding physical microstructures is highly challenging due to the inherent complexity of mapping $4^\text{th}$-order tensors, whose directional stiffness projections $E(\theta)$ (b) exhibit heavily coupled normal and shear behaviors. To physically implement these complex effective parameters, state-of-the-art strategies typically rely on either local coordinate transformations of pentamode lattice networks (c)~\cite{buckmann2014elasto, buckmann2015mechanical}, or data-driven optimization of pixelated unit-cells (d)~\cite{wang2022mechanical}. The structures depicted here are scale-invariant and do not correspond to a specific physical realization.}
	\label{fig:RD_demo}
\end{figure}

To overcome these limitations, an alternative approach consists in directly \textcolor{blue}{mapping} discrete lattice-like networks, while preserving the local mechanical properties of the connecting elements~\cite{buckmann2014elasto,buckmann2015mechanical} (Fig.~\ref{fig:RD_demo}a). These so-called pentamode structures, initially proposed by Milton and Cherkaev~\cite{milton1995elasticity}, naturally exhibit fluid-like mechanical behavior~\cite{norris2009acoustic, martin2012phonon, zhang2020isotropic, kadic2014pentamode}, allowing an effective decoupling between the bulk modulus $B$ and the shear modulus $G$. Among their various demonstrations, Kadic \textit{et al.} showed that the achievable $B/G$ ratio depends critically on the slenderness of the connecting struts~\cite{kadic2012practicability}, which induces highly localized stress concentrations and potentially increases the risk of structural failure. In a similar context, Wang \textit{et al.} explored a machine learning approach to construct databases of microstructured unit cells, enabling advanced optimization of local mechanical properties~\cite{wang2022mechanical} (Fig.~\ref{fig:RD_demo}b). This strategy offers great flexibility to explore vast parameter spaces~\cite{mao2020designing, bastek2022inverting, zheng2023deep}. Optimal microstructures for each cell are selected through gradient-based optimization, allowing the generation of highly effective mechanical cloaks with complex unit cell architectures. However, the richness of these patterns also introduces significant variability during fabrication, contributing to discrepancies between theoretical predictions and experimental measurements, and requiring the use of additive manufacturing techniques.

This synthesis method naturally \textcolor{black}{leads to} the more general principle of topology optimization in mechanics~\cite{sigmund2013topology,wu2021topology}. These approaches aim to determine the optimal material distribution within a given domain, in order to achieve an ideal trade-off between performance, weight, and cost, while satisfying specific constraints such as stress limits or stiffness targets~\cite{amstutz2010topological,aage2017giga}. The main strength of these methods lies in their exploitation of the inherently local nature of the equations governing static solid mechanics, enabling rapid convergence toward high-performance structures through fine control over small-scale material distributions~\cite{zheng2016multiscale}.

Within this landscape, \textcolor{black}{several} complementary directions are particularly relevant\textcolor{blue}{.} Garnier \textit{et al.} use anisotropic reaction-diffusion guided by a diffusion tensor field to grow oriented lattice- and membrane-like structures that conform to boundaries and align with principal directions. Their emphasis is on oriented pattern growth and structural robustness, for example against buckling, not on a calibrated link from reaction–diffusion parameters to quantitative orthotropic stiffness components or on cloaking functions~\cite{garnier2022growth}. Tricard \textit{et al.} generate freely orientable microstructures through procedural texturing based on anisotropically filtered spatial noise with spatially graded spectra. This fabrication-oriented strategy programs a rigid-transverse response by steering local fiber-like orientations, but it does not provide a tensor calibration for static elasticity or address cloaking~\cite{tricard2020freely}. A recent, complementary line formalizes the distinction between reinforcement and true unbiased elastostatic cloaking, and achieves cloaks by optimizing against worst-case load sets with an energy-mismatch objective \cite{senhora2025unbiased}. This optimization-based route targets universality across loadings, whereas our goal here is a calibrated, decentralized generative model.

Our approach builds on this context with a fully decentralized strategy where complex structures emerge without central supervision. This philosophy has underpinned our previous studies on bandgap materials and on morphogenetic metasurfaces with locally emerging electromagnetic tensors \cite{chehami2023morphogenetic, fromenteze2023morphogenetic}. It motivates a route that is local and scalable. The generative law operates pointwise through reactions and couples only through short-range anisotropic diffusion. It requires no global adjoint solves and no iterative re-evaluation of performance to update the design. This algorithmic locality supports growth over large domains and high resolutions while keeping computational and fabrication complexity under control~\cite{liu2022growth}.

This article advances morphogenesis-inspired structuring from oriented pattern growth to quantitative tensor synthesis in static continuum mechanics. We show that anisotropic reaction–diffusion, combined with homogenization, enables local self-organization toward prescribed orthotropic elasticity tensors. The calibrated map from generative parameters to effective properties controls the magnitude of the elastic response, principal anisotropy, and shear, and does so without global optimization or feedback loops driven by response evaluation. We further provide a proof-of-principle experimental demonstration of mechanical cloaking. An orthotropy-preserving tensor transformation formulated in Mandel–Kelvin form defines a target field of elasticity tensors that we realize through decentralized growth, which restores far-field displacements around a cavity.

The Results section summarizes the reachable tensor space, the experimental demonstration of the mechanical cloak, and the generalization of the method to various loading scenarios and stress-guided optimization. The Methods section details the reaction–diffusion engine, the homogenization database and the tensor transformation. The Discussion clarifies implications and limitations, and explains how this local, scalable and non-iterative pipeline complements reaction-diffusion-grown oriented structures and procedurally synthesized orientable microstructures~\cite{saldivar2025mimicking}.

\section*{Results}

\subsection*{Morphogenetic Generation of Anisotropic Patterns}

The core of the decentralized strategy proposed in this work lies in the development of an anisotropic reaction-diffusion generative model. The control of the synthesized mechanical properties is enabled by the selection of chemical parameters and their local diffusion characteristics, which drive the \textcolor{blue}{growth} of self-organized forms exhibiting the desired geometric features. According to mechanisms envisioned in Alan Turing's final work~\cite{turing1952chemical}, the emergence of form results from \textcolor{blue}{coupling} chemical reactions and diffusion processes, whereby species migrate according to their concentration gradients~\cite{cross1993pattern}. Turing referred to these chemical species as \textit{morphogens}, emphasizing their ability to generate spatial structures through the dynamics of systems undergoing instabilities and bifurcations.

Considering morphogens $U_i$ whose concentrations are described in space and time, the general form of the anisotropic reaction-diffusion equations reads:
\begin{align}
	\frac{\partial U_i}{\partial t} = d_i\, \nabla \cdot (\underline{\underline{\mathbf{D}}} \nabla U_i) + f_i(U_1, \ldots, U_n),
\end{align}

\begin{figure}[H]
	\centering
	\includegraphics[width = 1\textwidth]{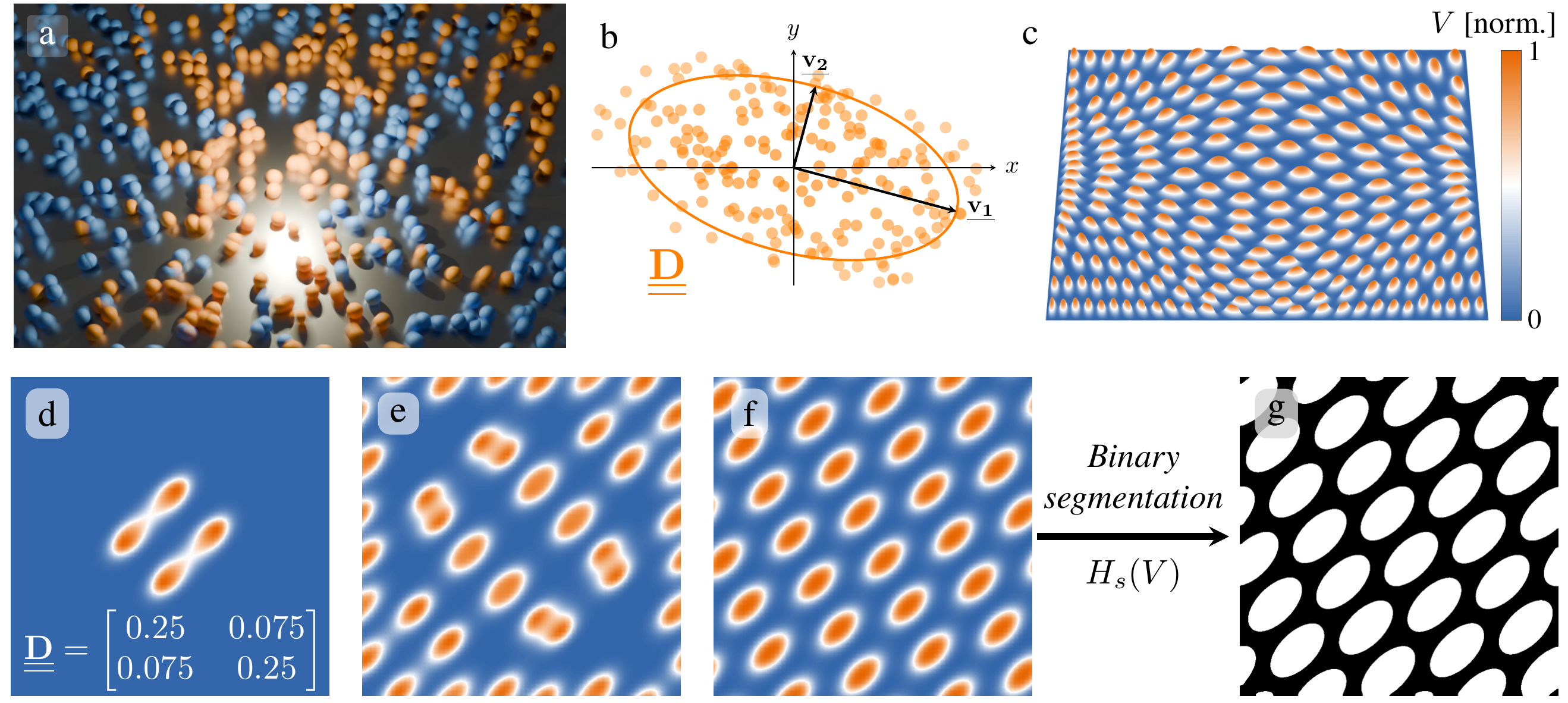}
	\caption{\color{blue} Generative metamaterial model based on anisotropic reaction-diffusion. We propose a self-organized generative model inspired by morphogenesis. (a) This process simulates the interaction between two chemical species, or morphogens, capable of reacting and converting into one another. (b) These morphogens also undergo anisotropic diffusion, locally governed by a tensor $\underline{\underline{\mathbf{D}}}$ that defines preferred directions of migration. (c) This model enables the emergence of self-organized patterns shaped by purely local constraints on type, size, and orientation. Emergence of a self-organized periodic pattern: (d--g) The choice of reaction and diffusion parameters enables the growth of elliptical patterns in the spatial concentration of morphogens and their segmentation into a porous microstructure. \textcolor{reveditor}{Panels (c--f) share the same colormap encoding the normalized morphogen concentration $V$ [norm.] (colorbar shown in (c)). Panel (g) shows the binary segmentation obtained by thresholding $V$ at a fixed value $s$.}}
	\label{fig:RD_croissance}
\end{figure}

In this equation, $U_i$ denotes the local concentration of the $i$-th species (or morphogen), $d_i$ is a coefficient controlling the diffusion intensity for each species, and $\underline{\underline{\mathbf{D}}}$ is a diffusion tensor (a symmetric positive-definite matrix) that encodes the local anisotropy of transport and whose orientation may vary in space~\cite{witkin1991reaction}. The term $\nabla \cdot (\underline{\underline{\mathbf{D}}} \nabla U_i)$ thus models anisotropic diffusion, that is, the tendency of species to migrate more easily along preferred directions. The term $f_i(U_1, ..., U_n)$ encompasses all local interactions (reactions) between the various species. The combination of reaction mechanisms with anisotropic diffusion ultimately enables the emergence of self-organized patterns (Fig.~\ref{fig:RD_croissance}).

Particular \textcolor{black}{clarification on} the notational conventions used throughout this work \textcolor{black}{is provided} to ensure clarity in the algebraic developments of continuum mechanics. Scalar quantities are written in regular font (e.g., $d_i$), while all tensorial quantities are typeset in bold. Among them, vectors are denoted with a single underscore, such as $\bm{\underline{\sigma}}$, second-order tensors with a double underscore, such as $\mathbf{\underline{\underline{D}}}$, and fourth-order tensors used here to represent elasticity law are written using blackboard bold font, for instance $\mathbb{C}$. These conventions are intended to facilitate the tracking of fourth-order tensors when reformulated into matrix form.

To provide a demonstration that is both simple and tractable, the generative model is reduced to the interaction of only two morphogens, $U$ and $V$, following the reaction $1U + 2V \rightarrow 3V$. This specific case corresponds to the Gray--Scott model~\cite{gray1983autocatalytic}, generalized here to include anisotropic diffusion:
\begin{align}
	\frac{\partial U}{\partial t} & = d_u \, \nabla \cdot \left( \underline{\underline{\mathbf{D}}} \nabla U \right) - UV^2 + f(1 - U) \\
	\frac{\partial V}{\partial t} & = d_v \, \nabla \cdot \left( \underline{\underline{\mathbf{D}}} \nabla V \right) + UV^2 - (f + k)V
\end{align}

The diffusion coefficients $d_u$ and $d_v$ control the characteristic size of the generated patterns, while the parameters $f$ (feed) and $k$ (kill) determine their nature (spots, labyrinths, etc.). The tensor $\underline{\underline{\mathbf{D}}}$ imposes the anisotropy and orientation of the resulting structures. Finally, the continuous concentration fields are converted into monophase porous structures by applying a simple thresholding operation to the distributions (Fig.~\ref{fig:RD_croissance}). \textcolor{blue}{A detailed analysis based on a linearization of the model is presented in the \textit{Supplementary Information}, which justifies the link between diffusion and the spatial frequency of the emerging patterns~\cite{murray2007mathematicalII, krause2021modern}.}

This generative model \textcolor{blue}{yields stable} patterns that self-organize in accordance with the imposed local constraints. To demonstrate the generation of mechanical structures with locally controlled elastic properties, it is first necessary to construct databases that link the desired physical properties to the local morphogenetic parameters governing the structuring process.

\subsection*{Homogenization and Mechanical Property Mapping}

Building on our previous work in electromagnetism~\cite{fromenteze2023morphogenetic}, we extend the morphogenetic approach to the framework of static continuum mechanics. The objective is to demonstrate the ability of this formalism to locally synthesize the desired stiffness tensors, thereby enabling control over the stress--strain relationships.

To connect morphogenetic parameters with effective properties, we construct a structured database of stiffness tensors $\mathbb{C}$, obtained for a wide range of microstructures generated by varying the parameters of the anisotropic reaction-diffusion model. Each structure is characterized by a controlled porosity level, anisotropy, and inclusion orientations computed over a periodic domain, allowing homogenization of the mechanical properties and mapping of morphogenetic parameter sets to unique effective tensors.

The modeling of mechanical phenomena involves the manipulation of high-order tensors. In two dimensions, the strain $\bm{\underline{\underline{\epsilon}}}$ and stress $\bm{\underline{\underline{\sigma}}}$ fields are second-order tensors. The linear constitutive law is written as
\begin{align}
	\bm{\underline{\underline{\sigma}}} = \mathbb{C} : \bm{\underline{\underline{\epsilon}}},
\end{align}
where $\mathbb{C}$ denotes the fourth-order elasticity tensor. Although the physical quantities involved in mechanics are generally intuitive, the direct handling of such mathematical objects is often cumbersome. To make this formalism more tractable for numerical implementation, we adopt the Mandel--Kelvin representation, which involves vectorizing the second-order tensors $\bm{\underline{\underline{\sigma}}}$ and $\bm{\underline{\underline{\epsilon}}}$, thereby reformulating the elasticity tensor as a matrix linking these quantities. The constitutive relation then becomes
\begin{align}
	\bm{\underline{\sigma}} = \mathbf{\underline{\underline{C}}}\, \bm{\underline{\epsilon}},
\end{align}
where $\bm{\underline{\epsilon}}$ and $\bm{\underline{\sigma}}$ denote the vectorized forms of the strain and stress fields, and $\mathbf{\underline{\underline{C}}}$ is the equivalent stiffness matrix in Mandel--Kelvin notation. This change of basis preserves all physical symmetries while simplifying numerical manipulation and mechanical property optimization. \textcolor{blue}{While} Voigt notation is common in the literature for such problems, we favor the Mandel--Kelvin convention here, as it guarantees the self-adjointness of tensors in matrix form. A detailed definition of the algebraic quantities used is provided in the Methods section of this article, and in a developed version in the \textit{Supplementary Information}.

\begin{figure}[H]
	\centering
	\includegraphics[width = 0.65\linewidth]{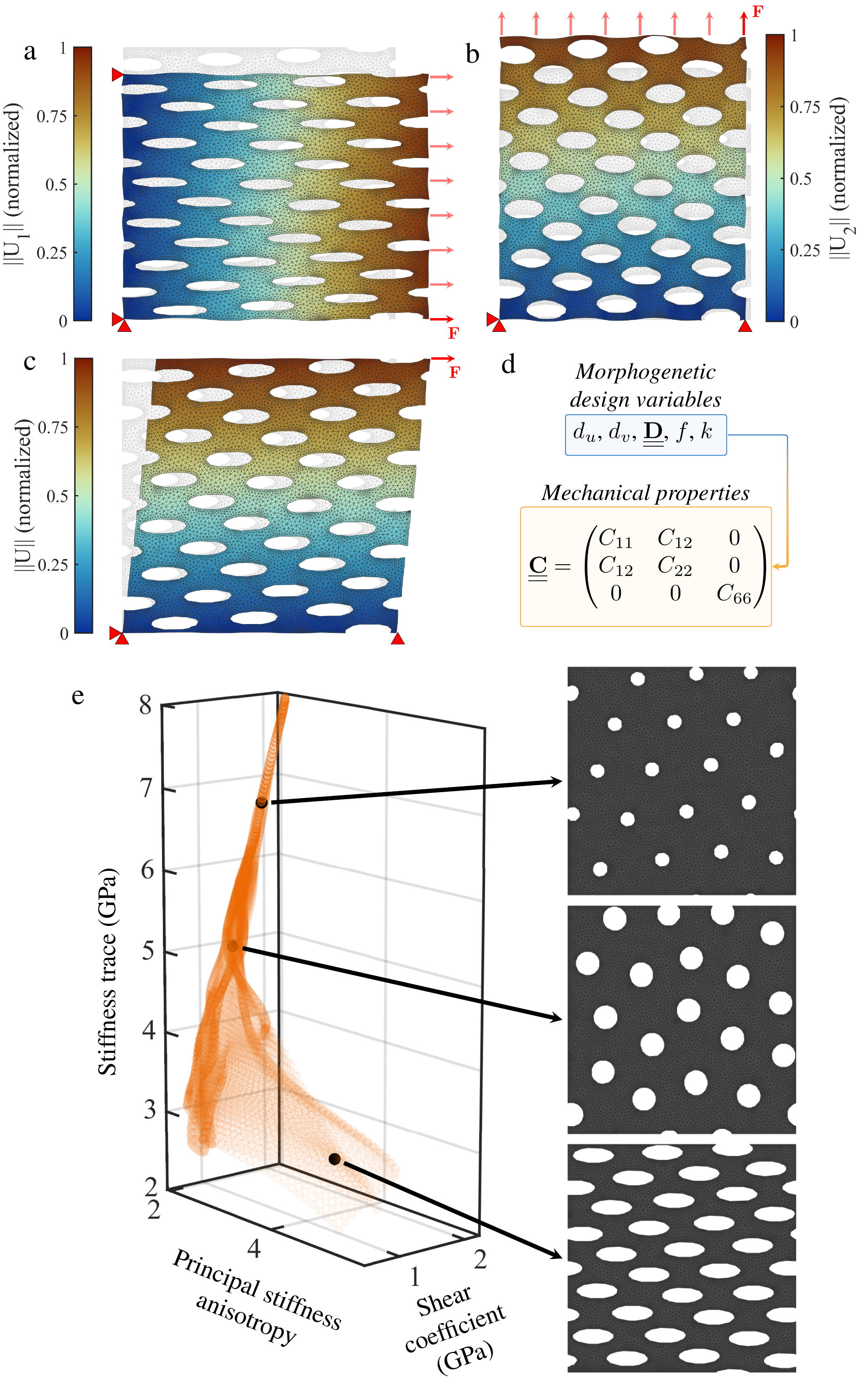}
	\caption{ \color{blue} Extraction of effective elastic properties from periodic supercells. (a) Horizontal uniaxial tension, (b) vertical uniaxial tension, and (c) simple shear are applied in finite-element simulations with periodic boundary conditions.
		(d) The homogenized stiffness is assembled in Mandel--Kelvin form under an orthotropic assumption.
		(e) Design space of the effective stiffness: defined by the principal normal anisotropy (ratio of the two largest eigenvalues, associated with normal modes), the shear eigenvalue $\lambda_3 = 2C_{66}$, and the stiffness trace $\mathrm{Tr}(\mathbf{\underline{\underline{C}}})$. Periodic geometries associated with three black markers are depicted. \textcolor{reveditor}{Periodic supercells are generated on a $101\times101$ pixel grid. The microstructures shown in (a--d) are scale-invariant and not associated with specific physical dimensions.} Source data are provided as a Source Data file. }
	\label{fig:Caracterisation}
\end{figure}

To establish a direct link between the morphogenetic parameters controlling growth and the effective mechanical properties of the resulting structures, a database was generated by combining morphogenetic and mechanical simulations (Fig.~\ref{fig:Caracterisation}). \textcolor{revtwo}{All mechanical field visualizations presented in this Article and in the Supplementary Information employ the perceptually uniform Roma colormap \cite{crameri2020misuse}, selected to ensure accessibility for readers with color vision deficiencies.}

The finite-difference generative model naturally produces 2D periodic patterns, whose type and characteristic size are determined by the reaction parameters ($f$, $k$) and the diffusion coefficients ($d_u$, $d_v$), respectively, which are kept constant in this study. The anisotropy, orientation, and porosity of the generated structures are controlled by the choice of the diffusion tensor $\underline{\underline{\mathbf{D}}}$ and the binarization threshold $s$, both of which directly influence the elastic coefficients obtained after homogenization. Each simulated periodic cell contains identical, aligned elliptical inclusions, allowing the stiffness tensor to be simplified into an orthotropic model. Three distinct mechanical simulations are systematically performed on each cell to accurately extract the corresponding average mechanical properties.

Anticipating the fabrication of prototypes by machining polyoxymethylene (POM) plates, the samples were simulated assuming a Young's modulus of $E = 3$~GPa and a Poisson's ratio of $\nu = 0.35$ (Young modulus was confirmed by tensile tests using an extensometer according the NF EN ISO 527-2 standard). The intermediate steps linking the simulation outputs to the coefficients of the orthotropic stiffness matrices $\mathbf{\underline{\underline{C}}}$ are detailed in the Methods section. Figure~\ref{fig:Caracterisation}e illustrates the range of achievable mechanical parameters with this generative model. \textcolor{revtwo}{The three axes are defined with combinations of the eigenvalues $\lambda_1$, $\lambda_2$, $\lambda_3$ of the $3\times3$ $\mathbf{\underline{\underline{C}}}$-matrices, which are strictly invariant under coordinate transformation.} \textcolor{blue}{The} overall stiffness is directly related to the porosity level of the \textcolor{blue}{patterns}. This quantity is represented by the stiffness trace, corresponding to the sum of the diagonal elements of each orthotropic matrix $\mathbf{\underline{\underline{C}}}$\textcolor{revtwo}{, equal to $\lambda_1 + \lambda_2 + \lambda_3$ and spanning $[2.31,\;7.95]$~GPa across the database}. The degree of anisotropy is captured by the balance between the first two eigenvalues of $\mathbf{\underline{\underline{C}}}$, which directly depends on the aspect ratio of the elliptical inclusions\textcolor{revtwo}{; the ratio $\lambda_1/\lambda_2$ ranges from $1.93$ to $5.72$}. Naturally, the porosity level also affects anisotropy, as small inclusions relative to the supercell size have only a marginal influence on the effective elastic response. Finally, the shear behavior corresponds to the \textcolor{revtwo}{third eigenvalue $\lambda_3 = 2C_{66}$, spanning $[0.47,\;2.24]$~GPa}. \textcolor{blue}{A more comprehensive visualization of the achievable mechanical properties and their dependence on generative parameters is provided in the \textit{Supplementary Information}.}

Each sample within this mechanical parameter space is directly associated with a specific set of morphogenetic coefficients. This enables the generation of larger and more complex structures with locally tailored mechanical properties. These characteristics result from the progressive self-organization of morphogens, which adapt to the locally imposed morphogenetic constraints defined by carefully selected spatial distributions of control parameters. In contrast to other approaches in the scientific literature~\cite{buckmann2015mechanical,wang2022mechanical}, this work demonstrates the ability to generate structures without any requirement for order or multi-scale organization of elementary motifs, which confers strong resilience to fabrication variability~\cite{aage2017giga}.

A demonstration of mechanical cloaking is presented in the following section to illustrate the ability of the proposed approach to independently control local stiffness and anisotropy.

\newpage

\subsection*{Design and Implementation of Mechanical Cloaking}

\begin{figure}[H]
	\centering
	\includegraphics[width = 0.7\linewidth]{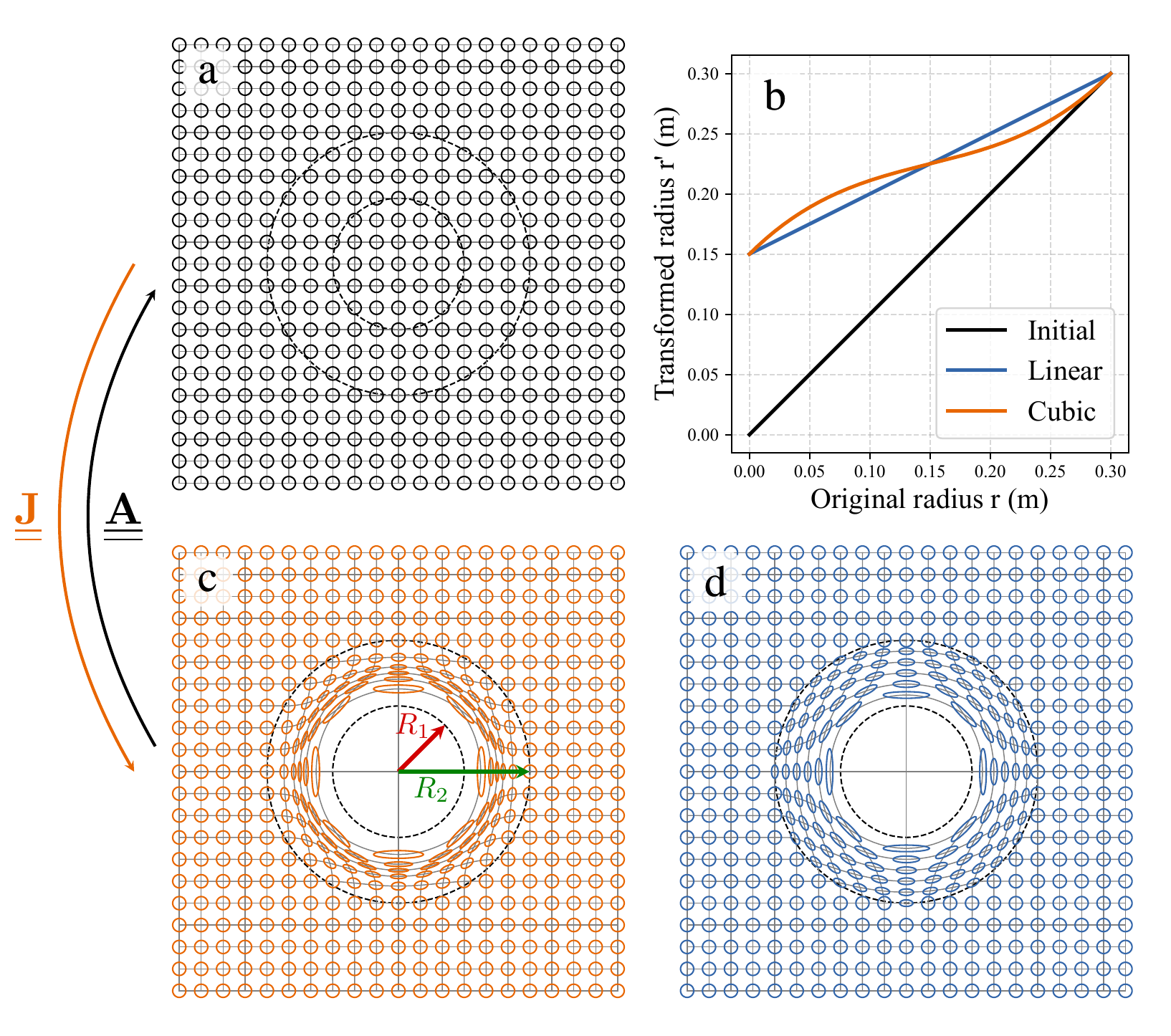}
	\caption{Design of a mechanical cloak. \textcolor{reveditor}{(a)} An initially isotropic domain is transformed around a central cavity, with the cloak defined between radii $R_1$ and $R_2$. Following~\cite{kadic2020elastodynamic}, a cubic mapping (c) provides smoother transitions compared to the linear reference case (d), as illustrated in (b): the cubic transformation concentrates material near $R_1$ more gradually, reducing the singularity of required tensor values at the inner boundary. Applying the inverse Jacobian $\mathbf{\underline{\underline{A}}} = \mathbf{\underline{\underline{J}}}^{-1}$ maps the deformed virtual space back to the physical domain, yielding anisotropic and spatially varying elasticity tensors that guide stress around the cavity. \textcolor{reveditor}{For the experimental case: the studied domain is the $130 \times 130$~mm$^2$ central square region of the $190 \times 130$~mm$^2$ plate; inner cavity radius $R_1 = 19.5$~mm; outer cloak radius $R_2 = 2R_1 = 39$~mm.}}
	\label{fig:Transformation}
\end{figure}

The concept of cloaking was originally introduced in the context of electromagnetism~\cite{pendry2006controlling, schurig2006metamaterial}, \textcolor{black}{redirecting} incident wavefronts around a target object \textcolor{black}{to render} it invisible. This approach relies on defining a spatial transformation that bends space around the region to be concealed (Fig.~\ref{fig:Transformation}).

We consider a spatial transformation that maps each point with coordinates $(x, y)$ in the initial space to a point $(u, v)$ in the transformed space. The Jacobian of this transformation is then written as
\begin{align}
	\mathbf{\underline{\underline{J}}} =
	\begin{pmatrix}
		\frac{\partial u}{\partial x} & \frac{\partial u}{\partial y} \\
		\frac{\partial v}{\partial x} & \frac{\partial v}{\partial y}
	\end{pmatrix},
\end{align}
and locally describes the stretching and rotation induced by the transformation at each point.


To ensure that the displacement and stress fields outside the cloaking region remain unperturbed, the design of a mechanical cloak relies on a spatial transformation that deforms the reference domain so as to redirect stress trajectories around the obstacle. In this transformed space, the material remains homogeneous and isotropic, but the geometry is described by a spatially varying metric tensor that \textcolor{blue}{redirects} the stress trajectories around the cavity. Mapping this configuration back to the undeformed physical domain through the inverse Jacobian $\mathbf{\underline{\underline{A}}} = \mathbf{\underline{\underline{J}}}^{-1}$ \textcolor{blue}{yields} effective stiffness tensors~\cite{milton2007modifications}. \textcolor{blue}{This} transformation prescribes a spatial distribution of anisotropic and inhomogeneous elasticity tensors, denoted $\mathbf{\underline{\underline{C}}}'$, which are related to the original stiffness tensor $\mathbf{\underline{\underline{C}}}$ by:
\begin{equation}
	\mathbf{\underline{\underline{C}}}' = (\det\mathbf{\underline{\underline{A}}})^{-1}\, \mathbf{\underline{\underline{T}}} \, \mathbf{\underline{\underline{C}}} \, \mathbf{\underline{\underline{T}}}^{T},
\end{equation}
where $\mathbf{\underline{\underline{J}}}$ denotes the Jacobian matrix of the spatial transformation, and $\mathbf{\underline{\underline{T}}}$ is the associated transformation matrix, constructed from the components of $\mathbf{\underline{\underline{A}}}$ in accordance with the Mandel--Kelvin formalism. A more detailed derivation of this expression is synthesized in the Methods section, and detailed in the \textit{Supplementary Information}.

The previously constructed database then enables the identification of morphogenetic parameters that promote the local emergence of stiffness tensors closely matching the transformed targets. An experimental demonstration is presented in the next section, based on displacement field measurements under uniaxial loading, to validate the tensorial control capabilities offered by this generative framework.

\subsection*{Validation}

The proposed generative model, inspired by biological morphogenesis, enables the emergence of complex structures that self-organize progressively, without \textcolor{blue}{gradient-based} optimization, toward target elastic properties defined at every point in space. This capability is demonstrated through the following cloaking experiment, inspired by former works~\cite{buckmann2015mechanical,wang2022mechanical}. The reference plate is generated by applying identical generative parameters throughout the domain, starting from an isotropic diffusion tensor $\mathbf{\underline{\underline{D}}}$ and a thresholding parameter $s = 0.8$, \textcolor{blue}{creating} a structure with high porosity.

\begin{figure}[H]
	\centering
	\includegraphics[width = 1\linewidth]{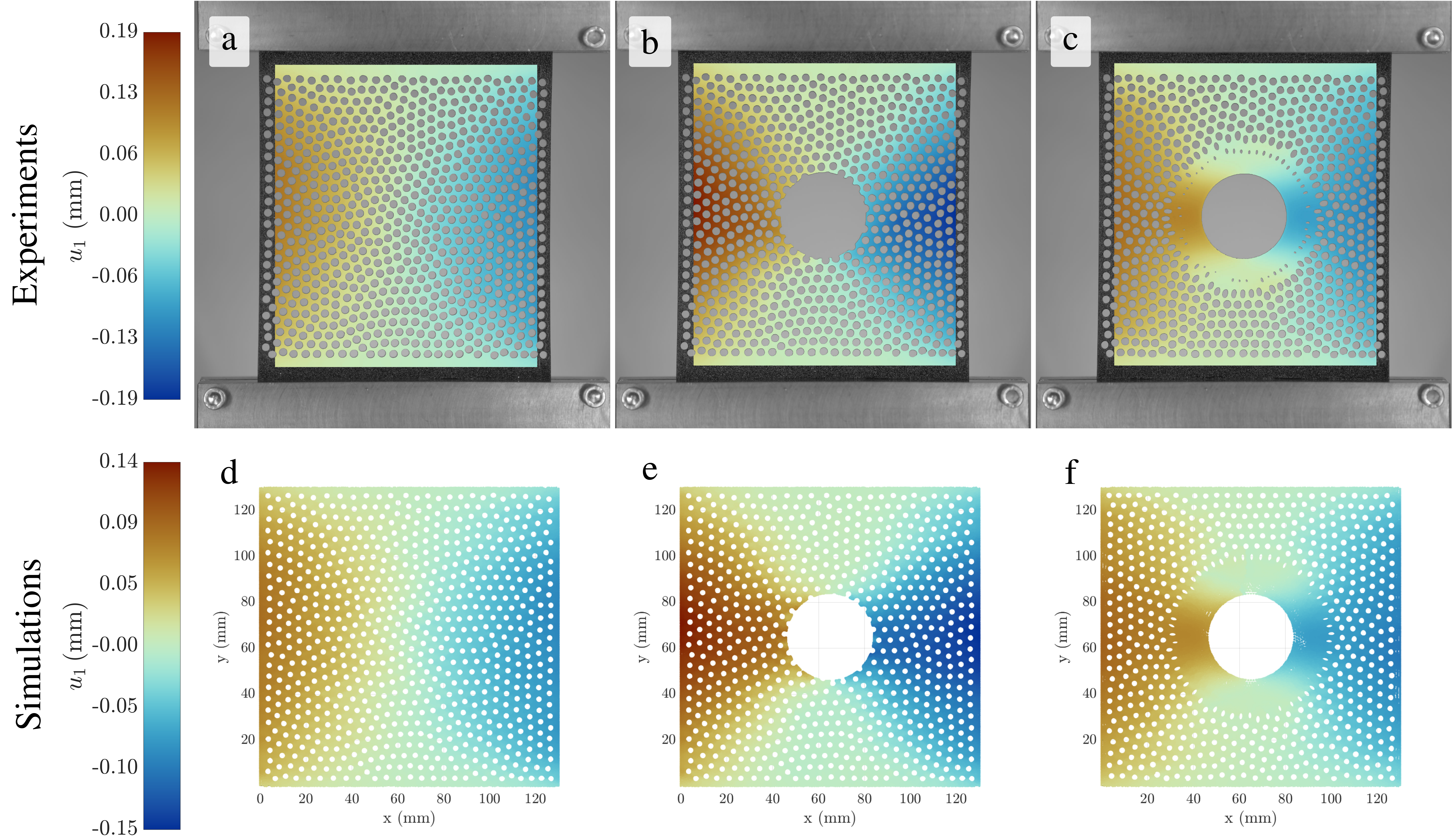}
	\caption{\color{blue} Comparison of experimental and simulated horizontal displacement fields ($u_1$). The experimental results (top row) are presented for the reference plate (a), the plate with a central cavity of radius $R_1$ (b), and the plate with the mechanical cloak of radius $R_2 = 2\,R_1$ (c). The corresponding simulation results (bottom row) are shown for the same configurations (d, e, f). Plate dimensions are $190 \times 130$~mm$^2$.}
	\label{fig:U1}
\end{figure}

\begin{figure}[H]
	\centering
	\includegraphics[width = 1\linewidth]{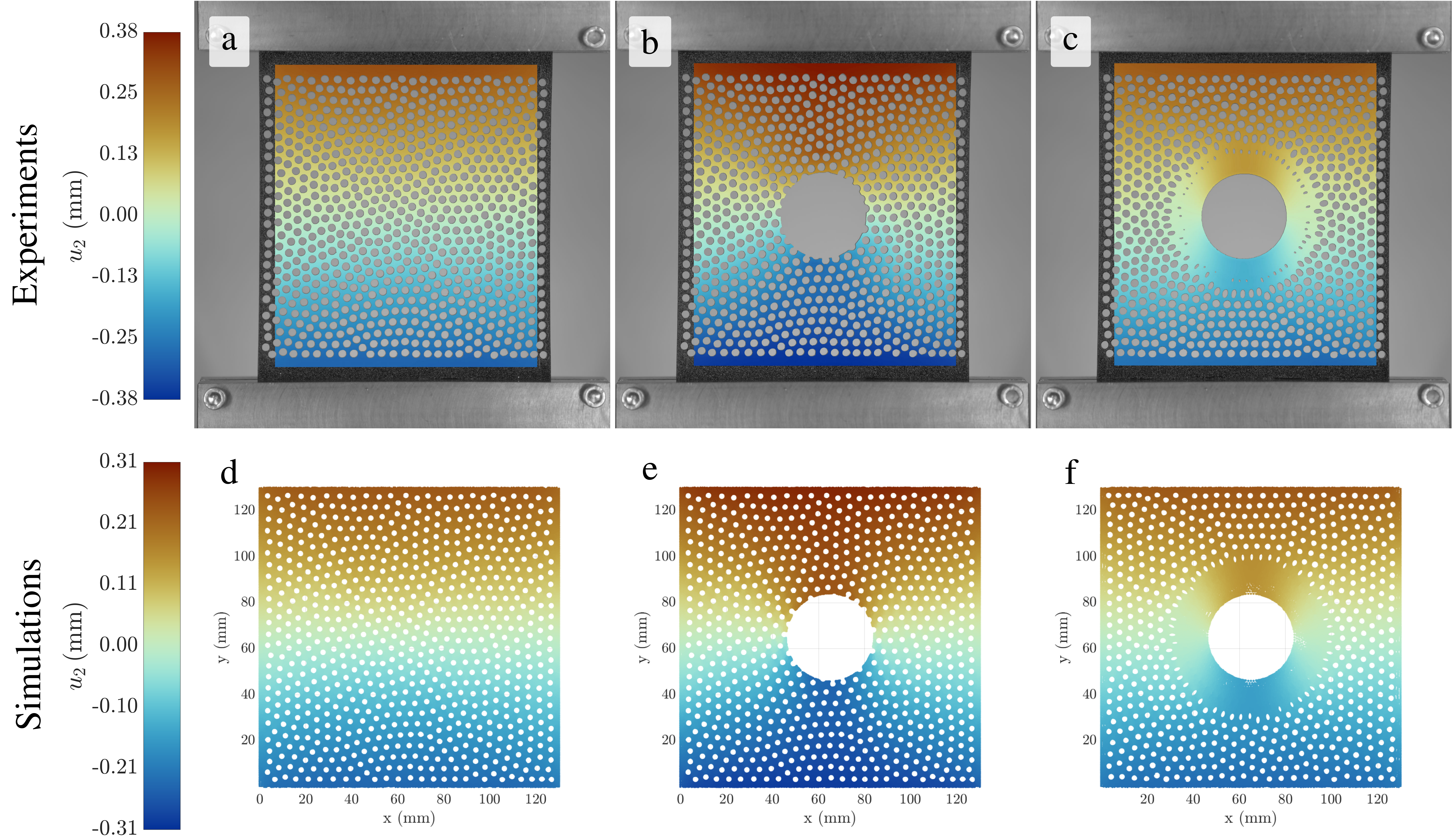}
	\caption{\color{blue} Comparison of experimental and simulated vertical displacement fields ($u_2$). The experimental results (top row) are presented for the reference plate (a), the plate with a central cavity of radius $R_1$ (b), and the plate with the mechanical cloak of radius $R_2 = 2\,R_1$ (c). The corresponding simulation results (bottom row) are shown for the same configurations (d, e, f). Plate dimensions are $190 \times 130$~mm$^2$.}
	\label{fig:U2}
\end{figure}

\textcolor{blue}{The outcomes of this validation process are summarized in Figs.~\ref{fig:U1} and \ref{fig:U2}, which respectively present the horizontal ($u_1$) and vertical ($u_2$) displacement fields. Each figure provides a direct comparison between experimental measurements (top row) and numerical predictions (bottom row) across the three studied configurations: the homogeneous reference plate, the plate with a central cavity of radius $R_1$, and the cloaked specimen with a functionalized annulus up to $R_2 = 2\,R_1$.}

\textcolor{blue}{Qualitatively, the experimental results demonstrate the effectiveness of the mechanical cloak. In the absence of cloaking, the cavity induces significant distortions in the displacement fields (Figs.~\ref{fig:U1}b and \ref{fig:U2}b), which are characteristic of stress concentration. By applying the spatial transformation within the annular region, these disturbances are largely mitigated, and the displacement fields outside the cloak (Figs.~\ref{fig:U1}c and \ref{fig:U2}c) are restored to a state closely resembling the continuous reference case (Figs.~\ref{fig:U1}a and \ref{fig:U2}a).}

\textcolor{blue}{To quantify this performance, we employ an efficiency metric defined as the relative deviation between the displacement fields of the reference plate, assumed isotropic and homogeneous, and those measured in the tested configuration:}
\begin{equation}
	\Delta =
	\frac{\left\| \boldsymbol{\underline{u}} - \boldsymbol{\underline{u}}_0 \right\|_{L^2(\Omega_s)}}
	{\left\| \boldsymbol{\underline{u}}_0 \right\|_{L^2(\Omega_s)}} ,
	\label{eq:metrique}
\end{equation}
\textcolor{blue}{where the evaluation domain $\Omega_s$ is restricted to the region outside the cloak to strictly assess the restoration of the external fields.}

\textcolor{blue}{The effectiveness of the morphogenetic approach in synthesizing controlled tensors is quantitatively confirmed by the close agreement between experimental measurements and 3DEXPERIENCE simulations. Experimentally, the presence of the cloak reduces the relative displacement error from $\Delta_{\text{hole}} = 39.2$~\% to $\Delta_{\text{cloak}} = 10.7$~\%, achieving a $3.66\times$ improvement factor. These values align remarkably well with numerical predictions, where the error decreases from $42.0$~\% to $10.8$~\% ($3.89\times$ reduction). This consistency across both horizontal and vertical fields reflects the robustness of the generative method in mitigating the distortions introduced by the central cavity.}

Despite a minor discrepancy in the peak displacements along both axes, most plausibly attributable to small deviations from ideal testing conditions, \textcolor{blue}{the distribution shows close agreement with experimental measurements}. For instance, the $u_1$ displacement fields in the reference case (Fig.~\ref{fig:U1}a and Fig.~\ref{fig:U1}d) exhibit the same slight asymmetry, which is attributed to an imperfect homogenization coupled to the Poisson effect under tensile loading.

\textcolor{blue}{Benchmarking these results against state-of-the-art mechanical cloaks~\cite{buckmann2015mechanical, wang2022mechanical} highlights the sensitivity of performance metrics to specific test configurations. For transparency, we note that our experimental fields were symmetrized to facilitate comparison with the literature, compensating for the rigid-body motion induced by the lower clamp. While omitting this adjustment would yield more optimistic metrics ($\Delta_{\text{hole}} = 28.1$~\% and $\Delta_{\text{cloak}} = 5.8$~\%, a $4.8\times$ reduction factor), we intentionally report the more conservative values ($\Delta_{\text{hole}} = 39.2$~\% and $\Delta_{\text{cloak}} = 10.7$~\%) associated with the rigid-body motion compensation to align with the standard reference frames of other studies. This choice also reveals the inherent sensitivity of commonly adopted metrics to loading conditions. Despite these conservative metrics and the simplicity of the generated structures, our results demonstrate competitive performance compared to lattice-transformation~\cite{buckmann2015mechanical} and numerically optimized~\cite{wang2022mechanical} architectures, especially considering the strong agreement between experimental and numerical results (Table~\ref{tab:metriques}).}

\newcommand{\twoline}[2]{\begin{tabular}{@{}c@{}}#1\\#2\end{tabular}}

\begin{table}[H]
	\centering
	\setlength{\tabcolsep}{4pt}
	\renewcommand{\arraystretch}{1.10}
	\caption{\textcolor{reveditor}{Cloaking performance across studies. Reported are the relative displacement errors $\Delta$ measured outside the cloak for the holed plate ($\Delta_{\text{hole}}$) and with the cloak ($\Delta_{\text{cloak}}$), the improvement factor $R$ (larger $R$ indicates better cloaking), and the consistency factor $\beta = (R^\text{Sim} - R^\text{Exp}) / R^\text{Exp}$ ($\beta = 0$ indicates perfect agreement between simulation and experiment). “Sim.” and “Exp.” denote simulation and experiment.}}
	\label{tab:metriques}
	\resizebox{\textwidth}{!}{%
	\begin{tabular}{|c c c | c c c c | c|}
		\hline
		Study                                 & \twoline{Method}{Geometry} & \twoline{Load}{Boundary cond.}             & Type &
		$\Delta_{\text{hole}}$ (\%)                    & $\Delta_{\text{cloak}}$ (\%)                          & $R = \displaystyle\frac{\Delta_{\text{hole}}}{\Delta_{\text{cloak}}}$ & \textcolor{reveditor}{$\beta$}                \\
		\hline
		\multirow{2}{*}{\cite{buckmann2015mechanical}} & Lattice transformation                                & Lateral pressure                                                      & Sim.          & 738   & 22   & 33.6 & \multirow{2}{*}{\textcolor{reveditor}{0.22}}\\
		                                               & Double trapezoids                            & Sliding edges                                                & Exp.          & 714   & 26   & 27.5 & \\
		\hline
		\multirow{2}{*}{\cite{wang2022mechanical}}     & Data-driven lattice                                   & Compression                                                           & Sim.          & 17.2  & 4.0  & 4.30 & \multirow{2}{*}{\textcolor{reveditor}{0.94}}\\
		                                               & Pixelated cells                              & Free edges                                                   & Exp.          & 15.16 & 6.86 & 2.21 & \\
		\hline
		\multirow{2}{*}{This work}                     & Morphogenesis                                         & Tension                                                               & Sim.          & 42.0  & 10.8 & 3.89 & \multirow{2}{*}{\textcolor{reveditor}{0.06}}\\
		                                               & Aperiodic ellipses                           & Free edges                                                   & Exp.          & 39.2  & 10.7 & 3.66 & \\
		\hline
	\end{tabular}}%
	\vspace{2pt}
	{\footnotesize For each study, the first row reports simulation (Sim.) results with the method and loading condition. The second row reports experimental (Exp.) results with the geometry type and boundary condition.}
\end{table}

We propose to evaluate, for each case, an improvement factor on the displacement error $R = \Delta_{\text{hole}}/\Delta_{\text{cloak}}$, which serves both to assess the effect of cloaking and to capture fabrication and measurement dispersions when moving from numerical studies to experimental validation. It is again important to note that these metrics cannot be directly compared across studies, which involve different geometries and loading conditions. However, they can be used to emphasize the robustness of the proposed approaches by confronting numerical and experimental results~(Table~\ref{tab:metriques}).

Computing the parameter $\beta  = (R^\text{Sim} - R^\text{Exp}) / R^\text{Exp}$, which represents the relative bias on the improvement factor $R$, pinpoints the very good agreement between simulations and measurements in this work. This advantage is notably linked to the simplicity of the generated elliptical shapes. Unlike reference approaches that rely on locally thinned geometries requiring additive manufacturing technologies~\cite{buckmann2015mechanical, wang2022mechanical}, the single-scale geometries obtained here enabled laser machining. Moreover, the departure from periodic architectures afforded by the self-organizing nature of the proposed technique facilitates homogenization of the synthesized tensors in the absence of periodic boundary conditions~\cite{kumar2020inverse}.

These results highlight the combined effect of local control over stiffness and anisotropy in mitigating the deformations caused by the central cavity. The ability of the system to self-replicate elliptical shapes and self-organize under complex boundary conditions, through the introduction of partial disorder, also enables a remarkably simple experimental demonstration compared to existing mechanical cloaking approaches~\cite{buckmann2014elasto, buckmann2015mechanical,wang2022mechanical}. The quality of this match is primarily attributed to the simplicity of the generated geometries, which do not require pixelated microstructures or thin connecting elements, resulting in lower fabrication variability and less stringent requirements on compatible manufacturing processes. The introduction of partial disorder also enhances robustness with respect to such variability, as there are no dominant symmetry axes that could be broken~\cite{man2013isotropic, froufe2017band, gkantzounis2017hyperuniform, liu2022growth}.

\textcolor{blue}{This capability to independently control local stiffness and anisotropy through a single-scale, self-organized structure offers an alternative to pixelated or lattice-based designs.}

\color{blue}

\subsection*{Generalization across loading scenarios and scales}

\begin{figure}[h!]
	\centering
	\includegraphics[width=0.6\textwidth]{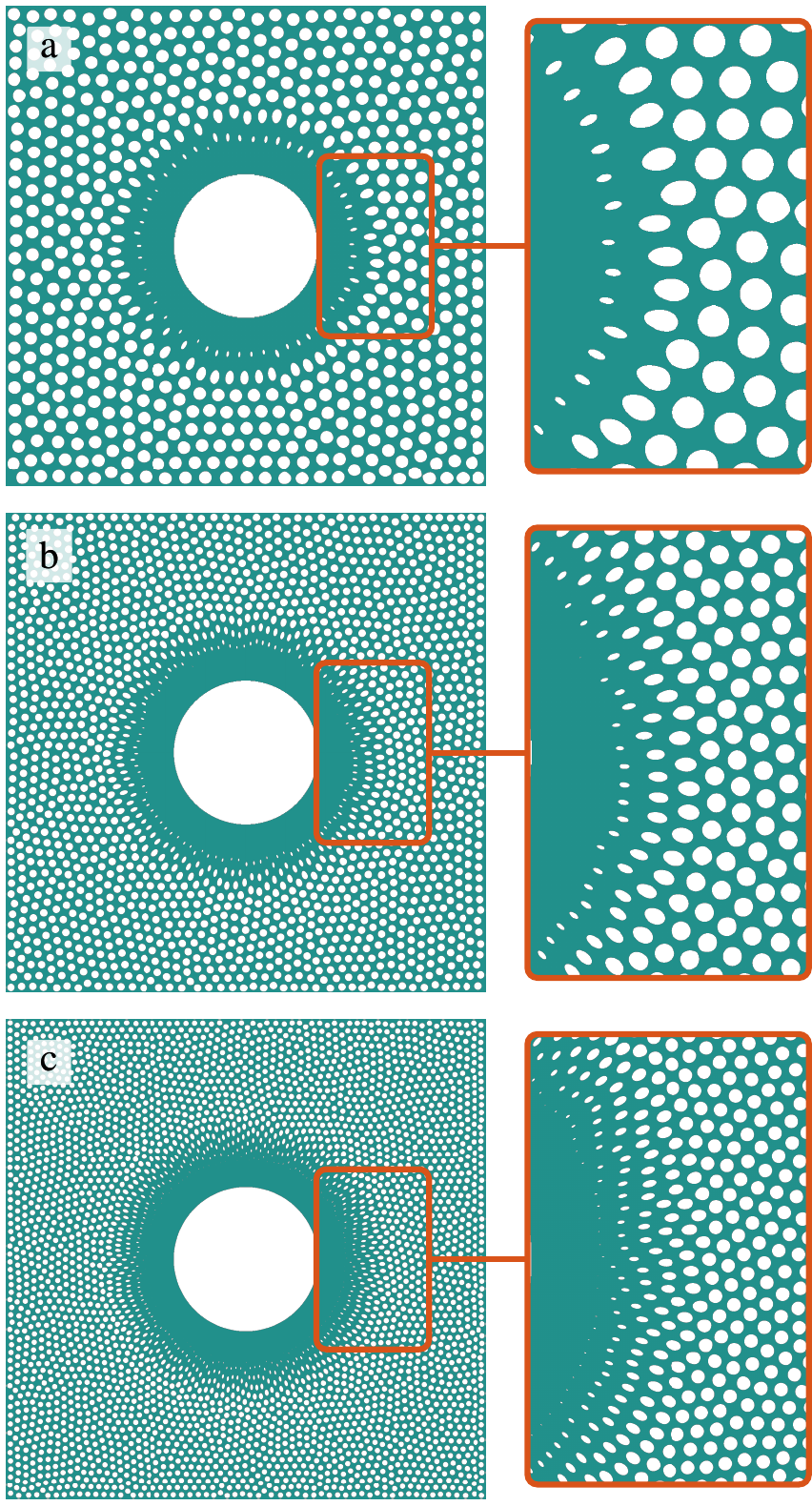}
	\caption{Scalability of the generative model. The generation-domain size can be changed directly, with self-organization preserving morphology as resolution increases (a - $603^2$ pixels, b - $1005^2$ pixels, c - $1407^2$ pixels). These patterns are scale-invariant and not tied to specific physical dimensions.}
	\label{fig:scalable}
\end{figure}

The generative model proposed in this work is inherently scalable. In our experimental demonstration, generations were performed over square domains of $603^2$ pixels, which provide favorable conditions for homogenization while yielding pore sizes suitable for our laser fabrication. To illustrate scalability, Fig.~\ref{fig:scalable} depicts three generations with varying domain sizes, for which we provide animations of the decentralized growth process for the mechanical cloak at varying grid resolutions \textcolor{blue}{(see \textit{Supplementary Information})}.

For completeness, we also report generation times excluding visualization for four square domains: $201^2$, $603^2$, $1005^2$, and $1407^2$ pixels.  The corresponding times were 71.3 s, 144.3 s, 268.2 s, and 488.4 s, measured on a workstation with a 12th-Gen Intel Core i7-12700 (2.10 GHz) CPU, 64 GB RAM, and an NVIDIA GeForce RTX 3090 GPU. A least-squares fit of $t(P)=t_0+aP$ with $P=N^2$ (number of grid points) yields $t_0=61.96$ s and $a=2.1347\times10^{-4}$ s/pixel with $R^2=0.9987$, consistent with near-linear scaling and a constant overhead from initialization. For very small domains, executing fully on the GPU and minimizing one-off costs can markedly reduce the constant term $t_0$. However, the most relevant regime for this work is large-scale generation, where the linear-in-$P$ behavior dominates.

Because all updates arise from pointwise reactions and short-range diffusions using fixed-size stencils, the algorithmic cost per iteration scales as $\mathcal{O}(P)$ with $P=N^2$ grid points. This locality avoids repeated global solves and naturally supports parallelization, which in turn preserves scalability as the design domain and resolution grow.

\begin{figure}[H]
	\centering
	\includegraphics[width = 1\textwidth]{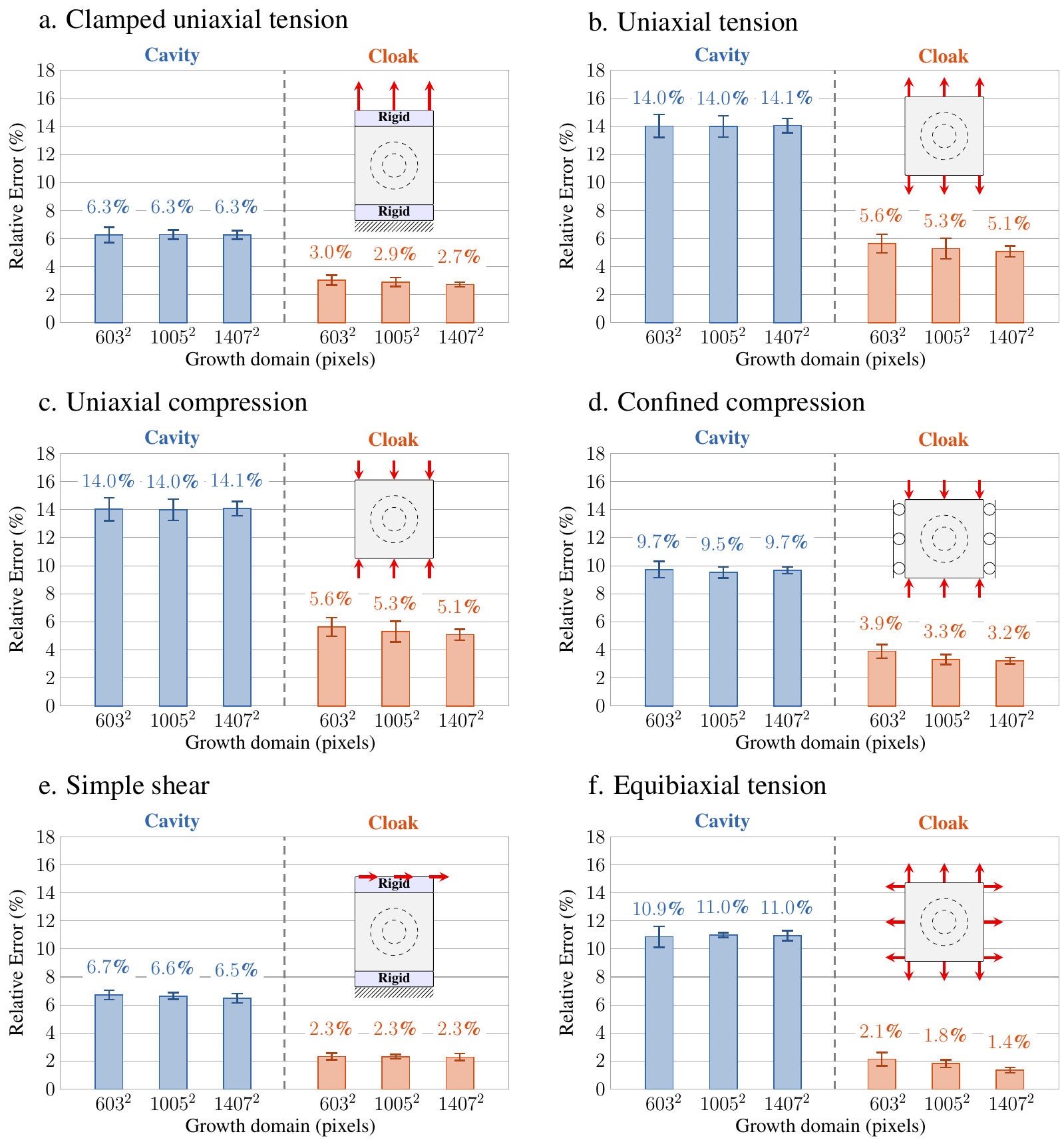}
	\caption{\color{blue} Cloaking performance across six loading modes and three structural resolutions. The assessed cases cover tension (a, b, f), compression (c, d), and shear (e) loading conditions. Bar charts compare the relative displacement error $\Delta$ for the reference (no cavity), the holed plate, and the cloaked plate, for three growth domains of $603^2$, $1005^2$, and $1407^2$ pixels and 10 independent random seeds. Error bars represent $\pm$ one standard deviation ($n = 10$ independent seeds). All simulations are performed by imposing displacements represented by the red arrows on each test type. The metrics presented here are computed without top and bottom padding zones around the structures, justifying the differences in the case of clamped uniaxial tension compared to the previous studies. Source data are provided as a Source Data file.}
	\label{fig:Simulations_extra}
\end{figure}

To demonstrate the versatility of our generative model coupled with the proposed transformational methods, we present results for six distinct loading cases (Fig.~\ref{fig:Simulations_extra}). A key challenge in this context is to generate a single structure capable of acting as a cloak under multiple loading scenarios. In the linear elastic regime considered here, the stiffness matrix depends solely on the geometry and material properties, implying that the response in tension and compression should be symmetric. Consequently, we rely on a single database of elasticity tensors determined under periodic boundary conditions in tension (see \textit{Supplementary Information} for a comparison with compression). The generated structure can thus be conserved for all loading modes. These simulations were performed using the Abaqus mechanical solver, employing a 2D tetrahedral mesh under a plane stress assumption.

To evaluate the evolution of cloaking performance with respect to structural resolution, tests were conducted on growth domains of $603^2$, $1005^2$, and $1407^2$ pixels, respectively. To further demonstrate the robustness of our generative models to initial growth conditions, 10 different random seeds were used for each domain size, resulting in the emergence of structures with distinct, self-organized pore distributions. For each domain size and random seed, three structures (reference, cavity, and cloak) were generated to determine the relative error of the displacement fields, revealing cloaking performance in a manner analogous to the numerical and experimental tests presented previously.

The results first attest to the effectiveness of this cloaking method across various mechanical loads, achieving a systematic reduction in the mean relative error by a factor of 2 to 8 depending on the case. Considering plates of identical physical dimensions for all simulations, increasing the growth domain size jointly increases the number and decreases the size of the generated pores (Fig.~\ref{fig:scalable}). This increased resolution has no notable effect on the relative displacement error for the cavity case but generally leads to better cloaking performance, converging toward displacement fields closer to those of the reference without a central cavity. These results reflect an improved capability to synthesize the desired elasticity tensors at a local scale. Indeed, the method relies on a database of elasticity tensors determined under periodic boundary conditions (Fig.~\ref{fig:Caracterisation}). The increased resolution, combined with the self-organizing capabilities of our generative model, ensures lower perturbation from neighboring elements with different geometries and orientations.

To demonstrate the robustness of the proposed generative model, each triplet of structures was generated 10 times from 10 random seeds, yielding plates with similar appearance but systematically different self-organized pore distributions. The results presented in Fig.~\ref{fig:Simulations_extra} are determined by averaging the relative errors for each triplet, with error bars representing $\pm$ one standard deviation. This statistical study reveals both the low dependence of the results on the initial generation conditions and a reduction in variance with increasing growth size for several loading modes.

It is important to note a specific distinction regarding the clamped uniaxial tension case. The experimental setup involves spatial padding at the top and bottom of the structure to accommodate the grips. This padding is not included in the equivalent simulations, where all domains are simply square with specified dimensions. This difference leads to relative errors that differ from the experimental case presented earlier, despite similar loading conditions. This discrepancy highlights a crucial point regarding the positioning of this work within the scientific literature: while loading conditions influence cloaking metrics, the shape of the simulated domain and the presence of significant non-porous zones can also greatly influence the results. Beyond a quantitative and absolute positioning, which seems difficult given the lack of real standardization, the extension of this work to other loading configurations nevertheless demonstrates a systematic reduction in relative errors, validating the effectiveness of cloaking in each studied case.

Beyond the flexibility of loading conditions, these simulations emphasize two central properties of this generative model: the results are robust, with a low impact of initial generation conditions, and scalable, with native adaptation to larger domains, which simultaneously improves homogenization conditions and cloaking efficiency in most presented cases.
\textcolor{revtwo}{The corresponding improvement factors $R = \Delta_{\text{hole}} / \Delta_{\text{cloak}}$ for all six configurations and three growth domain resolutions are reported in Supplementary Table~1.}

\subsection*{One-shot stress-guided morphogenetic growth}

While the cloaking application demonstrates the ability to grow structures fulfilling a specific kinematic transformation, the generative model can also be harnessed for structural optimization. In contrast to classical topology optimization methods, which typically rely on iterative processes involving repeated evaluations of the mechanical equilibrium, we propose a "one-shot" approach capable of growing optimized structures in seconds. This method is particularly advantageous for high fill-fraction applications (here volume fraction $V_f = 80\%$). In this regime, classical topology optimization typically converges toward bulk solids, effectively erasing the microstructure. By contrast, the morphogenetic approach enforces a characteristic length scale determined by the reaction-diffusion wavelength, guaranteeing the persistence of a structured, load-bearing architecture even at high densities. Detailed analysis is provided in the \textit{Supplementary Information}, developing the link between the generation parameters and the periodicity of the emergent geometries.

The optimization relies on a single initial simulation of a homogeneous, isotropic domain under the target load to extract the mechanical blueprint: the stress tensor field $\bm{\underline{\underline{\sigma}}}$. Guided by this field, the morphogenetic growth is modulated to minimize the deformation energy $E = \frac{1}{2} \int \bm{\underline{\underline{\sigma}}} : \bm{\underline{\underline{\varepsilon}}} \, dV$. The local density of the structure is first adapted by linking the reaction-diffusion binarization parameter $s$ to the trace of the stress tensor, $\text{Tr}(\bm{\underline{\underline{\sigma}}})$ (see Methods for mapping details). High-stress regions inhibit pore formation, increasing local stiffness, while low-stress regions allow for larger porosity, yielding a graded isotropic structure.
To leverage the full tensorial nature of the stress field, a second level of modulation aligns the local anisotropy of the pores with the principal stress directions. By mapping the anisotropy of the stress tensor to that of the diffusion tensor $\underline{\underline{\mathbf{D}}}$, emerging pores elongate along the principal stress direction, thereby stiffening the structure along critical load paths.

\begin{figure}[H]
	\centering
	\includegraphics[width=1\textwidth]{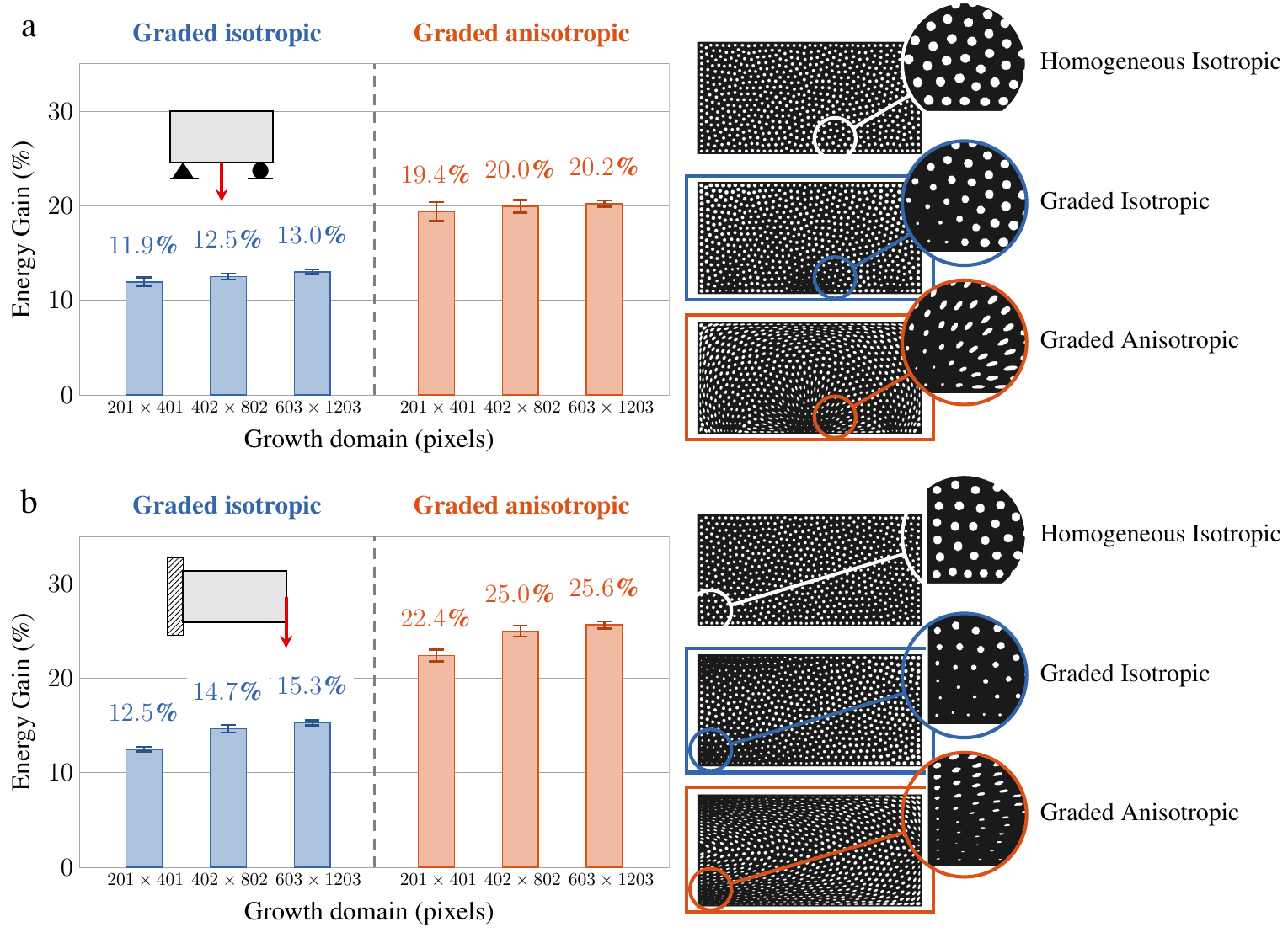}
	\caption{\color{blue} Stress-guided structural optimization for high fill-fraction metamaterials ($V_f = 80\%$). (a) Distributed three-point bending case ("Bridge"): comparison of potential energy reduction for Graded Isotropic (density only) and Graded Anisotropic (density + orientation) structures relative to a uniform reference. The visualized structures correspond to the median performance for the intermediate resolution ($402 \times 802$ px). (b) Cantilever bending case: similar analysis showing up to 25\% energy reduction for the fully guided structures. Data points represent the mean over 10 independent random seeds, and error bars indicate $\pm 1$ standard deviation ($n = 10$). Source data are provided as a Source Data file. }
	\label{fig:optim_2_cases}
\end{figure}

We assessed this approach on two distinct loading cases presented in Fig.~\ref{fig:optim_2_cases}: a distributed three-point bending (bridge-like structure, panel a) and a cantilever bending scenario (panel b).
In both configurations, we compare the mechanical potential energy of the generated structures against a reference uniform isotropic geometry. All structures are constrained to an identical fill fraction of $V_f = 80\%$.
For the bridge case (Fig.~\ref{fig:optim_2_cases}a), the density modulation alone ("Graded Isotropic") yields a significant performance improvement, reducing the deformation energy by over 10\%. Activating the anisotropic guiding ("Graded Anisotropic") provides an additional gain, demonstrating the efficiency of the microstructural alignment.
Similar trends are observed for the cantilever case (Fig.~\ref{fig:optim_2_cases}b), where energy reductions reach up to 25\% for the guided structures. Displacement fields are presented in the \textit{Supplementary Information}.
These results are reported as the average over 10 random generation seeds for an intermediate domain size ($402 \times 802$ pixels), with error bars indicating one standard deviation. The low dispersion confirms the robustness of the process: despite the stochastic nature of morphogenesis, the mechanical performance remains deterministic.
Moreover, the approach is scalable: increasing the domain size leads to finer microstructures that better approximate the target constitutive tensors (see \textit{Supplementary Information} detailed results of varying growth domain size).

The mechanical efficiency of the proposed morphogenetic framework is assessed by positioning our results against established topology optimization and inverse design strategies. A critical distinction lies in the targeted volume fraction regime ($V_f \approx 80\%$) and the computational paradigm employed. Standard topology optimization methods targeting bone-like infills, such as those proposed by Wu \textit{et al.}~\cite{wu2017infill}, rely on iterative finite element loops to strictly minimize compliance. While these approaches define the theoretical upper bound for stiffness maximization, they incur significant computational costs. In contrast, our stress-guided growth achieves a 25\% reduction in strain energy (relative to an isotropic baseline of equivalent mass) through a single-shot generative process. Although this performance remains bounded by the mathematical optimum achievable via iterative solvers, it offers a near-instantaneous approximation of the optimal material distribution. Furthermore, compared to data-driven inverse design frameworks for spinodoid metamaterials developed by Kumar \textit{et al.}~\cite{kumar2020inverse}, which leverage deep learning to map target tensors to microstructural parameters, our approach operates without the training overhead of neural networks. While Kumar \textit{et al.} demonstrate precise anisotropy control primarily in medium-density regimes ($V_f \in [0.3, 0.5]$), our method successfully induces strong functional anisotropy within a high-density quasi-solid medium ($V_f \geq 0.8$). This capability to micro-structure the bulk material allows for significant stress flux redirection and stiffness enhancement while bypassing the complexity of black-box training or heavy iterative optimization.

\color{black}

\section*{Discussion}

This demonstration highlights the \textcolor{blue}{synthesis of} a database linking a morphogenetic design parameter space to \textcolor{blue}{elasticity tensors with tailored} stiffness and anisotropy. Leveraging this generative model enabled the design of a mechanical invisibility cloak \textcolor{blue}{to reduce} the effect of a circular cavity in a \textcolor{blue}{plate} under tensile loading. The simplicity of the generated geometries, free from any imposed regularity, \textcolor{blue}{yields} excellent agreement between the anticipated deformation reduction in simulation and experimental results. \color{blue}
Beyond the tensile cloak, the extension to six distinct loading scenarios demonstrates three properties of the approach: scalability, versatility, and robustness.

Scalability is inherent to the decentralized nature of the generative model. \textcolor{revtwo}{A distinguishing feature of our method is its departure from classical optimization formulations. Rather than defining a global objective function iteratively coupled to a mechanical equilibrium solver, our morphogenetic approach operates as a single-pass forward generative model. It is designed as a scalable and autonomous alternative to centralized optimization methods, circumventing the need for iterative global evaluations of mechanical performance during the generative process.} All updates result from pointwise reactions and short-range diffusions, without global adjoint solves or centralized gradient loops. As a result, the end-to-end computational complexity scales linearly with the number of grid points, is trivially parallelizable, and remains robust when the design domain and resolution grow. This stands in sharp contrast to topology-optimization pipelines (and data-driven variants) that repeatedly solve global equilibrium problems and accumulate design variables with mesh refinement, which can limit scalability and sensitivity to fabrication variability.

Robustness arises from the introduction of partial disorder and the self-organizing properties of the reaction-diffusion engine. The system adapts to arbitrary boundaries and heterogeneous targets through spatially varying diffusion tensors and thresholds, letting microstructures self-organize under purely local prescriptions. This self-organization facilitates the synthesis of orthotropic elasticity tensors with prescribed eigenstructure, without imposing periodicity, thin struts, or pixelated unit cells. In practice, this improves manufacturing robustness: the patterns self-organize with partial disorder and avoid the extreme geometric ratios required by pentamode-like or direct lattice-transformation approaches~\cite{fang2022programmable, meng2022deployable}.

Versatility is achieved through the development of two complementary design strategies. First, a transformational method based on an adaptation of Jacobian operators to the Mandel-Kelvin tensor formalism allows a single generated structure to provide effective cloaking across multiple loading scenarios.
Second, a stress-guided generative approach demonstrates the ability to solve structural optimization problems by locally adapting the microstructure to the mechanical demand. The present strategy complements transformation-based cloaks implemented via pentamode lattices, and data-driven unit-cell libraries optimized cell by cell.
Whereas those families rely on highly structured micro-architectures, our approach achieves comparable field steering with purely local updates and minimal geometric constraints. It also connects to recent oriented-infill and bio-inspired growth methods: instead of de-homogenizing an optimized tensor field, it is proposed to grow it. In this regard, it shares the "one-shot" philosophy of recent topology optimization advances, while extending them to functionally graded anisotropic media.
Finally, the locality of the generative law makes the 3D transfer direct. Replacing the 2D diffusion tensor by its 3D counterpart and adopting 3D homogenization yields orthotropic $6\times6$ Mandel–Kelvin stiffness matrices with identical control handles (principal axes, eigenvalues, and porosity). No change of paradigm is required, only voxelization and a 3D database. \textcolor{revtwo}{However, transitioning to physical 3D fabrication poses specific manufacturing challenges. The Turing mechanism naturally generates closed inclusions, which complicate subtractive manufacturing and restrict the use of additive processes relying on resin or powder beds (e.g., stereolithography or selective laser sintering), as uncured material would remain trapped within the bulk. To overcome this limitation, several avenues are currently being explored, encompassing both the synthesis of modified geometries to enforce pore connectivity, and the evaluation of alternative additive manufacturing technologies.} On that basis, we anticipate applications well beyond axisymmetric cloaks. Delivering these functions will require extending the transformation formalism used here, still orthotropy-preserving, but generalized to non-cylindrical mappings and richer target manifolds, which we will develop in subsequent work.

\color{black}

\section*{Methods}

\newcommand{\meth}[1]{\noindent\textbf{#1.}\ }
\newcommand{\chunkgap}{\par\medskip}

\subsection*{Reaction-Diffusion Model and Numerical Implementation}

\meth{Generative model and rationale}
The generation of the proposed mechanical microstructures is based on an anisotropic generative model inspired by the Gray--Scott system~\cite{gray1983autocatalytic}, a realization of the reaction-diffusion framework originally proposed by Alan Turing~\cite{turing1952chemical} to describe morphogenetic mechanisms~\cite{xiang2022turing, paul2024widespread}. In contrast to the classical isotropic formulation, our approach generalizes the model by incorporating an anisotropic diffusion tensor, enabling precise local control over the orientation and shape of the emerging patterns.
\chunkgap

\meth{Governing equations and anisotropic tensor}
The spatiotemporal dynamics of the system are governed by the following pair of reaction-diffusion equations:
\begin{align}
	\frac{\partial U}{\partial t} & = d_u \nabla \cdot (\underline{\underline{\mathbf{D}}} \nabla U) - UV^2 + f(1-U)   \\
	\frac{\partial V}{\partial t} & = d_v \nabla \cdot (\underline{\underline{\mathbf{D}}} \nabla V) + UV^2 - (f + k)V
\end{align}
where $U$ and $V$ denote the local concentrations of two interacting morphogens. The coefficients $d_u$ and $d_v$ control the characteristic spatial scales of the resulting patterns, while $f$ and $k$ determine their morphology (e.g., spots, labyrinths, etc.).

Anisotropic diffusion is modeled using a symmetric positive-definite tensor $\underline{\underline{\mathbf{D}}}$, which dictates the preferred directions of morphogen migration. To ensure local mass conservation, this tensor is normalized to have unit trace:
\begin{align}
	\underline{\underline{\mathbf{D}}} =
	\begin{pmatrix}
		D_{xx} & D_{xy} \\ D_{yx} & D_{yy}
	\end{pmatrix} =
	\underline{\underline{\mathbf{R}}}
	\begin{pmatrix}
		\lambda & 0 \\ 0 & 1 - \lambda
	\end{pmatrix}
	\underline{\underline{\mathbf{R}}}^T
\end{align}
where $\lambda \in [0,1]$ controls the degree of anisotropy, and $\underline{\underline{\mathbf{R}}}$ is a rotation matrix that defines the local principal orientation along a local angle $\theta$. In all simulations, we used fixed values: \(f=0.035\), \(k=0.08\), \(d_u=4\), and \(d_v=1\).

\chunkgap

\meth{Discretization, stencil, and boundary conditions}
The numerical implementation of the model is based on an explicit Euler scheme using finite differences with periodic boundary conditions, enabling efficient and easily parallelizable computations. The discrete anisotropic Laplacian is evaluated using a tailored convolution mask, defined as follows:
\begin{align}
	\mathbf{\underline{\underline{L}}} =
	\begin{bmatrix}
		-\frac{D_{xy}}{4} & \frac{D_{yy}}{2} & \frac{D_{xy}}{4}  \\
		\frac{D_{xx}}{2}  & -1               & \frac{D_{xx}}{2}  \\
		\frac{D_{xy}}{4}  & \frac{D_{yy}}{2} & -\frac{D_{xy}}{4}
	\end{bmatrix}
\end{align}
where the components $D_{xx}$, $D_{yy}$, and $D_{xy}$ are computed from the eigenvalues of the diffusion tensor and the local orientation angle.

The operator $\nabla \cdot (\underline{\underline{\mathbf{D}}} \nabla U)$ is numerically approximated by a local convolution:
\begin{align}
	\nabla \cdot (\underline{\underline{\mathbf{D}}} \nabla U)|_{i,j} \approx \sum_{p=-1}^1 \sum_{q=-1}^1 L_{p,q} \, U_{i+p,j+q}
\end{align}
where $L_{p,q}$ denotes the coefficients of the convolution mask $\underline{\underline{L}}$ applied to the neighborhood of each point in the domain. This operation is naturally suited to grid-based discretization and vectorized implementation. By default, periodic boundary conditions are assumed, ensuring continuity of the generated patterns and preventing edge artifacts. These conditions are well-suited for generating supercells for homogenization and analyzing periodic motifs. This approach enables the controlled growth of anisotropic patterns tailored to specific mechanical constraints, defined through carefully designed spatial distributions of morphogenetic parameters.
\chunkgap

\subsection*{Computational workflow and mechanical homogenization}

\meth{Pattern synthesis and meshing}
We generate oriented, anisotropic morphogenetic patterns using a Gray–Scott reaction–diffusion model in which the Laplacian is made anisotropic via a rotated diffusion tensor. Model parameters control the principal direction (angle $\theta$) and the degree of anisotropy. Simulations are \textcolor{blue}{run to} 80 000 iterations (\textcolor{blue}{taking} a few seconds \textcolor{blue}{and ensuring stationarity}), then the resulting scalar field is thresholded into a binary mask representing the microstructure. To simplify subsequent analysis, the structures are oriented so that their principal axes align with the simulation axes. This yields an orthotropic effective response with negligible coupling between axial and shear components. From the binary mask, we extract external and pore boundaries (MATLAB \textcolor{reveditor}{R2022b} -- including Image Processing Toolbox, MathWorks, Natick, MA, USA) and simplify them (Douglas–Peucker) to obtain a compact polygonal description. A high-quality triangular mesh is then generated with the mesh2d library~\cite{engwirda2014locally}, edited to ensure periodic boundary conditions of the edge nodes.
\chunkgap

\meth{Homogenization and load cases}
Effective properties are computed via numerical homogenization on a periodic Representative Volume Element~\cite{okereke2013virtual}. We programmatically write Abaqus input files (.inp) that specify the mesh, a linear elastic material, and periodic boundary constraint equations (multi-point constraints tying opposite faces and corner nodes). Three displacement-driven load cases are solved in Abaqus \textcolor{reveditor}{2024}~\cite{simulia_abaqus} (Dassault Systèmes, Vélizy-Villacoublay, France): uniaxial tension along $x$ and $y$, and in-plane shear (Fig.~\ref{fig:Caracterisation}). The Abaqus command-line interface executes each job, and the Abaqus Python API (odbAccess) exports nodal displacements and reaction forces as CSV files. From corner-node reactions and displacements, we compute the homogenized orthotropic parameters $(E_1, E_2, \nu_{12}, \nu_{21}, G_{12})$ and assemble the plane-stress stiffness tensor $\underline{\underline{\mathbf{C}}}$.
\chunkgap

\meth{Post-processing and availability}
Across parameter sweeps (anisotropy, threshold, and $\theta$), results are aggregated into a structured database. Central and axial symmetries of the design space are exploited, and Savitzky–Golay filtering (MATLAB Signal Processing Toolbox) is applied to obtain smooth maps of material parameters. From these, we reconstruct $\underline{\underline{\mathbf{C}}}$ consistently, compare eigenvalues and principal directions, and derive bulk ($B$) and shear ($G$) modulus maps. Robustness is ensured through automated meshing checks, periodic boundary verification, basic validity tests, and retry logic for simulation convergence. The entire workflow is implemented end-to-end in MATLAB and Abaqus, with scripts, dependencies, and templates provided (MATLAB Image Processing and Signal Processing Toolboxes, Abaqus CLI and Python/odbAccess) to facilitate reproduction and reuse.
\chunkgap

\color{blue}

\subsection*{Tensor Formalism in Mandel--Kelvin Notation}

\meth{Constitutive framework and Mandel--Kelvin vectorization}
The modeling of elastic mechanical behavior involves the use of high-order tensors. In the framework of linear elasticity, the stress tensor $\underline{\underline{\bm{\sigma}}}$ and strain tensor $\underline{\underline{\bm{\epsilon}}}$ are related through a fourth-order elasticity tensor $\mathbb{C}$ via $\underline{\underline{\bm{\sigma}}} = \mathbb{C} : \underline{\underline{\bm{\epsilon}}}$. To facilitate numerical treatment, we adopt the Mandel--Kelvin representation, which vectorizes symmetric second-order tensors while preserving self-adjointness (unlike Voigt notation). The constitutive relation then becomes
\begin{align}
	\underline{\bm{\sigma}} = \underline{\underline{\mathbf{C}}}\, \underline{\bm{\epsilon}},
\end{align}
where $\underline{\bm{\epsilon}}$ and $\underline{\bm{\sigma}}$ denote the vectorized forms of the strain and stress fields, and $\underline{\underline{\mathbf{C}}}$ is the equivalent stiffness matrix. A detailed definition of the vectorization scheme and algebraic properties is provided in the \textit{Supplementary Information}.

\chunkgap
\meth{Orthotropic specialization}
We restrict ourselves to a plane-stress orthotropic model without shear–normal couplings, expressed in the material's principal directions:
\begin{align}
	\underline{\underline{\mathbf{C}}} =
	\begin{pmatrix}
		C_{11} & C_{12} & 0      \\
		C_{21} & C_{22} & 0      \\
		0      & 0      & C_{66}
	\end{pmatrix},
\end{align}
with entries determined from Young's moduli $E_1$, $E_2$, Poisson's ratios $\nu_{12}$, $\nu_{21}$, and shear modulus $G_{12}$. This decoupled form holds only in the principal frame; rotation reintroduces coupling terms. Complete expressions relating $C_{ij}$ to material parameters are given in the \textit{Supplementary Information}.

\chunkgap
\meth{Tensor transformation framework}
For rotations, the fourth-order elasticity tensor transforms via the Bond matrix $\underline{\underline{\mathbf{B}}}(\underline{\underline{\mathbf{R}}})$~\cite{bond1943mathematics, mehrabadi1990eigentensors}, which acts directly on vectorized quantities in Mandel--Kelvin space:
\begin{align}
	\underline{\underline{\mathbf{C}}}' = \underline{\underline{\mathbf{B}}}\,\underline{\underline{\mathbf{C}}}\,\underline{\underline{\mathbf{B}}}^T.
\end{align}
For general spatial transformations described by a Jacobian $\underline{\underline{\mathbf{J}}}$, we use the inverse $\underline{\underline{\mathbf{A}}} = \underline{\underline{\mathbf{J}}}^{-1}$ and define the transformation matrix $\underline{\underline{\mathbf{T}}}(\underline{\underline{\mathbf{A}}})$ constructed from the components of $\underline{\underline{\mathbf{A}}}$ according to Mandel--Kelvin conventions:
\begin{align}
	\underline{\underline{\mathbf{T}}}(\underline{\underline{\mathbf{A}}}) =
	\begin{pmatrix}
		A_{11}^2             & A_{12}^2             & \sqrt{2}A_{11}A_{12}      \\
		A_{21}^2             & A_{22}^2             & \sqrt{2}A_{21}A_{22}      \\
		\sqrt{2}A_{11}A_{21} & \sqrt{2}A_{12}A_{22} & A_{11}A_{22}+A_{12}A_{21}
	\end{pmatrix}.
\end{align}
The transformed constitutive relation becomes
\begin{align}
	\underline{\underline{\mathbf{C}}}' = \frac{1}{\det(\underline{\underline{\mathbf{A}}})}\,
	\underline{\underline{\mathbf{T}}}(\underline{\underline{\mathbf{A}}})\,
	\underline{\underline{\mathbf{C}}}\,
	\underline{\underline{\mathbf{T}}}(\underline{\underline{\mathbf{A}}})^T .
\end{align}
If $\underline{\underline{\mathbf{A}}}$ is symmetric positive definite, $\underline{\underline{\mathbf{T}}}$ is symmetric, ensuring that $\underline{\underline{\mathbf{C}}}'$ remains self-adjoint (a necessary condition for physical consistency). A comprehensive development of the transformation framework is provided in the \textit{Supplementary Information}.

\chunkgap
\meth{Cloaking formalism and radial/cubic mappings}
In this work, the reference medium is homogeneous and orthotropic. To preserve orthotropy, we use a cylindrical mapping in the $(r,\varphi)$ frame with a diagonal $\underline{\underline{\mathbf{A}}}=\mathrm{diag}(A_r,A_\varphi)$, which suppresses normal–shear couplings in Mandel--Kelvin form. The Jacobian reads
\begin{align}
	\underline{\underline{\mathbf{J}}} =
	\begin{pmatrix}
		\frac{\partial r^\prime}{\partial r} & 0                  \\
		0                                    & \frac{r^\prime}{r}
	\end{pmatrix}.
\end{align}
A linear mapping $r^\prime = (R_2 - R_1) r + R_1$ is simple but introduces slope discontinuities. Following~\cite{kadic2020elastodynamic}, a cubic mapping
\begin{align}
	r^\prime = a_3 r^3 + a_2 r^2 + a_1 r + a_0,
\end{align}
with constraints
\begin{align}
	f(0) = R_1,\;\; f(R_2)=R_2,\;\; f^\prime(0)=1,\;\; f^\prime(R_2)=1,
\end{align}
leads to
\begin{align}
	a_0 = R_1,\quad a_1=1,\quad a_2 = -\tfrac{3 R_1}{R_2^2},\quad a_3 = \tfrac{2 R_1}{R_2^3}.
\end{align}
This ensures smoother boundary matching. Following empirical observations, a slight quasi-adiabatic adjustment ($f^\prime(0) = 1.2$) improved cloak performance.

\color{black}

\color{blue}
\subsection*{Inferring generative parameters from target tensors}
\meth{Overview}
Building on the orthotropy-preserving tensor transformation introduced in the previous subsection, we convert the target field of elasticity matrices $\underline{\underline{\mathbf{C}}}'(x,y)$ (Mandel--Kelvin, plane stress) into spatially varying morphogenetic parameters that drive the anisotropic reaction–diffusion growth~\cite{bastek2022inverting}. The conversion proceeds by summarizing $\underline{\underline{\mathbf{C}}}'$ into a small set of invariants measurable on our database of homogenized reaction–diffusion patterns, nearest-neighbor assignment in this database with tunable weights, and spatial regularization followed by a snap-back to the discrete parameter grid used at fabrication time.

To anchor the transform, we work relative to a default isotropic, high-porosity host material. In Mandel–Kelvin (plane stress) notation, its orthotropic stiffness is
\[
	\mathbf{C}_{\mathrm{0}} =
	\begin{bmatrix}
		1.221 & 0.446 & 0     \\
		0.446 & 1.221 & 0     \\
		0     & 0     & 0.807
	\end{bmatrix}\ \mathrm{GPa}.
\]
The default pattern outside the cloak uses \(\lambda=0.5\) (isotropic generation) and \(s=0.8\) (segmentation threshold controlling porosity). Within the cloak annulus, \(\lambda\) and \(s\) vary pointwise. This modulates local anisotropy and porosity of the generated elliptical motifs while the other controls remain unchanged. Under this baseline, the orthotropy-preserving transform maps the target field \(\mathbf{C}'(x,y)\) to \(\lambda(x,y), s(x,y),\text{ and } \theta(x,y)\) that drive growth.

\chunkgap
\meth{Database and restricted operating window}
We rely on a precomputed database that maps reaction–diffusion controls to effective orthotropic stiffness matrices $\underline{\underline{\mathbf{C}}}$~\cite{bonfanti2020automatic, ha2023rapid}. Each entry is indexed by a reaction–diffusion anisotropy parameter $\lambda$ that controls the ellipticity of the generated patterns, and a segmentation threshold $s$ that defines porosity. To avoid the merging of elliptical pores, we restrict the search to a practical anisotropy window (here $0.5$–$0.85$).

\meth{Feature space for matching}
For each candidate and for each target point, we work with in-plane principal quantities of the orthotropic tensors (see Fig.~\ref{fig:Caracterisation}).
We extract three features for each tensor: the stiffness trace $\mathrm{tr}$, defined as the sum of the two in-plane principal eigenvalues, the principal stiffness anisotropy, defined as the ratio of these two values, and the principal direction angle $\theta$, defined by the orientation of the main eigenvector.

\chunkgap
\meth{Weighted nearest-neighbor assignment}
At each point in the cloak, we select the database entry that minimizes a separable, scale-aware distance using only stiffness magnitude and principal orientation
\begin{equation}
	E = w_{\text{tr}}\bigl|\log \mathrm{tr}'-\log \mathrm{tr}\bigr|
	+ w_{\theta}\bigl|\theta'-\theta\bigr| ,
\end{equation}
where $\mathrm{tr}$ denotes the trace of the $2\times2$ normal block (a proxy for overall stiffness) and $\theta$ the angle of the principal eigenvector. We set $w_{\text{tr}}=0.6$ and $w_{\theta}=0.4$. The logarithmic distance on the trace balances multiplicative spreads, and the angle term aligns the principal directions. Anisotropy level and shear then follow implicitly from the selected entry. Outside the cloak and inside the cavity, indices are fixed to a single background entry to preserve a homogeneous host.

\chunkgap
\meth{Spatial regularization and snapping}
The raw index fields yield pixelwise maps of reaction–diffusion anisotropy and threshold. We first apply a mild Gaussian smoothing to each scalar map to avoid spurious high-frequency jumps that could hamper growth continuity. Then, to ensure manufacturability and consistency with the homogenized library, we snap the smoothed values back to the nearest points of the discrete restricted grid. This preserves continuity at the reaction–diffusion level while guaranteeing that each requested property corresponds to a realizable, characterized microstructure.

\chunkgap
\meth{Outputs to the reaction–diffusion engine}
The final outputs are spatial fields of reaction–diffusion anisotropy, binary segmentation level $s$ (porosity control), and diffusion orientation, here taken from the cloak’s local frame via the angle field $\theta_J$ defined by the polar map. These fields are bundled into a structure together with $(R_1,R_2)$ and the cloak mask, and directly feed the anisotropic reaction–diffusion simulator. In practice, the diffusion tensor at each grid point is built from the local angle and eigenvalues with unit-trace normalization, while the threshold sets the solid/void fraction at binarization. The result is a decentralized growth that self-organizes toward the target tensor field without any global optimization.
\color{black}

\subsection*{Cloak simulation}

\meth{Model and boundary conditions}
Static analyses were run in \textit{3DEXPERIENCE} SIMULIA~\cite{dassault_3dexperience} (Dassault Systèmes, Vélizy-Villacoublay, France) to mirror the tensile tests. The generative mask was extruded to a plate of thickness $t=3$~mm and modeled as a linear elastic solid (POM, $E=3$~GPa, $\nu=0.35$). To emulate the jaws, a rigid body was tied to each end of the plate over the screwed areas. The bottom rigid body was fully constrained, and a vertical tensile force $F = 2\ \text{kN}$ was applied at the mass-center reference point of the top rigid body, suppressing spurious moments and reproducing the experimental kinematics \textcolor{reveditor}{(Supplementary Figure~24)}.

\chunkgap
\meth{Discretization and outputs}
The plate was meshed with 10-node quadratic tetrahedra (C3D10). For the cloaked specimen \textcolor{reveditor}{(Supplementary Figure~24)}, the mesh comprised 215{,}103 elements. Mesh density was chosen to resolve ligament widths and hole fillets, with convergence checked on the displacement fields. A single static step was solved and the nodal displacements on the camera-facing surface were post-processed to extract $u_1$ (horizontal) and $u_2$ (vertical) for one-to-one comparison with DIC and with the reference/holed configurations in the Results.


\subsection*{Experimental Setup and Data Acquisition}

\meth{Specimen fabrication and fixturing}
The experiments \textcolor{blue}{were} carried out on a \textcolor{reveditor}{SYNTAX 100 tensile testing machine (3R)}. Polyoxymethylene (POM) plates, $3$~mm thick and with dimensions $190 \times 130$ mm$^2$, \textcolor{blue}{were} machined using a \textcolor{reveditor}{ILP Laser} $130$~W $\text{CO}_2$ laser. The central circular cavity has radius $R_1 = 0.15 \times 130~\mathrm{mm} = 19.5~\mathrm{mm}$, and the cloak radius is $R_2 = 2R_1 = 39~\mathrm{mm}$.

Custom grips were designed for this experiment, allowing each end of the plate to be screwed onto a steel bar with dimensions $130 \times 10 \times 4  \text{mm}^3$, fixed at five points per plate.
\chunkgap

\meth{Loading protocol and digital image correlation}
This \textcolor{blue}{setup} ensures proper gripping of the plates during tensile testing. The stiffness of the frame and the absence of slippage in the jaws \textcolor{blue}{improved} the reproducibility of the tests, enforcing zero horizontal displacement at the plate edges. After a pre-load of $110$~N, used to position the plate, a force of $2$~kN (specified in NF EN ISO 527-2 standard) \textcolor{blue}{was} applied vertically through the grip \textcolor{reveditor}{(Supplementary Figure~25)}. The displacement fields $u_1$ and $u_2$ \textcolor{blue}{were} measured using digital image correlation, based on two photographs taken before and after the application of the force. This procedure is largely inspired by the protocol described in the supplementary information of Wang \textit{et al.}\cite{wang2022mechanical}, here using a Nikon D780 camera with a Nikkor 70--200~mm lens. Image correlation is performed using an augmented Lagrangian method, with code freely available online, developed by Yang and Bhattacharya\cite{yang2019augmented} \textcolor{reveditor}{(AL-DIC v4.2, available at \url{https://data.caltech.edu/records/0x0xk-78097})}. Displacement tracking is facilitated by spraying each plate with a light, sparse speckle pattern using matte white spray paint. Reproducibility of the results is ensured by employing a ring light to control illumination conditions.
\chunkgap


\color{blue}

\subsection*{Stress-driven parameter mapping}

\meth{Overview}
Unlike the cloaking approach which relies on a geometric transformation to define target stiffness tensors, the structural optimization strategy leverages the stress field from a preliminary homogeneous simulation \textcolor{reveditor}{(Supplementary Figure~26a)} to spatially modulate the morphogenetic parameters. This process avoids the computational cost of iterative optimization.

\meth{Numerical optimization setup}
The structural optimization strategies are evaluated on a domain of physical dimensions $200 \times 100$ mm$^2$, composed of a verifiable isotropic linear elastic material ($E=3$ GPa, $\nu=0.35$). In contrast to the kinematic cloaking problem where displacements were imposed, here we apply force-controlled boundary conditions with a total load magnitude of $F=2$ kN.
Two distinct configurations are considered. The first is a distributed three-point bending test ("Bridge"), where the bottom corners are supported over 15\% of the width (left: pin, right: roller) and the load is distributed over the central 10\% of the bottom edge. The second is a cantilever beam, where the left edge is fully clamped, and the load is applied downwards distributed over the central 10\% of the right edge.

\chunkgap
\meth{Density modulation via stress invariants}
The local fill fraction of the structure is controlled by the reaction-diffusion threshold parameter $s$. We impose a direct mapping between the local potential energy density ansatz and the local porosity. Specifically, we define a normalized stress invariant metric $\bar{\tau} \in [0,1]$ derived from the trace of the stress tensor: $\bar{\tau} \propto \text{tr}(\bm{\underline{\underline{\sigma}}})$. The threshold $s$ is then linearly interpolated between a lower bound $s_{\text{dense}}=0.3$ (favoring solid material) and an upper bound $s_{\text{void}}=0.8$ (favoring voids):
\begin{equation}
	s = s_{\text{void}} - (s_{\text{void}} - s_{\text{dense}}) \cdot \bar{\tau}.
\end{equation}
\textcolor{reveditor}{(Supplementary Figure~26b)}. This mapping generates a "graded isotropic" structure where material accumulation naturally mirrors the potential energy density map, stiffening the most loaded areas.

\chunkgap
\meth{Anisotropic guiding via diffusion tensor}
To further optimize the structure, we break the isotropy of the reaction-diffusion process by guiding the growth orientation. The local anisotropy of the pattern is controlled by the diffusion tensor $\underline{\underline{\mathbf{D}}}$ of the morphogens, which is constructed to share the spectral basis (eigenvectors) of the stress tensor $\bm{\underline{\underline{\sigma}}}$.
Furthermore, the magnitude of the diffusion anisotropy $\lambda$ is scaled by the directionality of the local stress state \textcolor{reveditor}{(Supplementary Figure~26c)}. In regions dominated by uniaxial or shear stresses (high deviatoric component), $\lambda$ is driven towards $0.9$ to induce strong elongation of the pores orthogonal to the principal stress. Conversely, in hydrostatic regions, $\lambda$ remains close to $0.5$ (isotropic growth). This creates a "graded anisotropic" architecture that effectively channels the stress flow \textcolor{reveditor}{(Supplementary Figure~26d)}.


\color{black}

\section*{Data availability}
\textcolor{reveditor}{The data generated in this study are available in the paper and its Supplementary Information. Raw data supporting all display items are provided as Source Data files (one Excel file per figure or table). The morphogenetic generation and mechanical homogenization code is openly available at \url{https://github.com/thomasfromenteze/Morphogenetic_Mechanical_Metamaterials_2D} and archived on Zenodo (DOI: \href{https://doi.org/10.5281/zenodo.17107007}{10.5281/zenodo.17107007}). Source data are provided with this paper.}

\section*{Code availability}
All custom code for the 2D morphogenetic generator, homogenization pipeline, and analysis/plotting is openly available at
\url{https://github.com/thomasfromenteze/Morphogenetic_Mechanical_Metamaterials_2D}.
An archived snapshot corresponding to the version used in this study (Version~v1) is preserved on Zenodo with the DOI \href{https://doi.org/10.5281/zenodo.17107007}{10.5281/zenodo.17107007}. The archived package is labeled as Software (MATLAB and Python) and licensed under CC~BY~4.0.


\section*{Acknowledgements}
The authors thank Professor Michael Okereke (University of Greenwich) for his clear Abaqus tutorial videos, which facilitated our workflow automation.

\section*{Funding}
This work is supported by the LABEX $\Sigma$-LIM (ANR-10-LABX-0074-01), the French National Research Agency (ANR JCJC MetaMorph ANR-21-CE42-0005), and the CNRS through the MITI interdisciplinary programs via its Exploratory Research program. The authors also acknowledge the financial support of the IUT du Limousin. Thomas Fromenteze acknowledges support from the Institut Universitaire de France (IUF).

\section*{Author contributions}
T.F. proposed the initial idea and, together with P.M. and V.P., developed the conceptual foundations and scope of the project.
T.F. and P.M. defined the overall strategy for numerical and experimental validation.
T.F. developed the theory and code and conceived the cloaking demonstration, with periodic input from P.M.
T.F. extended the generative model to stress-guided structural optimization.
A.H. and P.M. conducted 3DEXPERIENCE simulations of the cloak.
A.H., P.M. and T.F. designed the experimental demonstration and set up correlation-based data acquisition.
T.F. performed data extraction and prepared the initial manuscript draft.
All authors discussed the results and contributed to manuscript revisions.

\section*{Competing interests}
\textcolor{reveditor}{The authors declare no competing financial or non-financial interests.}

\section*{Correspondence}
Correspondence should be addressed to Thomas Fromentèze: thomas.fromenteze@unilim.fr.

\clearpage

\setcounter{figure}{0}
\setcounter{table}{0}
\setcounter{section}{0}
\setcounter{equation}{0}

\begin{center}
{\Large\textbf{Supplementary Information}}\\[0.5em]
{\large\textit{Morphogenetic mechanical metamaterials: Emerging tensor properties from self-organized structures}}\\[0.3em]
{\normalsize Thomas Froment\`eze, Philippe Michaud, Ali Hassny, and Vincent Pateloup}
\end{center}

\vspace{1em}

\renewcommand{\figurename}{\textcolor{reveditor}{Supplementary Figure}}
\renewcommand{\tablename}{\textcolor{reveditor}{Supplementary Table}}

\color{blue}


\section*{\textcolor{reveditor}{Supplementary Note 1:} Tensor Formalism in Mandel--Kelvin Notation}

This section details the mathematical framework used to represent and manipulate the elastic properties of the generated metamaterials. We adopt the Mandel--Kelvin notation, which vectorizes second-order tensors while preserving the thermodynamic structure of the elasticity tensor. This formalism is particularly advantageous for applying coordinate transformations and for the numerical homogenization process.

\meth{Constitutive framework and Mandel--Kelvin vectorization}
The modeling of elastic mechanical behavior involves the use of high-order tensors. In the framework of linear elasticity, the relation between the stress tensor $\underline{\underline{\bm{\sigma}}}$ and the strain tensor $\underline{\underline{\bm{\epsilon}}}$ is expressed through a fourth-order elasticity tensor $\mathbb{C}$:
\begin{align}
	\underline{\underline{\bm{\sigma}}} = \mathbb{C} : \underline{\underline{\bm{\epsilon}}}.
\end{align}
In two dimensions, the tensors $\underline{\underline{\bm{\sigma}}}$ and $\underline{\underline{\bm{\epsilon}}}$ are symmetric second-order matrices ($2\times2$). To facilitate numerical and analytical treatment, it is convenient to recast this law into a compact matrix form. In the Mandel--Kelvin formalism, symmetric second-order tensors are vectorized as
\begin{align}
	\underline{\bm{\sigma}} =
	\begin{pmatrix}
		\sigma_{11},
		\sigma_{22},
		\sqrt{2}\sigma_{12}
	\end{pmatrix}^T
	=
	\begin{pmatrix}
		\sigma_{1},
		\sigma_{2},
		\sigma_{6}
	\end{pmatrix}^T ,
\end{align}
\begin{align}
	\underline{\bm{\epsilon}} =
	\begin{pmatrix}
		\epsilon_{11},
		\epsilon_{22},
		\sqrt{2}\epsilon_{12}
	\end{pmatrix}^T
	=
	\begin{pmatrix}
		\epsilon_{1},
		\epsilon_{2},
		\epsilon_{6}
	\end{pmatrix}^T .
\end{align}

\chunkgap
\meth{Orthotropic specialization and parameterization}
This representation leads to the constitutive law in vector form:
\begin{align}
	\underline{\bm{\sigma}} = \underline{\underline{\mathbf{C}}}\, \underline{\bm{\epsilon}},
\end{align}
where $\underline{\underline{\mathbf{C}}}$ is a symmetric $3\times3$ matrix obtained from $\mathbb{C}$. In the present work, we restrict ourselves to a plane-stress orthotropic model without shear–normal couplings:
\begin{align}
	\underline{\underline{\mathbf{C}}} =
	\begin{pmatrix}
		C_{11} & C_{12} & 0      \\
		C_{21} & C_{22} & 0      \\
		0      & 0      & C_{66}
	\end{pmatrix}.
\end{align}
Its entries are determined from three independent load cases:
\begin{align}
	\underline{\underline{\mathbf{C}}} =
	\begin{pmatrix}
		E_{1}/\alpha          & \nu_{12} E_{2}/\alpha & 0        \\
		\nu_{21} E_{1}/\alpha & E_{2}/\alpha          & 0        \\
		0                     & 0                     & 2 G_{12}
	\end{pmatrix},
\end{align}
with $\alpha = 1-\nu_{12}\nu_{21}$. We emphasize that this decoupled form, with zero normal-shear couplings, holds only when the constitutive matrix is expressed in the material’s principal directions. Away from the principal frame, a rotation reintroduces coupling terms in Mandel–Kelvin form.

\chunkgap
\meth{Tensor transformations from second- to fourth-order}
The transformation rules are best understood by starting with second-order tensors. For a pure rotation $\underline{\underline{\mathbf{R}}}$, the stress transforms as
\begin{align}
	\underline{\underline{\bm{\sigma}}}' = \underline{\underline{\mathbf{R}}}\,\underline{\underline{\bm{\sigma}}}\,\underline{\underline{\mathbf{R}}}^T .
\end{align}
For a fourth-order tensor such as $\mathbb{C}$, the transformation generalizes to~\cite{mehrabadi1990eigentensors}
\begin{align}
	C'_{ijkl} = R_{im}R_{jn}R_{kp}R_{lq}\,C_{mnpq}.
\end{align}
In Mandel--Kelvin form, this relation can be condensed using the so-called Bond matrix $\underline{\underline{\mathbf{B}}}(\mathbf{R})$~\cite{bond1943mathematics}, which acts directly on vectorized stresses and strains~\cite{mehrabadi1990eigentensors}:
\begin{align}
	\underline{\bm{\sigma}}' = \underline{\underline{\mathbf{B}}}\,\underline{\bm{\sigma}}, \qquad
	\underline{\bm{\epsilon}}' = \underline{\underline{\mathbf{B}}}\,\underline{\bm{\epsilon}}.
\end{align}

In two dimensions, the Mandel--Kelvin Bond matrix built from the rotation $\underline{\underline{\mathbf{R}}}$ (with entries $R_{ij}$) is
\begin{align}
	\underline{\underline{\mathbf{B}}}(\underline{\underline{\mathbf{R}}}) =
	\begin{pmatrix}
		R_{11}^2               & R_{12}^2               & \sqrt{2}\,R_{11}R_{12}     \\[2pt]
		R_{21}^2               & R_{22}^2               & \sqrt{2}\,R_{21}R_{22}     \\[2pt]
		\sqrt{2}\,R_{11}R_{21} & \sqrt{2}\,R_{12}R_{22} & R_{11}R_{22}+R_{12}R_{21}
	\end{pmatrix}.
\end{align}
By construction, $\underline{\underline{\mathbf{B}}}$ is orthogonal and symmetric in Mandel--Kelvin space, so rotations preserve inner products and the self-adjointness of the constitutive law. The corresponding elasticity matrix transforms as
\begin{align}
	\underline{\underline{\mathbf{C}}}' = \underline{\underline{\mathbf{B}}}\,\underline{\underline{\mathbf{C}}}\,\underline{\underline{\mathbf{B}}}^T.
\end{align}

\chunkgap
\meth{General mapping with Jacobian operator}
Beyond rotations, a general spatial transformation is described by a Jacobian $\underline{\underline{\mathbf{J}}}$. Since we map back from the transformed (virtual) space to the homogeneous physical space, we use the inverse Jacobian $\underline{\underline{\mathbf{A}}} = \underline{\underline{\mathbf{J}}}^{-1}$. The transformation of stresses is then
\begin{align}
	\underline{\underline{\bm{\sigma}}}' = \frac{1}{\det(\underline{\underline{\mathbf{A}}})}\,
	\underline{\underline{\mathbf{A}}}\,\underline{\underline{\bm{\sigma}}}\,\underline{\underline{\mathbf{A}}}^T.
\end{align}
By analogy with the Bond matrix, we define a matrix operator $\underline{\underline{\mathbf{T}}}(\underline{\underline{\mathbf{A}}})$ acting in Mandel--Kelvin space:
\begin{align}
	\underline{\underline{\mathbf{T}}}(\underline{\underline{\mathbf{A}}}) =
	\begin{pmatrix}
		A_{11}^2             & A_{12}^2             & \sqrt{2}A_{11}A_{12}       \\
		A_{21}^2             & A_{22}^2             & \sqrt{2}A_{21}A_{22}       \\
		\sqrt{2}A_{11}A_{21} & \sqrt{2}A_{12}A_{22} & A_{11}A_{22}+A_{12}A_{21}
	\end{pmatrix}.
\end{align}
The transformed constitutive relation then becomes
\begin{align}
	\underline{\underline{\mathbf{C}}}' = \frac{1}{\det(\underline{\underline{\mathbf{A}}})}\,
	\underline{\underline{\mathbf{T}}}(\underline{\underline{\mathbf{A}}})\,
	\underline{\underline{\mathbf{C}}}\,
	\underline{\underline{\mathbf{T}}}(\underline{\underline{\mathbf{A}}})^T .
\end{align}

\chunkgap
\meth{Compact Mandel--Kelvin form and symmetry conditions}
Starting from the classical definition
\begin{align}
	\sigma_{ij} = C_{ijkl}\,\epsilon_{kl},
\end{align}
the transformed fourth-order tensor reads
\begin{align}
	C'_{mnop} = \frac{1}{\det(\underline{\underline{\mathbf{A}}})}\,
	A_{i m}\,A_{j n}\,C_{ijkl}\,A_{k o}\,A_{l p}.
\end{align}
In Mandel--Kelvin vector form:
\begin{align}
	\underline{\bm{\sigma}}'            & = \tfrac{1}{\det(\underline{\underline{\mathbf{A}}})}\,
	\underline{\underline{\mathbf{T}}}(\underline{\underline{\mathbf{A}}})\,\underline{\bm{\sigma}},                                                \\
	\underline{\bm{\epsilon}}'          & = \underline{\underline{\mathbf{T}}}(\underline{\underline{\mathbf{A}}}^{-1})\,\underline{\bm{\epsilon}}, \\
	\underline{\underline{\mathbf{C}}}' & = \tfrac{1}{\det(\underline{\underline{\mathbf{A}}})}\,
	\underline{\underline{\mathbf{T}}}(\underline{\underline{\mathbf{A}}})\,\underline{\underline{\mathbf{C}}}\,\underline{\underline{\mathbf{T}}}(\underline{\underline{\mathbf{A}}})^T.
\end{align}
If $\underline{\underline{\mathbf{A}}}$ is symmetric positive definite (SPD), then $\underline{\underline{\mathbf{T}}}$ is symmetric, ensuring that $\underline{\underline{\mathbf{C}}}'$ remains self-adjoint, which is a necessary condition for physical consistency.

\chunkgap
\meth{Cloaking formalism and radial/cubic mappings}
In this work, the reference medium is homogeneous and orthotropic. To preserve orthotropy, we use a cylindrical mapping in the $(r,\varphi)$ frame with a diagonal $\underline{\underline{\mathbf{A}}}=\mathrm{diag}(A_r,A_\varphi)$, which suppresses normal–shear couplings in Mandel–Kelvin form. The Jacobian reads
\begin{align}
	\underline{\underline{\mathbf{J}}} =
	\begin{pmatrix}
		\frac{\partial r^\prime}{\partial r} & 0                  \\
		0                                    & \frac{r^\prime}{r}
	\end{pmatrix}.
\end{align}
A linear mapping $r^\prime = (R_2 - R_1) r + R_1$ is simple but introduces slope discontinuities. Following~\cite{kadic2020elastodynamic}, a cubic mapping
\begin{align}
	r^\prime = a_3 r^3 + a_2 r^2 + a_1 r + a_0,
\end{align}
with constraints
\begin{align}
	f(0) = R_1,\;\; f(R_2)=R_2,\;\; f^\prime(0)=1,\;\; f^\prime(R_2)=1,
\end{align}
leads to
\begin{align}
	a_0 = R_1,\quad a_1=1,\quad a_2 = -\tfrac{3 R_1}{R_2^2},\quad a_3 = \tfrac{2 R_1}{R_2^3}.
\end{align}
This ensures smoother boundary matching. In practice, a slight quasi-adiabatic adjustment ($f^\prime(0) = 1.2$) improved cloak performance.

\newpage
\section*{\textcolor{reveditor}{Supplementary Note 2:} Controlling emergent spatial frequencies with diffusion}

In this section, we provide the mathematical derivation relating the diffusion coefficients of the reaction-diffusion model to the characteristic length scale (wavelength) of the emerging patterns. This analysis, based on a linear stability analysis of the Turing system, justifies our ability to control the pore size and density of the metamaterials by adjusting the diffusion tensor eigenvalues.

\meth{Reaction kinetics and Gray-Scott model}
Considering a chemical reaction, the variables $u = [U]$ and $v = [V]$ represent the concentration of each morphogen varying in space and time. The definition of a first system of kinetic equations allows considering the effect of conversion between chemical species over time.
\begin{align}
	 & \frac{\mathrm{d}u}{\mathrm{d}t} = f(u,v)  \\
	 & \frac{\mathrm{d}v}{\mathrm{d}t} = g(u,v)
\end{align}

The functions $f$ and $g$ correspond to elementary reactions that depend on the probability of reactant encounter and their conversion. Our generative model is based on the Gray-Scott system~\cite{gray1983autocatalytic}:
\begin{align}
	U + 2V \overset{\kappa}{\rightarrow} 3V
\end{align}
\noindent where $\kappa$ is a reaction rate constant. The following development can however be adapted to other reaction rules. In these particular conditions, the variation of concentrations is directly translated into the chemical kinematics equation:
\begin{align}
	 & \frac{\mathrm{d}u}{\mathrm{d}t} = -\kappa \cdot uv^2  \\
	 & \frac{\mathrm{d}v}{\mathrm{d}t} = +\kappa \cdot uv^2
\end{align}
\chunkgap
\meth{Diffusion mechanisms and vector formalism}
Alan Turing imagined that an additional diffusion mechanism would allow the most sparsely populated areas to be occupied by morphogen migration~\cite{turing1952chemical}. To facilitate explanations, this diffusion is represented in a one-dimensional space:
\begin{align}
	\frac{\mathrm{d}u}{\mathrm{d}t} = f(u,v) + d_u \frac{\partial^2 u}{\partial x^2} \\
	\frac{\mathrm{d}v}{\mathrm{d}t} = g(u,v) + d_v \frac{\partial^2 v}{\partial x^2}
\end{align}

\noindent where $d_u$ and $d_v$ correspond to constants associated with diffusion, calculated by applying the Laplacian operator simplified here by a double spatial derivative. The diffusion phenomenon corresponds to a macroscopic and averaged view of random walks of morphogens. This model is finally composed of reaction and diffusion terms and seems capable of promoting the formation of spatial patterns under certain conditions. The mathematical justification of this part is more delicate. A vector formalism is considered for the rest of these developments. The model then takes the following form:
\begin{align}
	\begin{bmatrix}
		\displaystyle \frac{\mathrm{d} u}{\mathrm{d} t} \\[1em]
		\displaystyle \frac{\mathrm{d} v}{\mathrm{d} t}
	\end{bmatrix}
	=
	\begin{bmatrix}
		\displaystyle f(u,v) \\
		\displaystyle g(u,v)
	\end{bmatrix}
	+
	\begin{bmatrix}
		\displaystyle d_u \frac{\mathrm{d}^2 u}{\mathrm{d}x^2} \\[1em]
		\displaystyle d_v \frac{\mathrm{d}^2 v}{\mathrm{d}x^2}
	\end{bmatrix}
\end{align}

\chunkgap
\meth{Linear stability analysis}
This system of non-linear partial differential equations is studied by a perturbation method, implying here a first-order expansion around equilibrium solutions. Taking up the predator-prey analogy, there are many solutions where one population eventually dominates the other and fills the space until forming a stable and quasi-homogeneous set. The objective here is to slightly perturb these equilibria with spatial harmonics to try to force the system to emerge patterns (instabilities) whose spatial frequencies are compatible with its parameters. This approach amounts to linearizing the problem by considering a first-order expansion. We thus posit the relation $(u,v) = (u^*,v^*) + \epsilon (U,V)$, where $u^*$ and $v^*$ correspond to a quasi-homogeneous equilibrium point such that $f(u^*,v^*) = g(u^*,v^*) = 0$. The problem then takes the following form:
\begin{align}
	\begin{bmatrix}
		U_t \\
		V_t
	\end{bmatrix}
	=
	\underbrace{
		\begin{bmatrix}
			f_u & f_v  \\
			g_u & g_v
		\end{bmatrix}
	}_{\mathbf{J}}
	\begin{bmatrix}
		U \\
		V
	\end{bmatrix}
	+
	\underbrace{
		\begin{bmatrix}
			d_u & 0    \\
			0   & d_v
		\end{bmatrix}
	}_{\mathbf{D}}
	\begin{bmatrix}
		U_{xx} \\
		V_{xx}
	\end{bmatrix}
\end{align}

\chunkgap
\meth{Jacobian operator and spectral perturbations}
For these new notations adapted from the field of non-linear dynamic systems, $U_t$ and $V_t$ correspond to temporal derivatives of $U$ and $V$. $U_{xx}$ and $V_{xx}$ then represent the second derivatives with respect to $x$, resulting from the Laplacian calculation representing diffusion. This first-order approximation requires the calculation of the Jacobian operator $\mathbf{J}$, composed of the derivatives of $f$ and $g$ with respect to $u$ and $v$. Considering now that weak perturbations take the form
\begin{equation}
	U \propto e^{\lambda_k t} \cos\left(\frac{k \pi x}{L}\right)
\end{equation}

\noindent These solutions are studied for Neumann boundary conditions on a domain of length L.  The objective will be to determine if certain spatial frequencies can couple to the system by growing oscillations, translated by $\Re(\lambda_k)>0$.The injection of perturbations into the model and then the calculation of temporal derivatives ($U_t = \lambda_k\, U$) and spatial derivatives ($U_{xx} = -\left(\frac{k \pi}{L}\right)^2 U$) allows arriving at the following step:
\begin{align}
	\lambda_k
	\begin{bmatrix}
		U \\
		V
	\end{bmatrix}
	=
	\mathbf{J}
	\begin{bmatrix}
		U \\
		V
	\end{bmatrix}
	-
	\left(
	\frac{k\pi}{L}
	\right)^2
	\mathbf{D}
	\begin{bmatrix}
		U \\
		V
	\end{bmatrix}
\end{align}

\chunkgap
\meth{Eigenvalue problem formulation}
This expression then approaches the form of an eigenvalue problem by factoring the right-hand terms by the concentration of morphogens in space:
\begin{align}
	\lambda_k
	\begin{bmatrix}
		U \\
		V
	\end{bmatrix}
	=
	\mathbf{M}_k
	\begin{bmatrix}
		U \\
		V
	\end{bmatrix}
\end{align}

\noindent where $\mathbf{M}_k = \mathbf{J} - \left(\frac{k\pi}{L}\right)^2\mathbf{D}$ is a new matrix operator that finally allows arriving at the following formalism, considering the notation $\mathbf{I}$ for the identity matrix:
\begin{align}
	\left(
	\mathbf{M}_k
	- \lambda_k \mathbf{I}
	\right)
	\begin{bmatrix}
		U \\
		V
	\end{bmatrix}
	=
	0
\end{align}

\chunkgap
\meth{Characteristic equation derivation}
Non-trivial solutions for $(U,V)$ imply that the determinant of $\left(\mathbf{M}_k- \lambda_k \mathbf{I}\right)$ must be zero. Defining first the content of a matrix $\mathbf{M}_k$ such that:
\begin{align}
	\mathbf{M}_{k} =
	\begin{bmatrix}
		a & b  \\
		c & d
	\end{bmatrix}
\end{align}

\noindent the characteristic equation is given by:
\begin{align}
	\det(\mathbf{M}_k - \lambda_k \mathbf{I}) =
	\begin{vmatrix}
		a - \lambda_k & b              \\
		c             & d - \lambda_k
	\end{vmatrix} = 0
\end{align}

The development of the calculation implies that:
\begin{align}
	(a - \lambda_k)(d - \lambda_k) - bc         & = 0  \\
	\lambda_k^2 - (a + d) \lambda_k + (ad - bc) & = 0
\end{align}

Here, $a + d$ is the trace of the matrix $\mathbf{M}_{k}$ and $ad - bc$ is its determinant. Thus, we obtain the characteristic equation:
\begin{align}
	\lambda_k^2 - \text{tr}(\mathbf{M}_{k}) \lambda_k + \text{det}(\mathbf{M}_{k}) = 0
\end{align}

\chunkgap
\meth{Conditions for pattern emergence}
This polynomial gives the conditions for pattern emergence, certain modes of index $k$ associated with $\Re(\lambda_k)>0$ being able to ensure the appearance of a growing oscillation from a weak perturbation of the form $e^{\lambda_k t} \cos\left(\frac{k \pi x}{L}\right)$. To complete the demonstration, it is also necessary to guarantee that $\Re(\lambda_0)<0$, corresponding to the initial, stable and quasi-homogeneous state. The calculation of roots depends on the values of the operator $\mathbf{M}_k$, varying according to the Jacobian of the reaction functions and the diffusion terms:
\begin{align}
	\lambda_k = \frac{\text{tr}(\mathbf{M}_k) \pm \sqrt{\text{tr}(\mathbf{M}_k)^2 - 4\, \text{det}(\mathbf{M}_k)}}{2}
\end{align}

Surprisingly, the diffusion mechanism here perturbs the equilibrium and leads to the emergence of instabilities, the spatial modes considered being the eigenfunctions of the Laplacian operator. In many other examples of dynamic systems, diffusion is generally a vector of stability that smooths local concentrations and forces a return to the mean.  These counter-intuitive mechanisms were already identified as drivers of pattern emergence in Alan Turing's seminal contribution~\cite{turing1952chemical}. For more information, the reader can refer to~\cite{murray2007mathematicalII, krause2021modern}.

\newpage

\section*{\textcolor{reveditor}{Supplementary Note 3:} Analysis of six loading cases with the mechanical cloak}

\begin{figure}[H]
	\centering
	\includegraphics[width = 0.84\textwidth]{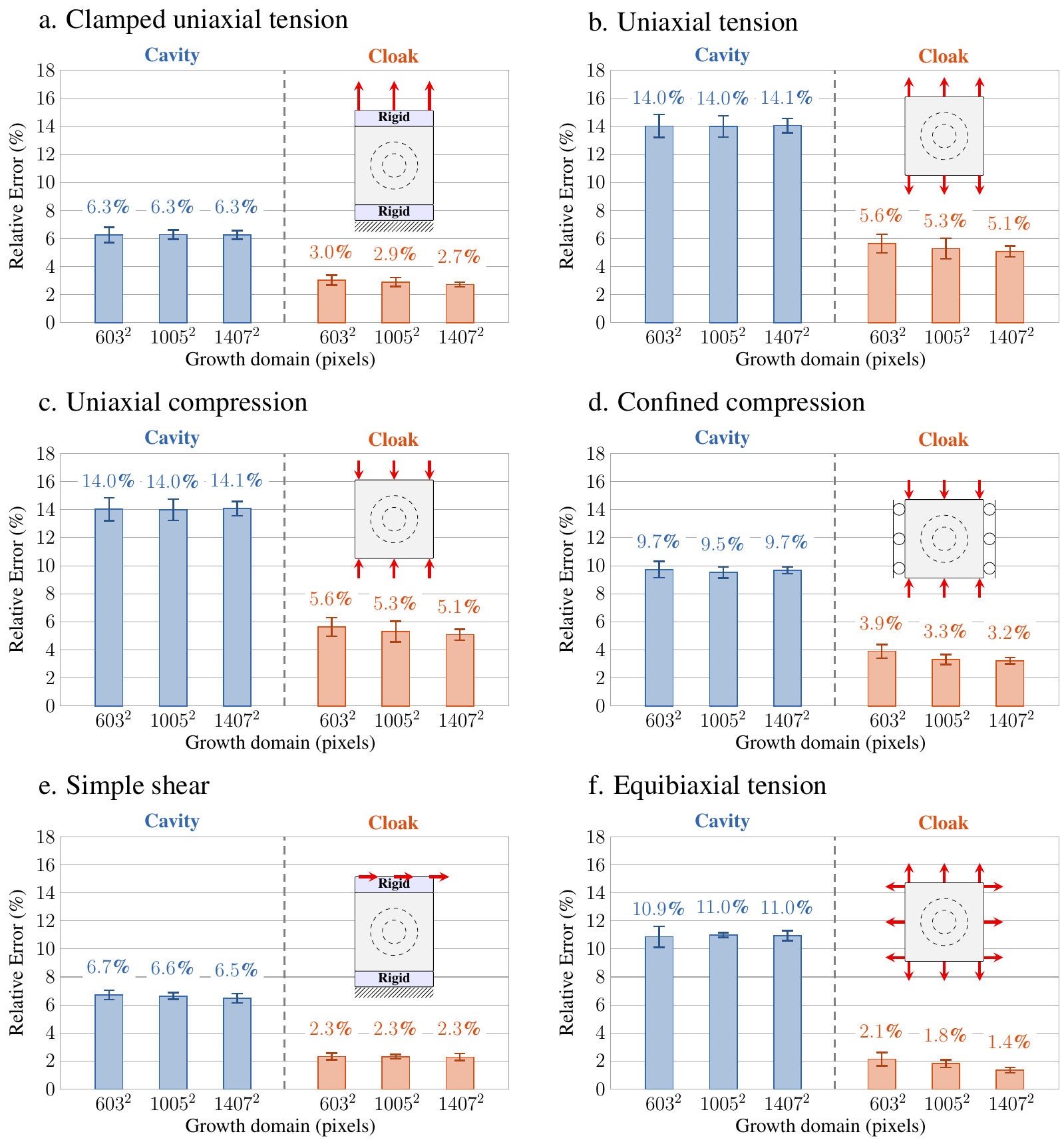}
	\caption{\color{blue} Summary of 6 loading modes, performed in each case for three growth domains of $603^2$, $1005^2$, and $1407^2$ pixels and 10 random seeds. The bar plots represent the mean results obtained in each case, associated with an error bar representing a variation of $\pm$ 1 standard deviation. All simulations are performed by imposing displacements represented by the red arrows on each test type.}
	\label{fig:Simulations_extra_synthese}
\end{figure}

In this section, various loading types are studied to demonstrate the effectiveness of the cloaking principle beyond the experimentally validated setup. We investigate six loading cases. For each case, simulations were performed using 10 different random seeds and 3 growth domain sizes ($603^2$, $1005^2$, and $1407^2$ pixels - Supplementary Figure~\ref{fig:Simulations_extra_synthese}).

The objective is twofold. First, to assess robustness to initial conditions by demonstrating low variance in results across different initializations. Second, to demonstrate the scalability of our techniques by showing automatic adaptation to different growth domain sizes. Since the final fabricated samples have a fixed size, changing the growth domain size effectively changes the inclusion density and reduces their relative size. Overall, the results indicate that performance tends to improve with an increasing quantity of generated pores, suggesting better control of the synthesized tensors.

It is important to note that while the generative model is defined on a pixel grid and is inherently scalable, the mechanical simulations presented here are conducted on a fixed physical domain size ($100 \times 100$ mm$^2$) for all resolutions. This choice ensures that the displacement fields, reported in millimeters, are directly comparable across different growth domain sizes ($603^2$, $1005^2$, $1407^2$ pixels). Consequently, increasing the grid resolution within this fixed physical window results in a finer microstructure with a higher density of smaller pores, rather than a larger macroscopic object. This approach allows us to isolate the effect of microstructural refinement on the homogenization quality and cloaking performance.

\textcolor{revtwo}{To facilitate comparison between the different loading conditions and contextualize these results alongside current literature, the quantitative performances of the morphogenetic cloaks are summarized in Table~\ref{tab:supp_metrics}. Consistent with the metric adopted in the main text, the improvement factor is defined as $R = \Delta_{\text{hole}} / \Delta_{\text{cloak}}$, where $\Delta$ denotes the mean relative displacement error. A higher $R$ indicates more effective cloaking. The three growth domain resolutions correspond to $603^2$, $1005^2$, and $1407^2$ pixels.}

\begin{table}[H]
    \centering
    \color{revtwo}
    \caption{\textcolor{revtwo}{Cloaking improvement factor $R = \Delta_{\text{hole}} / \Delta_{\text{cloak}}$ for the six assessed loading configurations and three growth domain resolutions. A higher $R$ indicates better cloaking performance.}}
    \label{tab:supp_metrics}
    \begin{tabular}{lccc}
        \toprule
        \textbf{Loading Case} & \textbf{$R$ ($603^2$ px)} & \textbf{$R$ ($1005^2$ px)} & \textbf{$R$ ($1407^2$ px)} \\
        \midrule
        Clamped uniaxial tension  & 2.07 & 2.16 & 2.30 \\
        Uniaxial tension          & 2.49 & 2.64 & 2.77 \\
        Uniaxial compression      & 2.49 & 2.64 & 2.77 \\
        Confined compression      & 2.49 & 2.88 & 3.00 \\
        Simple shear              & 2.90 & 2.87 & 2.84 \\
        Equibiaxial tension       & 5.09 & 6.07 & 8.08 \\
        \bottomrule
    \end{tabular}
\end{table}

\newpage

\subsection*{Clamped uniaxial tension}
This case corresponds to the experimental setup where the sample is clamped at both ends and stretched.

\begin{figure}[H]
	\centering
	\includegraphics[width=0.8\textwidth]{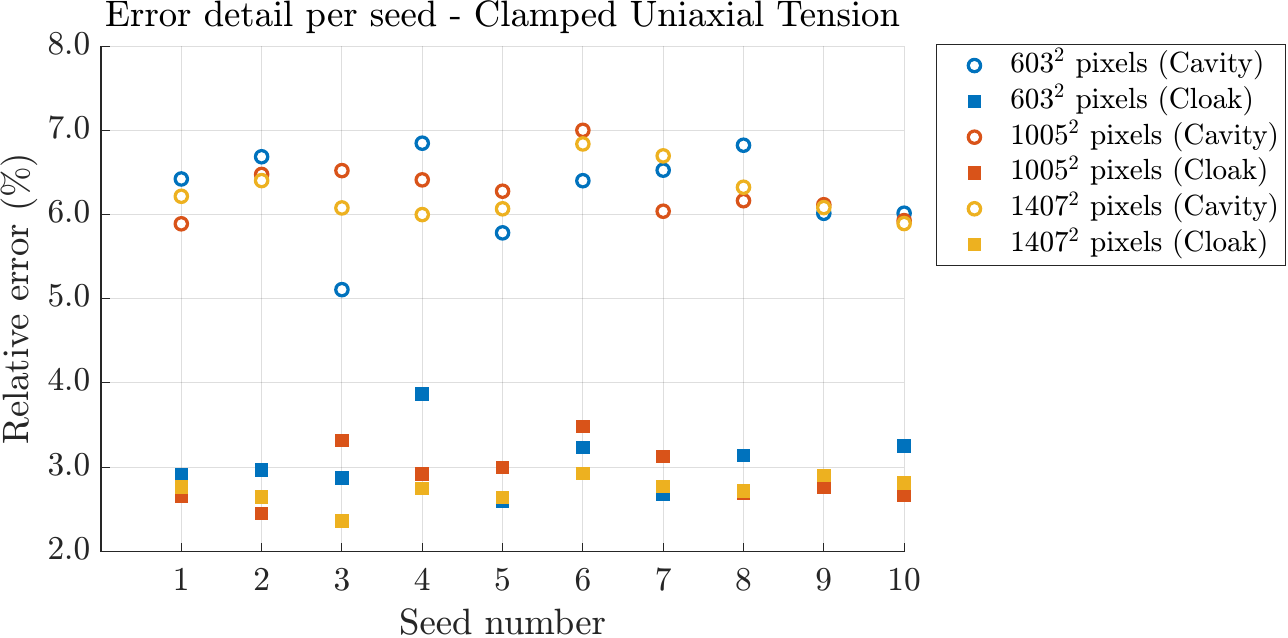}
	\caption{\color{blue} Relative errors obtained for the "Clamped uniaxial tension" loading, including three growth domain sizes and 10 different random seeds.}
	\label{fig:erreurs_essais_tension_reference}
\end{figure}

\newpage

\begin{figure}[H]
	\centering
	\includegraphics[width=0.8\textwidth]{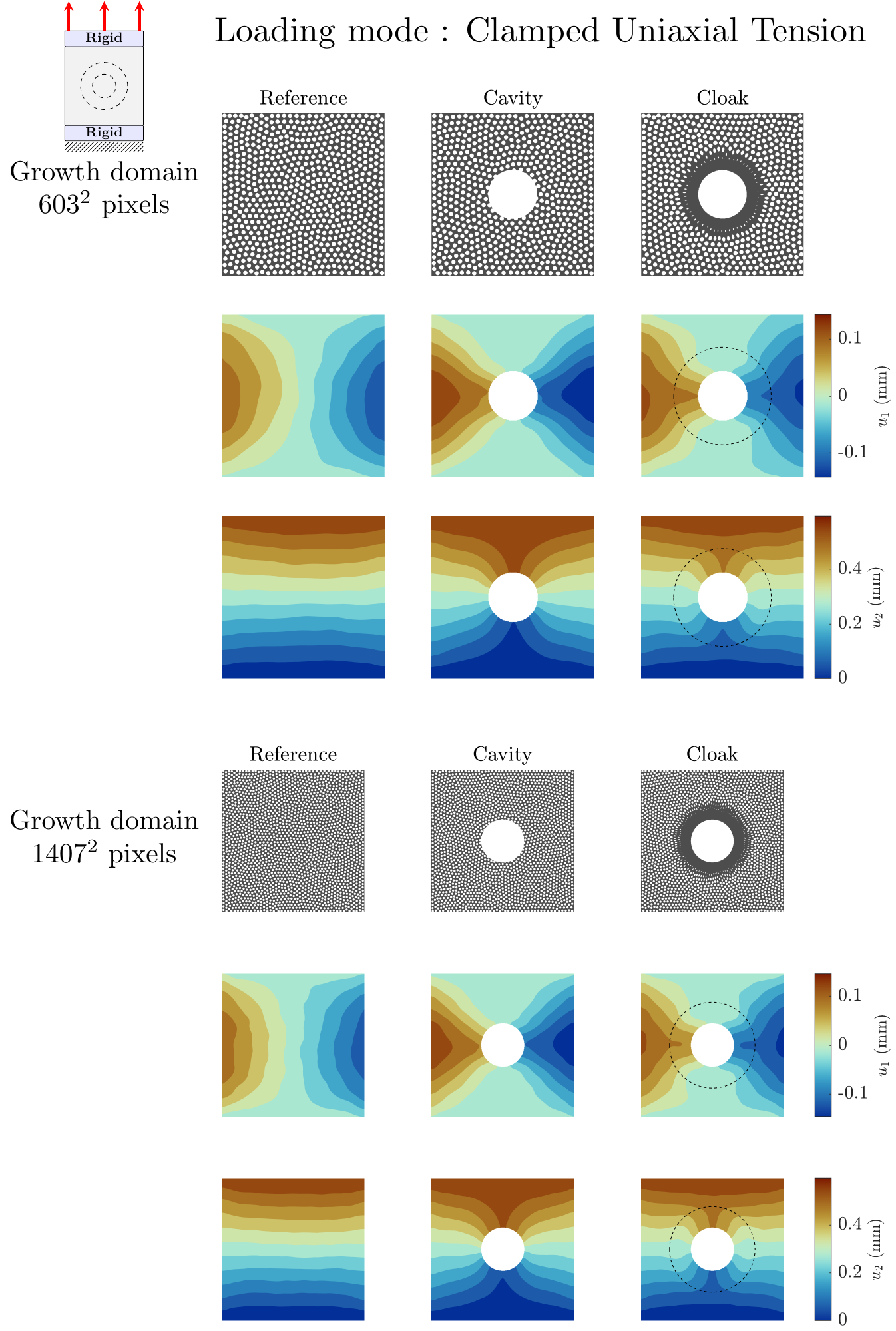}
	\caption{\color{blue} Displacement fields obtained for the "Clamped uniaxial tension" loading, for the first random seed value, from growth domains of $603^2$ pixels and $1407^2$ pixels.}
	\label{fig:champs_essais_tension_reference}
\end{figure}

\newpage

\subsection*{Uniaxial tension}
Standard uniaxial tension with free transverse boundaries, allowing for Poisson contraction.

\begin{figure}[H]
	\centering
	\includegraphics[width=0.8\textwidth]{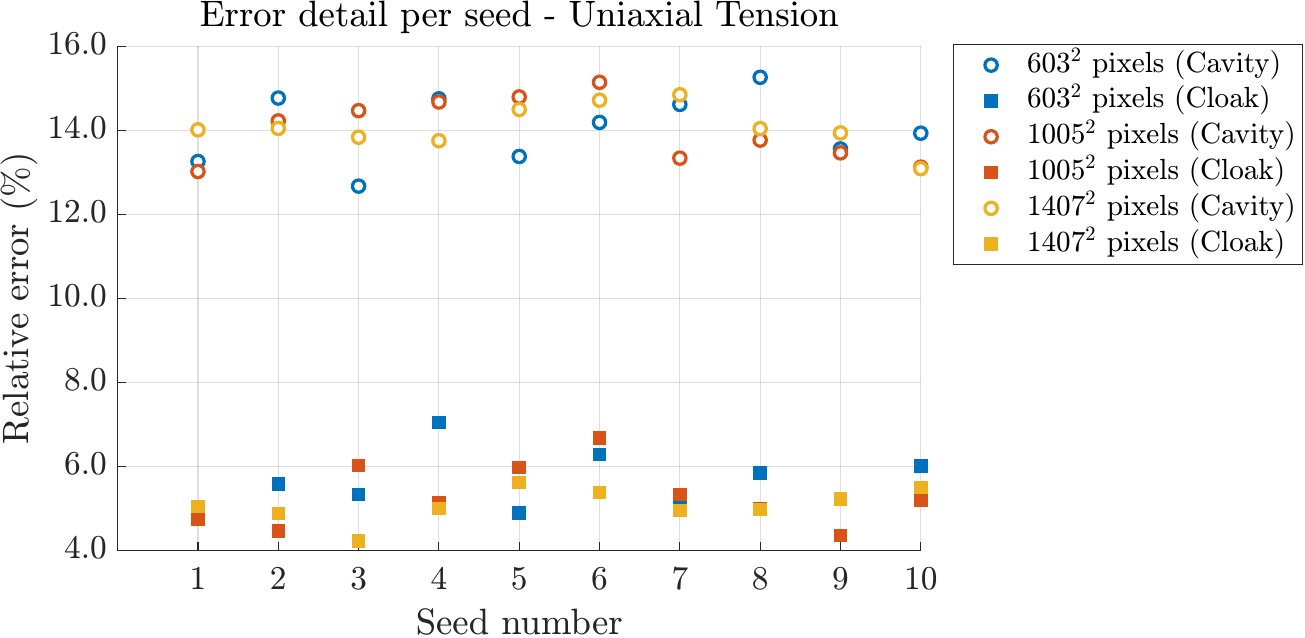}
	\caption{\color{blue} Relative errors obtained for the "Uniaxial tension" loading, including three growth domain sizes and 10 different random seeds.}
	\label{fig:erreurs_essais_tension_free}
\end{figure}

\newpage

\begin{figure}[H]
	\centering
	\includegraphics[width=0.8\textwidth]{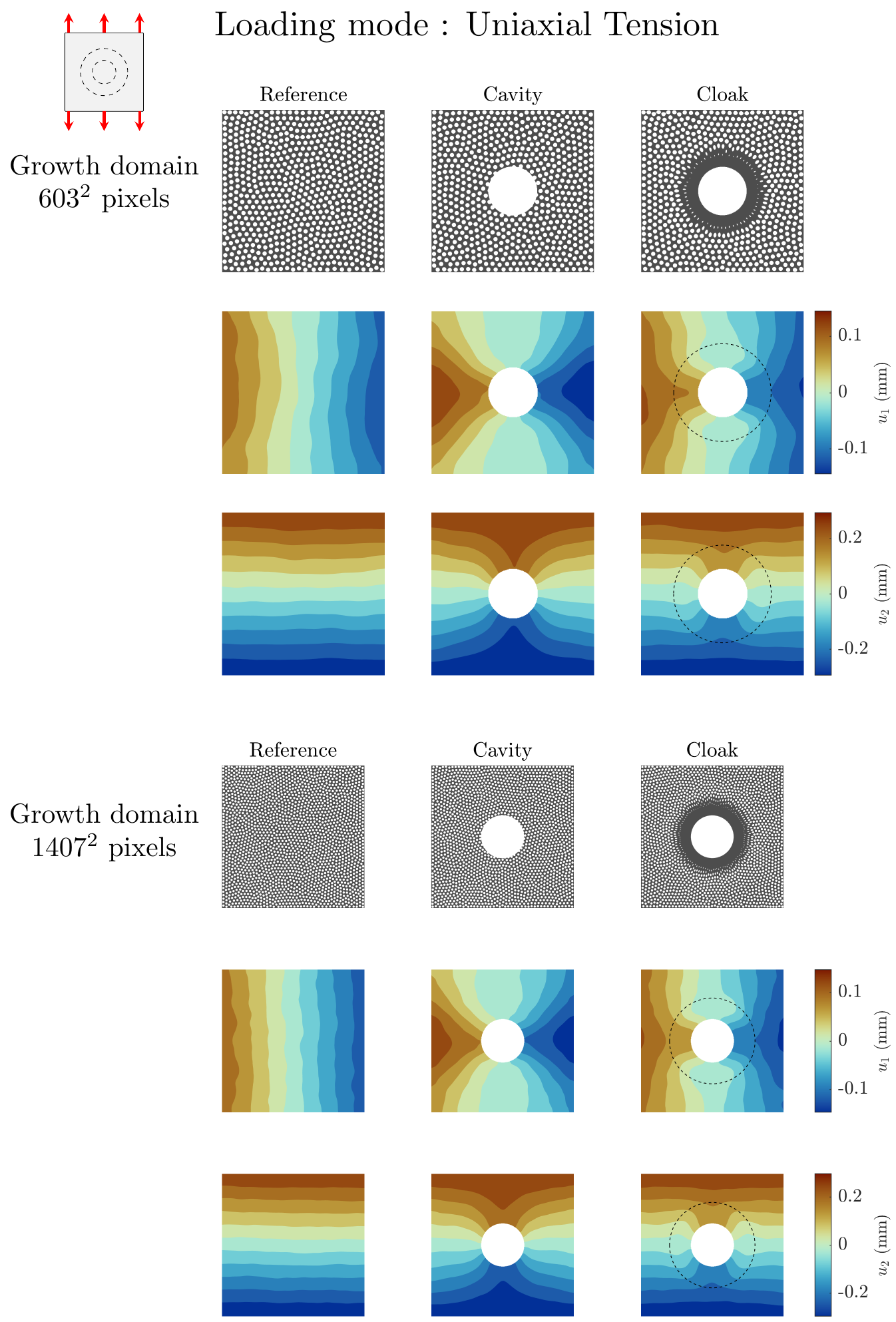}
	\caption{\color{blue} Displacement fields obtained for the "Uniaxial tension" loading, for the first random seed value, from growth domains of $603^2$ pixels and $1407^2$ pixels.}
	\label{fig:champs_essais_tension_free}
\end{figure}

\newpage

\subsection*{Uniaxial compression}
Standard uniaxial compression, symmetric to the tension case.

\begin{figure}[H]
	\centering
	\includegraphics[width=0.8\textwidth]{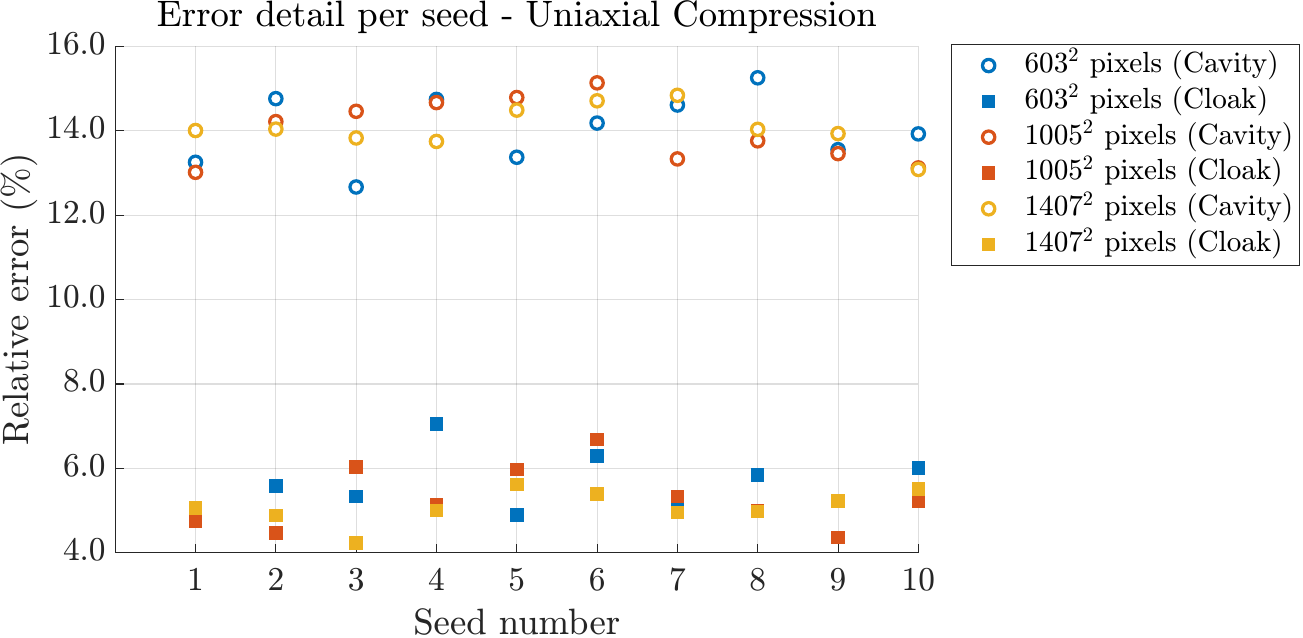}
	\caption{\color{blue} Relative errors obtained for the "Uniaxial compression" loading, including three growth domain sizes and 10 different random seeds.}
	\label{fig:erreurs_essais_compression_symmetric}
\end{figure}

\newpage

\begin{figure}[H]
	\centering
	\includegraphics[width=0.8\textwidth]{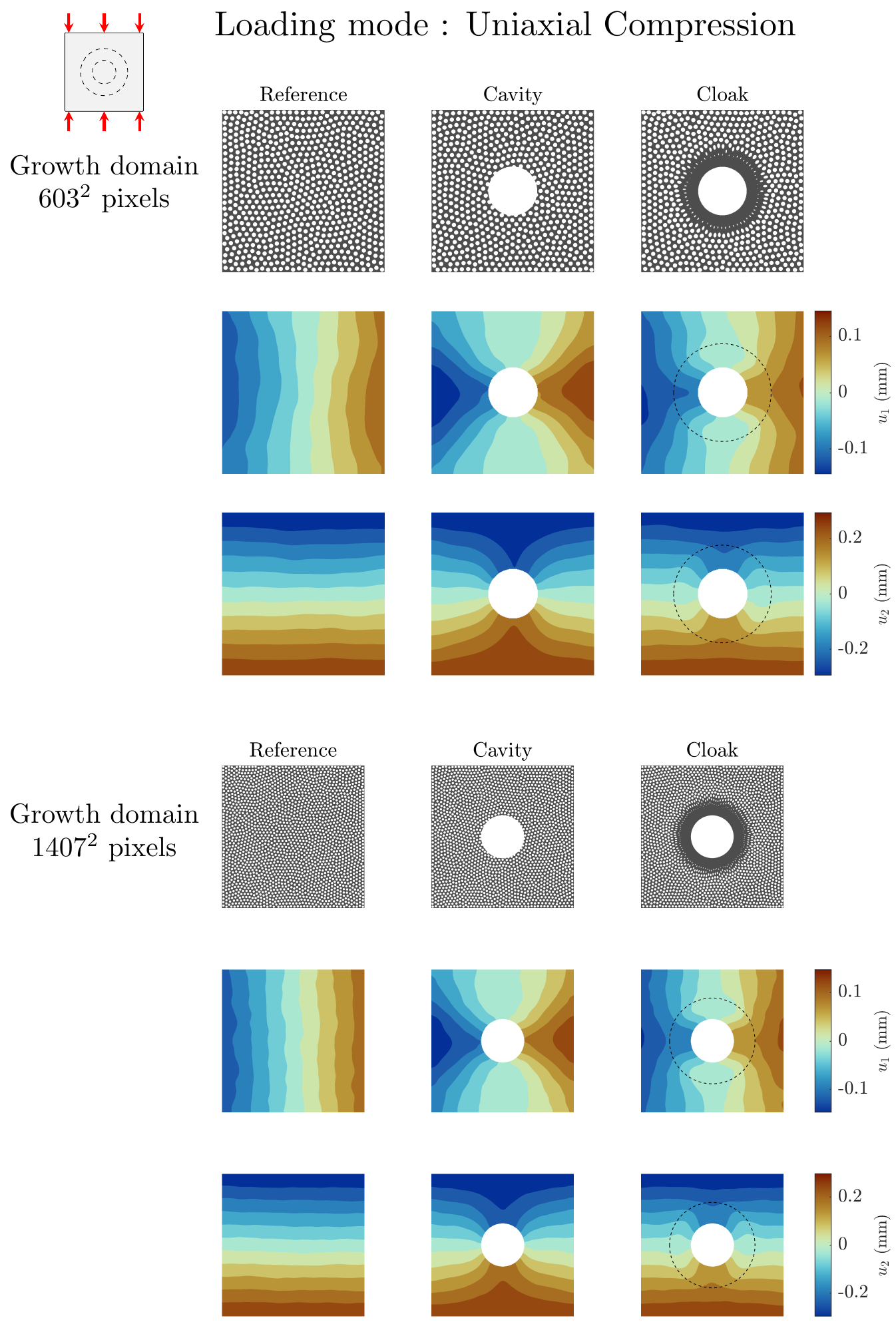}
	\caption{\color{blue} Displacement fields obtained for the "Uniaxial compression" loading, for the first random seed value, from growth domains of $603^2$ pixels and $1407^2$ pixels.}
	\label{fig:champs_essais_compression_symmetric}
\end{figure}

\newpage

\subsection*{Confined compression}
Compression with constrained transverse displacement (sliding boundary conditions), mimicking a confined environment.

\begin{figure}[H]
	\centering
	\includegraphics[width=0.8\textwidth]{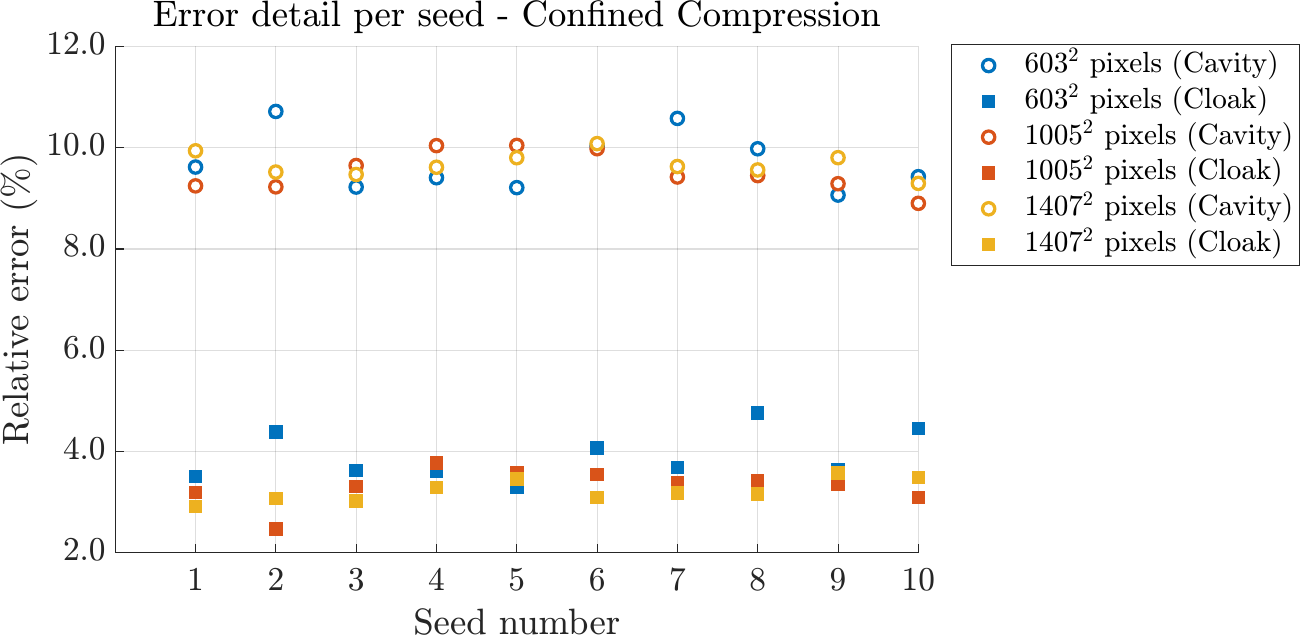}
	\caption{\color{blue} Relative errors obtained for the "Confined compression" loading, including three growth domain sizes and 10 different random seeds.}
	\label{fig:erreurs_essais_compression_sliding}
\end{figure}

\newpage

\begin{figure}[H]
	\centering
	\includegraphics[width=0.8\textwidth]{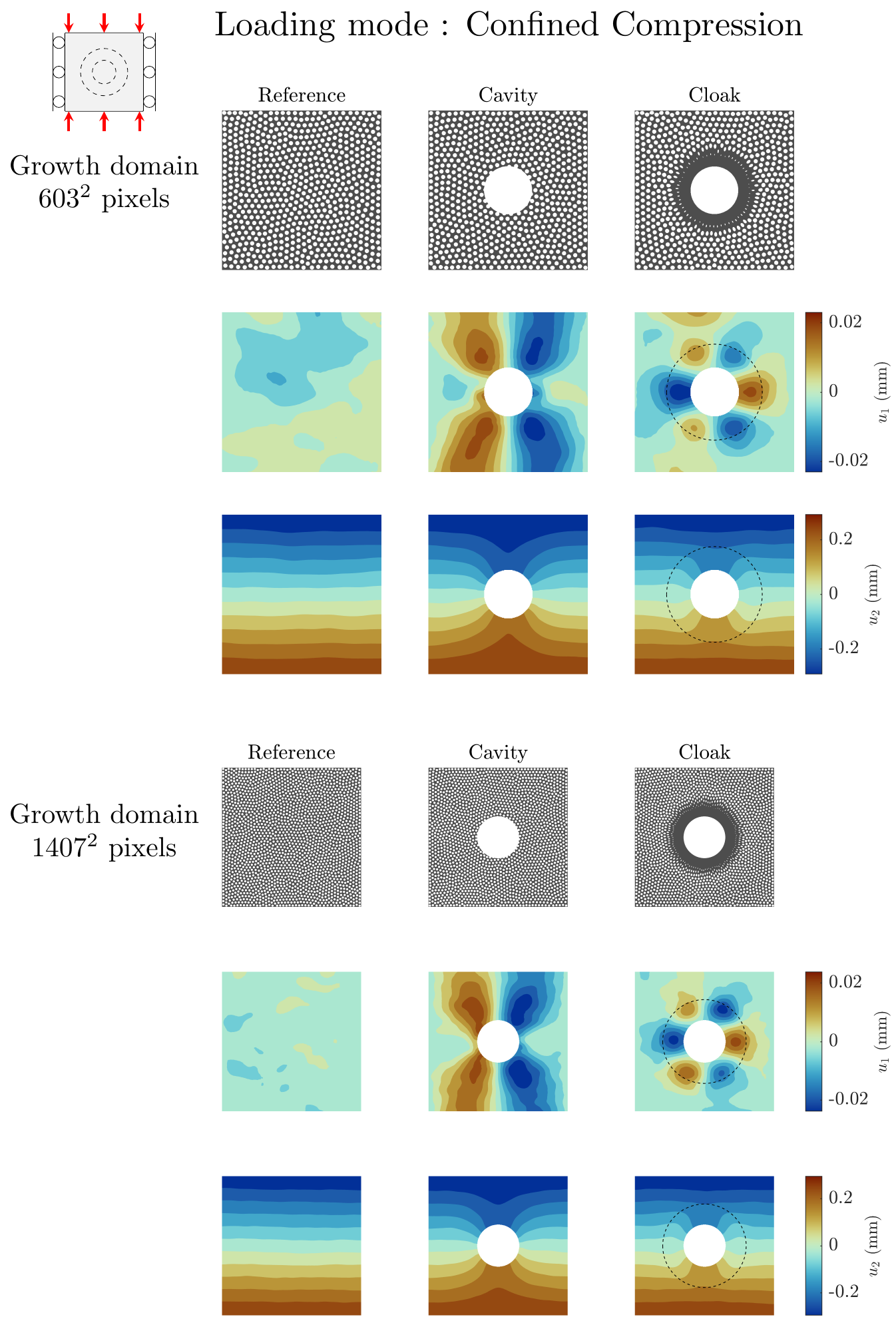}
	\caption{\color{blue} Displacement fields obtained for the "Confined compression" loading, for the first random seed value, from growth domains of $603^2$ pixels and $1407^2$ pixels.}
	\label{fig:champs_essais_compression_sliding}
\end{figure}

\newpage

\subsection*{Simple shear}
Simple shear loading applied to the domain.

\begin{figure}[H]
	\centering
	\includegraphics[width=0.8\textwidth]{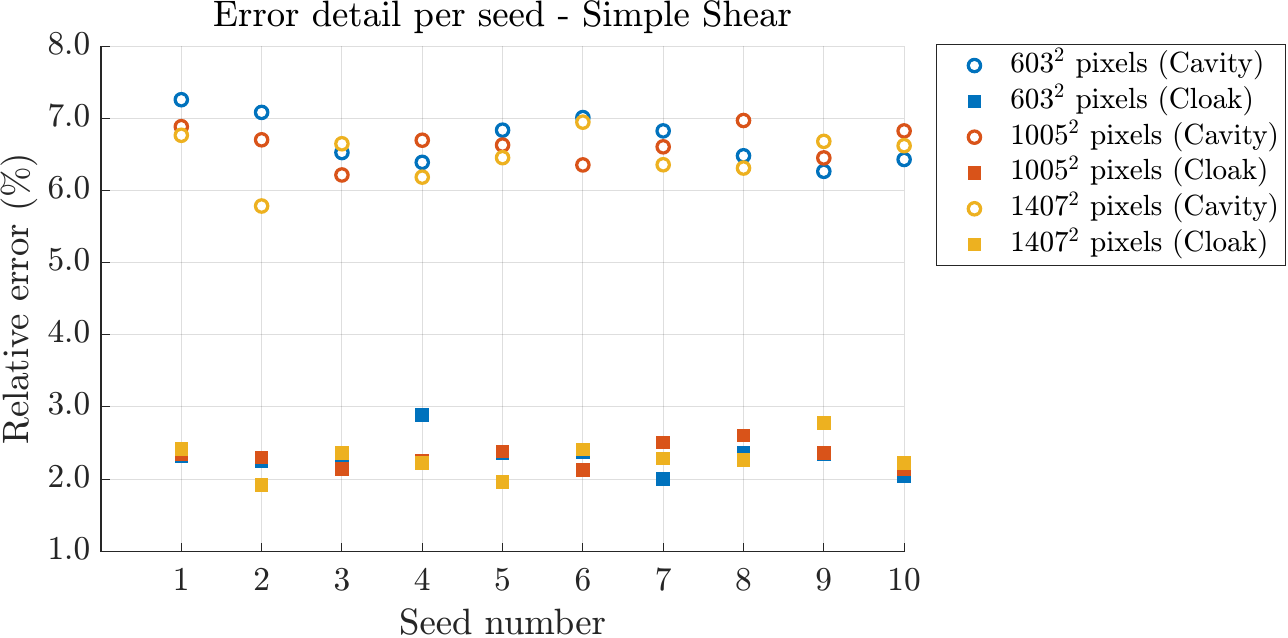}
	\caption{\color{blue} Relative errors obtained for the "Simple shear" loading, including three growth domain sizes and 10 different random seeds.}
	\label{fig:erreurs_essais_shear}
\end{figure}

\newpage

\begin{figure}[H]
	\centering
	\includegraphics[width=0.8\textwidth]{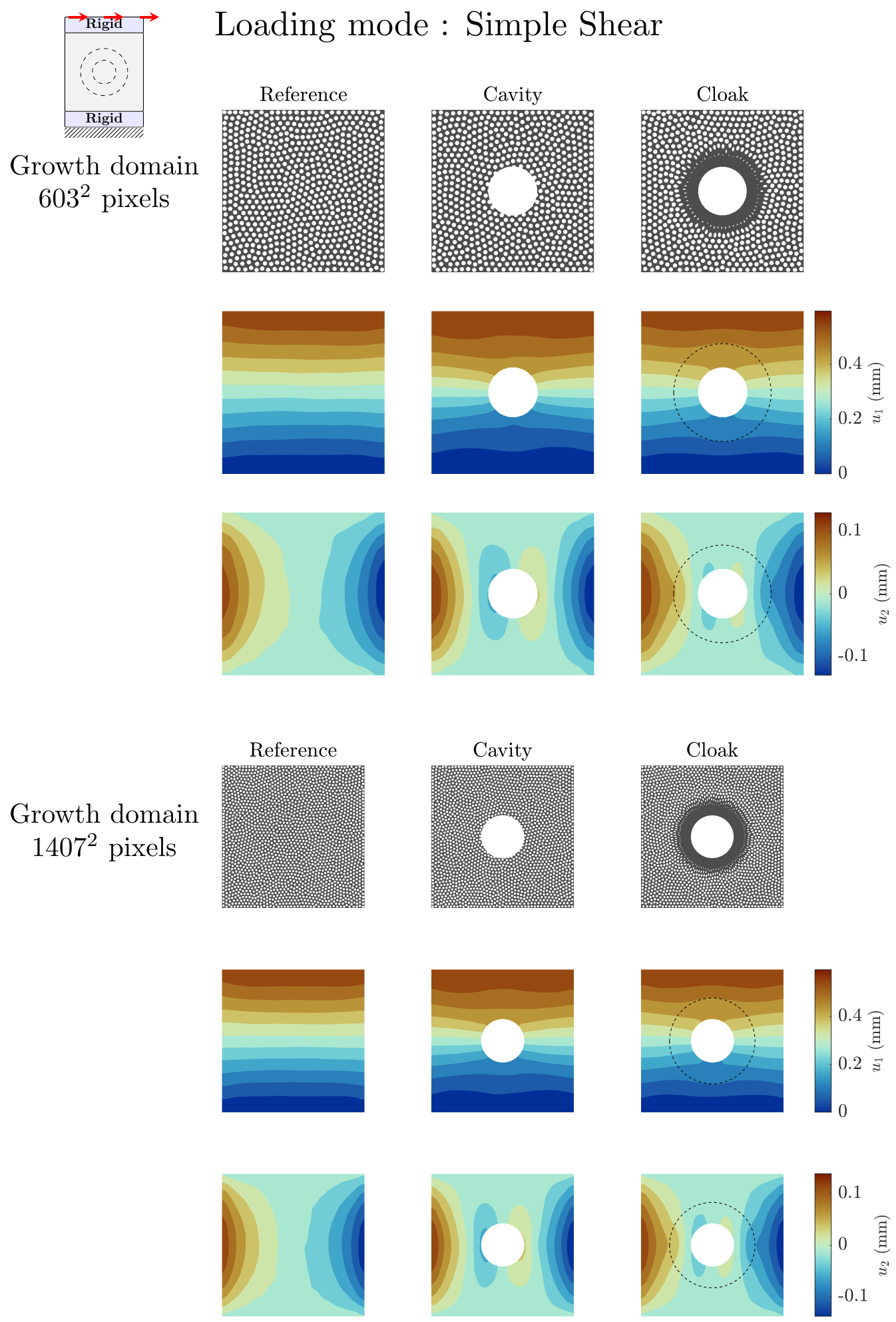}
	\caption{\color{blue} Displacement fields obtained for the "Simple shear" loading, for the first random seed value, from growth domains of $603^2$ pixels and $1407^2$ pixels.}
	\label{fig:champs_essais_shear}
\end{figure}

\newpage

\subsection*{Equibiaxial tension}
Simultaneous tension in both X and Y directions (dilating mode).

\begin{figure}[H]
	\centering
	\includegraphics[width=0.8\textwidth]{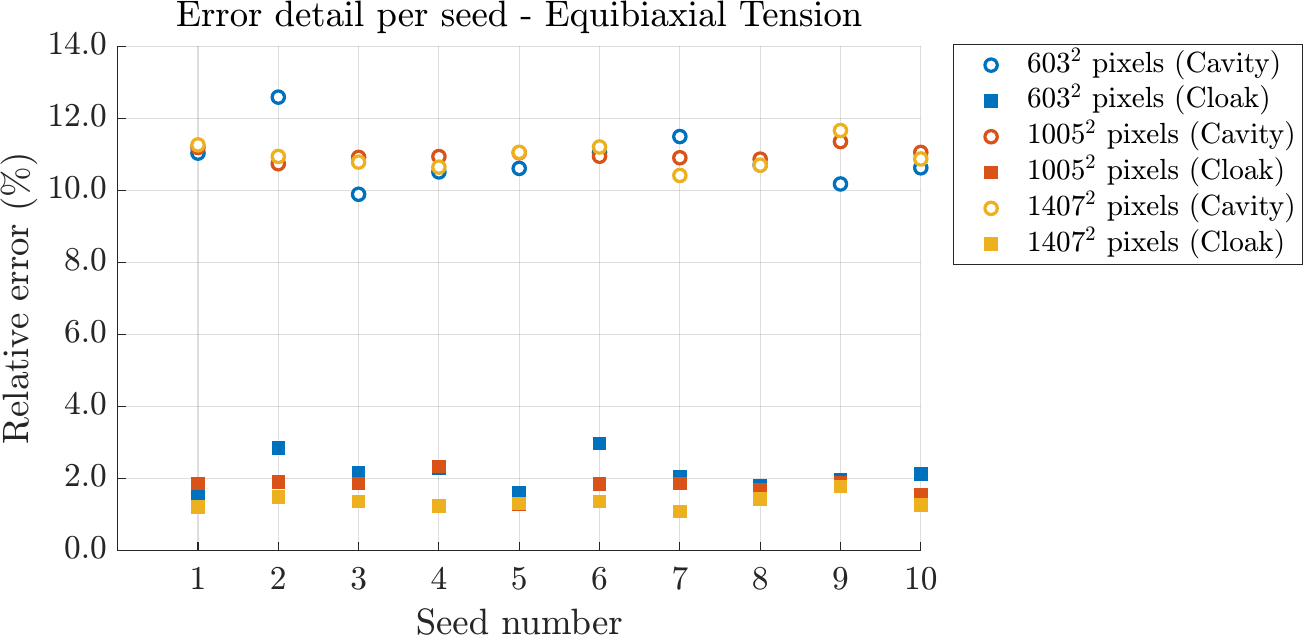}
	\caption{\color{blue} Relative errors obtained for the "Equibiaxial tension" loading, including three growth domain sizes and 10 different random seeds.}
	\label{fig:erreurs_essais_dilating}
\end{figure}

\newpage

\begin{figure}[H]
	\centering
	\includegraphics[width=0.8\textwidth]{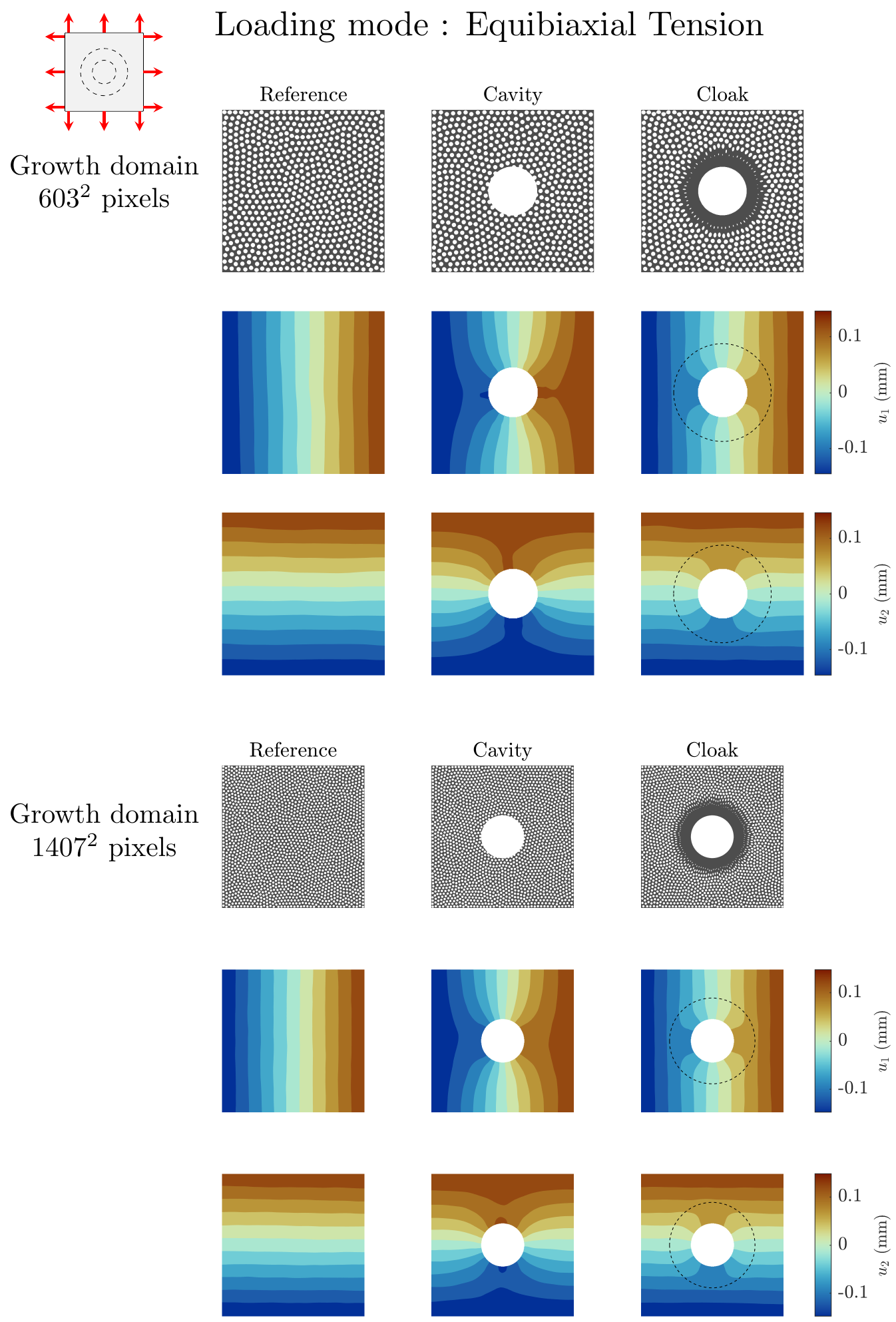}
	\caption{\color{blue} Displacement fields obtained for the "Equibiaxial tension" loading, for the first random seed value, from growth domains of $603^2$ pixels and $1407^2$ pixels.}
	\label{fig:champs_essais_dilating}
\end{figure}

\newpage

\section*{\textcolor{reveditor}{Supplementary Note 4:} Analysis of Stress-Field Guided Growth}

To evaluate the generalization capabilities of the proposed method, we apply the stress-guided growth strategy to two additional standard mechanical test cases: a distributed three-point bending scenario and a cantilever beam. These cases differ significantly from the training scenarios (tension/compression) and test the ability of the local growth rule to adapt to complex stress gradients. All simulations are performed on a $200 \times 100$ mm$^2$ domain with a linear elastic material ($E=3$ GPa, $\nu=0.35$) under a total load of 2 kN.

The local anisotropy of the generated structure is driven by the diffusion tensor's anisotropy ratio $\lambda$, which acts as a proxy for the elongation of the emerging patterns. To align the microstructure with the stress flow, $\lambda$ is linearly mapped from the local stress deviatoric component, ranging from $\lambda=0.5$ (isotropic growth) in hydrostatic regions to $\lambda=0.9$ (highly anisotropic) in regions dominated by uniaxial stress.

\subsection*{Distributed Three-point Bending}

This loading case corresponds to a "Bridge" configuration. The structure is supported at its bottom corners by two anchors, each covering 15\% of the total width (left corner fixed in X and Y, right corner fixed in Y only, allowing sliding). A vertical downward load is applied at the bottom center, distributed over 10\% of the width.

\begin{figure}[H]
	\centering
	\includegraphics[width=0.6\textwidth]{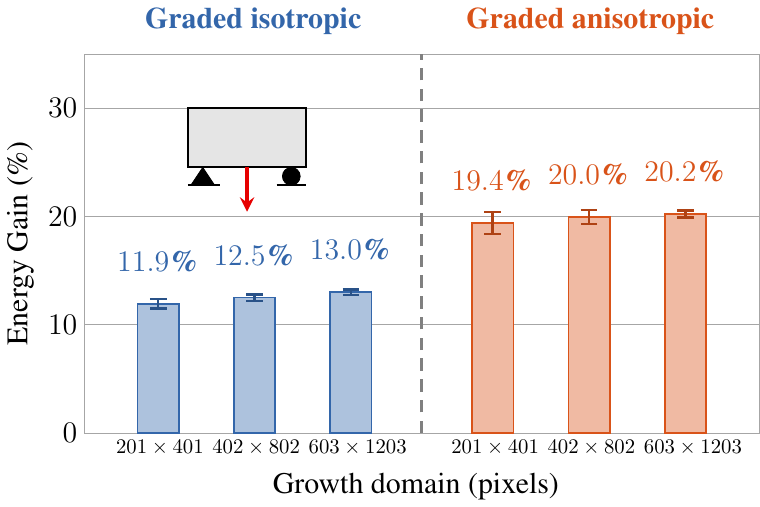}
	\caption{\color{blue} Strain Energy Reduction obtained for the "Distributed Three-point Bending" case, comparing the isotropic (variable density) and guided (anisotropic) structures against the reference. The "Energy Gain" reported corresponds to the relative reduction in potential energy compared to the reference, reflecting an increase in global stiffness.}
	\label{fig:energy_bridge}
\end{figure}

The displacement fields for representative simulations are shown below, illustrating the varying structural responses.

\begin{figure}[H]
	\centering
	\includegraphics[width=0.82\textwidth]{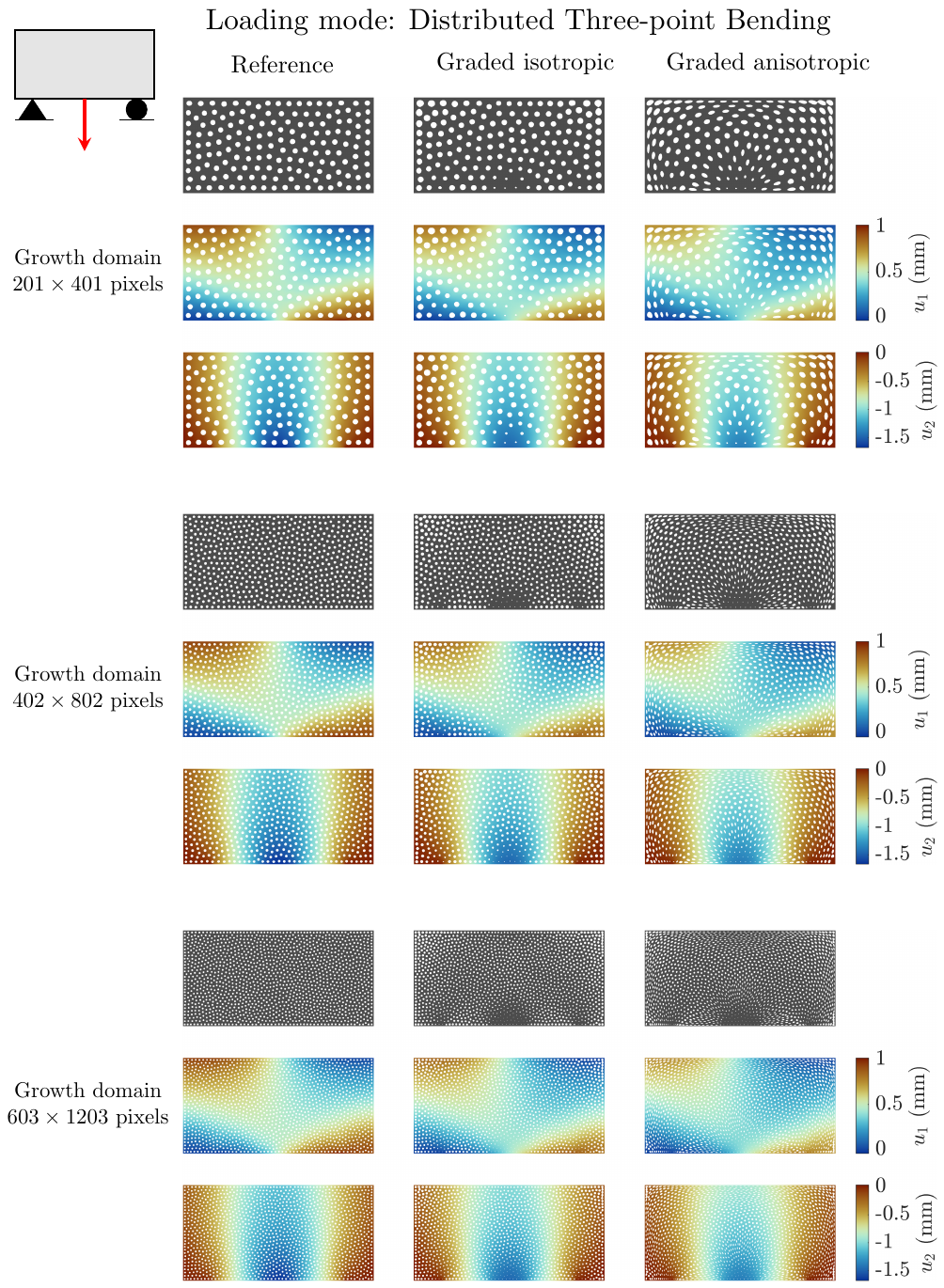}
	\caption{\color{blue} Displacement field magnitude maps for the Distributed Three-point Bending ("Bridge") case. The rows correspond to increasing growth domain resolutions. Columns compare: (Left) Uniform reference structure; (Middle) Graded Isotropic optimization, where local porosity is adapted to stress magnitude; (Right) Graded Anisotropic optimization, where pore orientation is additionally aligned with principal stress trajectories. The anisotropic guiding (right) \textcolor{revtwo}{promotes a directional material distribution that aligns with the dominant stress flow, which reduces} overall deflection more effectively than density modulation alone.}
	\label{fig:champs_bridge}
\end{figure}

\newpage

\subsection*{Cantilever Bending}

This case represents a cantilever beam scenario. The structure is fully clamped along its left edge (fixed displacement in X and Y). A vertical downward load is applied at the center of the right edge, distributed over 10\% of the domain height.

\begin{figure}[H]
	\centering
	\includegraphics[width=0.6\textwidth]{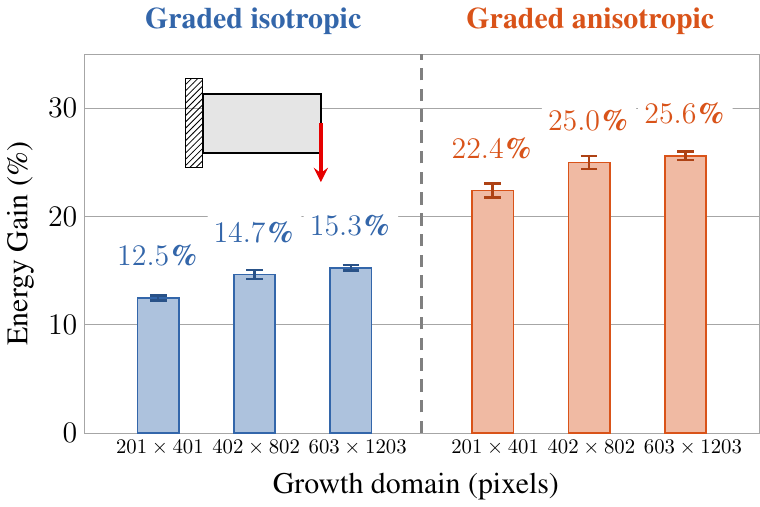}
	\caption{\color{blue} Strain Energy Reduction obtained for the "Cantilever Bending" case. The metric reflects the gain in structural stiffness relative to the reference design.}
	\label{fig:energy_cantilever}
\end{figure}

\begin{figure}[H]
	\centering
	\includegraphics[width=0.8\textwidth]{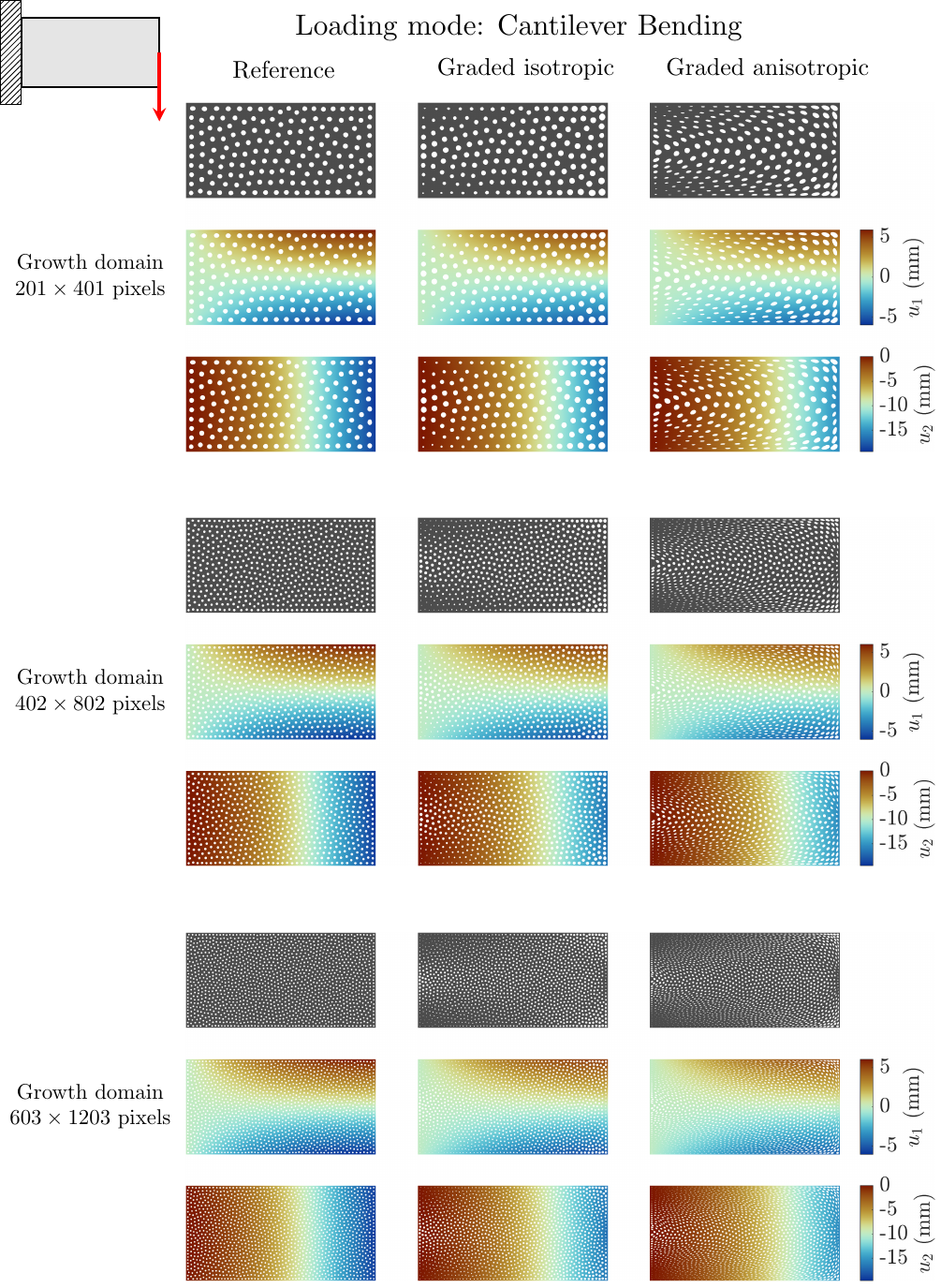}
	\caption{\color{blue} Displacement field magnitude maps for the Cantilever Bending case. Organization follows Fig.~\ref{fig:champs_bridge}: rows indicate increasing resolution, and columns compare Reference, Graded Isotropic (density only), and Graded Anisotropic (density + orientation) theories. The fully guided structures (right) exhibit naturally emerging stiffeners aligned with the bending moment distribution, minimizing tip displacement.}
	\label{fig:champs_cantilever}
\end{figure}

\section*{\textcolor{reveditor}{Supplementary Note 5:} Meshing and Numerical Accuracy}

To assess the reliability of our numerical predictions, we performed a convergence study concentrating on the mesh density. The simulations were conducted using the 3DEXPERIENCE platform (SIMULIA) with the Abaqus Standard solver. The domains were meshed using quadratic tetrahedral elements (C3D10).

\begin{figure}[H]
	\centering
	\includegraphics[width=0.8\textwidth]{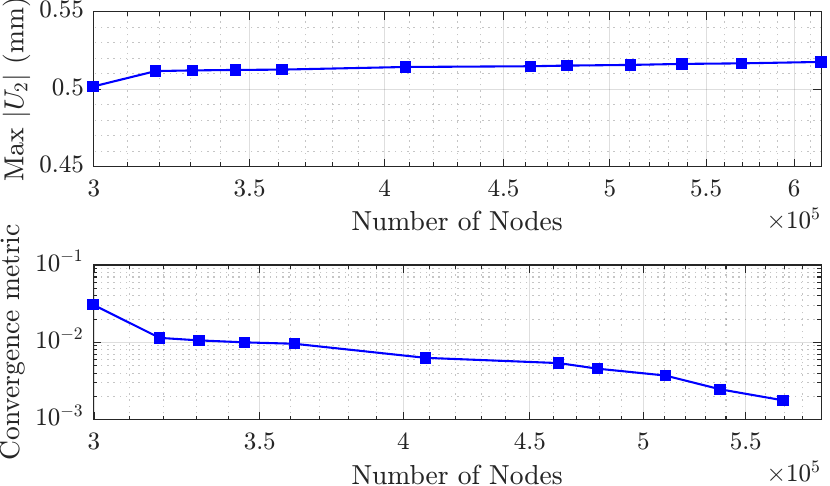}
	\caption{\color{blue} Mesh convergence study showing the relative error of the simulated displacement field with respect to a highly refined reference mesh, as a function of the number of elements.}
	\label{fig:convergence}
\end{figure}

The mesh convergence study confirms the numerical stability of the solution. The selected mesh density yields a relative error of 6.2\% between the simulated and experimental cloaking ratios, validating the accuracy of the 3D model with clamping boundary conditions.

\newpage
\section*{\textcolor{reveditor}{Supplementary Note 6:} Symmetrization of Displacement Fields}

In the experimental setup, the lower jaw is fixed while the upper jaw applies the tensile load. This asymmetry induces a global rigid body motion. To facilitate comparison with numerical simulations (often performed with symmetric boundary conditions) and to center the analysis on the deformation of the specimen rather than its translation, we applied a symmetrization procedure to the measured displacement fields.

\begin{figure}[H]
	\centering
	\includegraphics[width=1.0\textwidth]{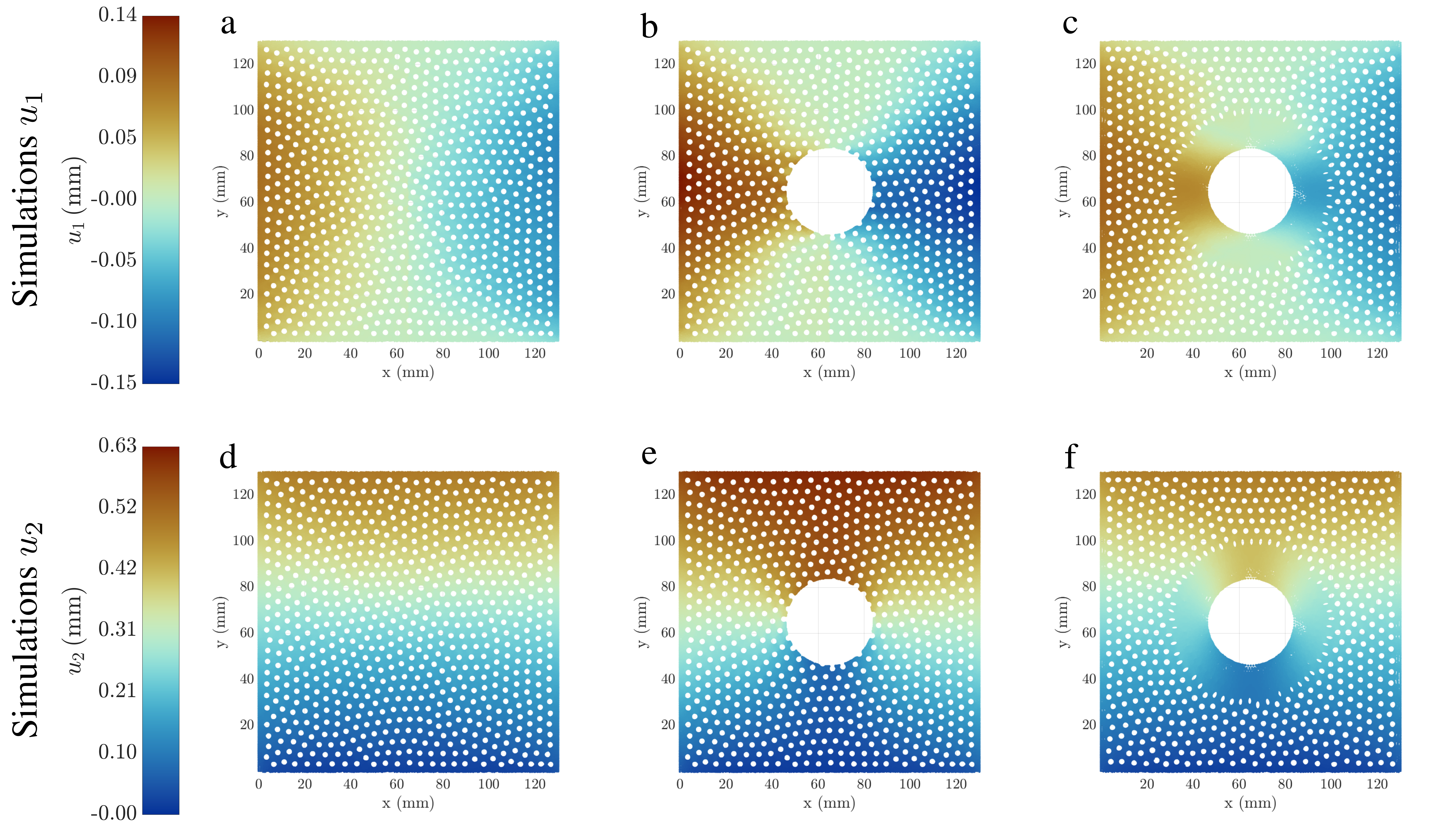}
	\caption{\color{blue} Visualization of the raw experimental vertical displacement fields ($u_2$) prior to symmetrization for the reference, cavity, and cloaked configurations, illustrating the effect of the fixed lower jaw.}
	\label{fig:raw_field_u2}
\end{figure}

These raw fields yield $\Delta_{\text{hole}} = 28.1\%$ and $\Delta_{\text{cloak}} = 5.8\%$. After symmetrization (removing the mean of $u_2$), the metrics obtained are $\Delta_{\text{hole}} = 39.2\%$ and $\Delta_{\text{cloak}} = 10.7\%$. Consequently, the improvement factor drops from $4.8\times$ for the raw data to $3.66\times$ after symmetrization, but provide metrics in conditions closer to those of the literature.

\newpage

\section*{\textcolor{reveditor}{Supplementary Note 7:} Experimental Strain Fields}

\begin{figure}[H]
	\centering
	\includegraphics[width=0.85\linewidth]{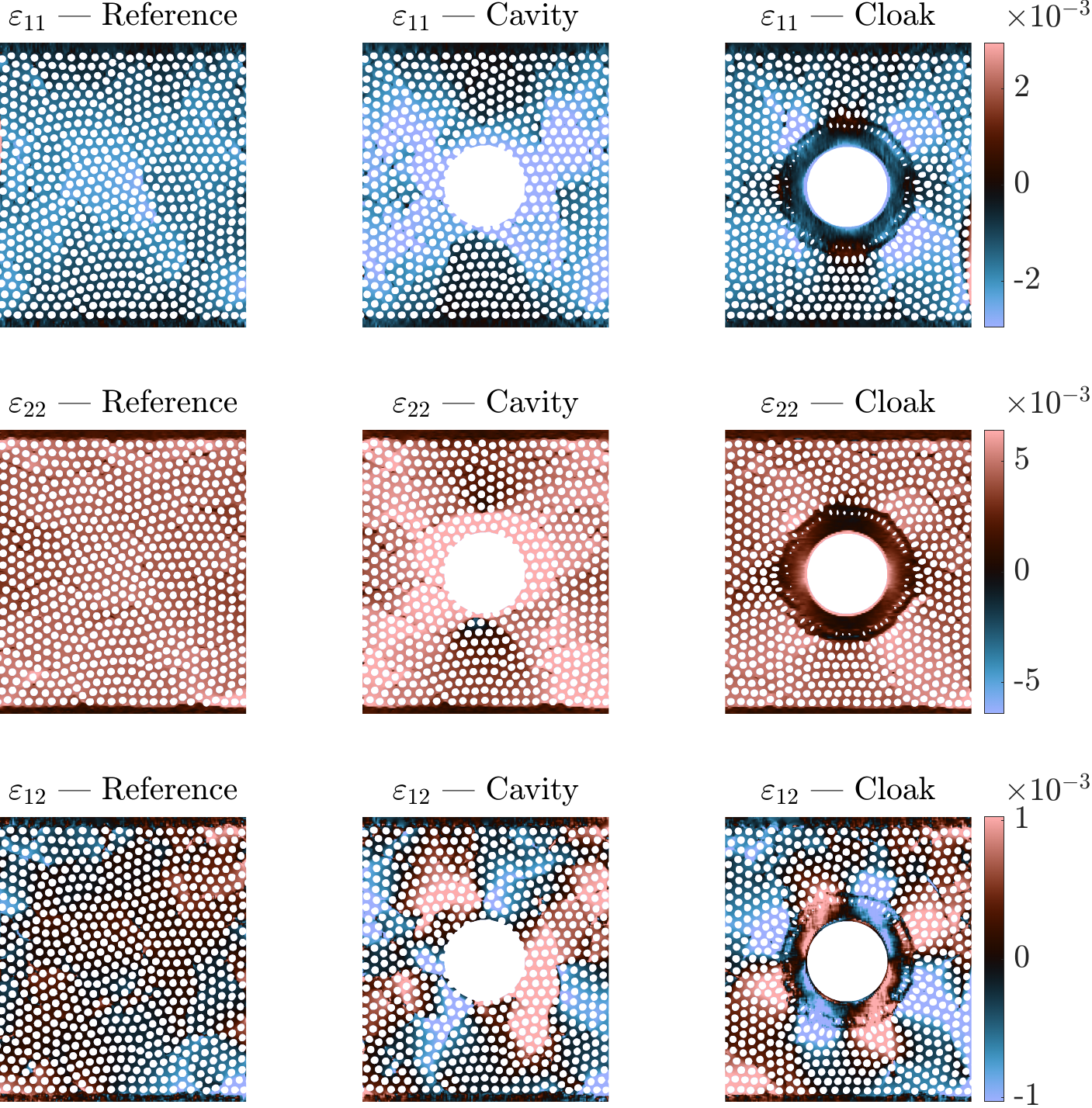}
	\caption{\color{revtwo} \textbf{Experimental strain fields.} Spatial distribution of the strain components ($\varepsilon_{xx}$, $\varepsilon_{yy}$, and $\varepsilon_{xy}$) obtained by spatial derivation of the experimental displacement fields. The fields correspond to the reference, uncloaked, and morphogenetic cloaked structures. Color scales are saturated to the $[1\%, 99\%]$ range to filter out local structural singularities while revealing the macroscopic behavior.}
	\label{fig:strain_fields_experiment}
\end{figure}

\textcolor{revtwo}{In addition to the macroscopic evaluation based on displacement fields, we computed the experimental strain tensor components ($\varepsilon_{xx}$, $\varepsilon_{yy}$, and $\varepsilon_{xy}$) by spatial derivation of the measured displacement fields. Figure~\ref{fig:strain_fields_experiment} presents these components for the reference plate, the uncloaked structure with a central cavity, and the morphogenetic cloak. To ensure consistent evaluation and colorblind accessibility, the strain maps are visualized using the \textit{managua} colormap~\cite{crameri2020misuse}, selected for its colorblind compatibility and its sequential dynamics better suited to sign-constrained normal strain components ($\varepsilon_{xx}$, $\varepsilon_{yy}$), in contrast to the diverging \textit{roma} colormap used for displacement fields. Furthermore, the color scales are intentionally saturated to the $[1\%, 99\%]$ percentile range. This strategy suppresses highly localized strain concentrations inherent to the porous architecture and edge effects, allowing the macroscopic mechanical behavior and the overall recovery of the unperturbed fields induced by the morphogenetic cloak to be clearly confirmed.}

\newpage

\section*{\textcolor{reveditor}{Supplementary Note 8:} Equivalence of Tension and Compression Databases}

A fundamental assumption of the linear elastic model employed in this work is that the mechanical response is independent of the sign of the applied displacement. Consequently, we validate that the stiffness tensors $\mathbb{C}$ characterizing the generated microstructures are identical whether determined under tensile or compressive periodic boundary conditions. This symmetry allows for the generation of a unique cloaking structure efficient with different boundary conditions.

\begin{figure}[H]
	\centering
	\includegraphics[width=1.0\textwidth]{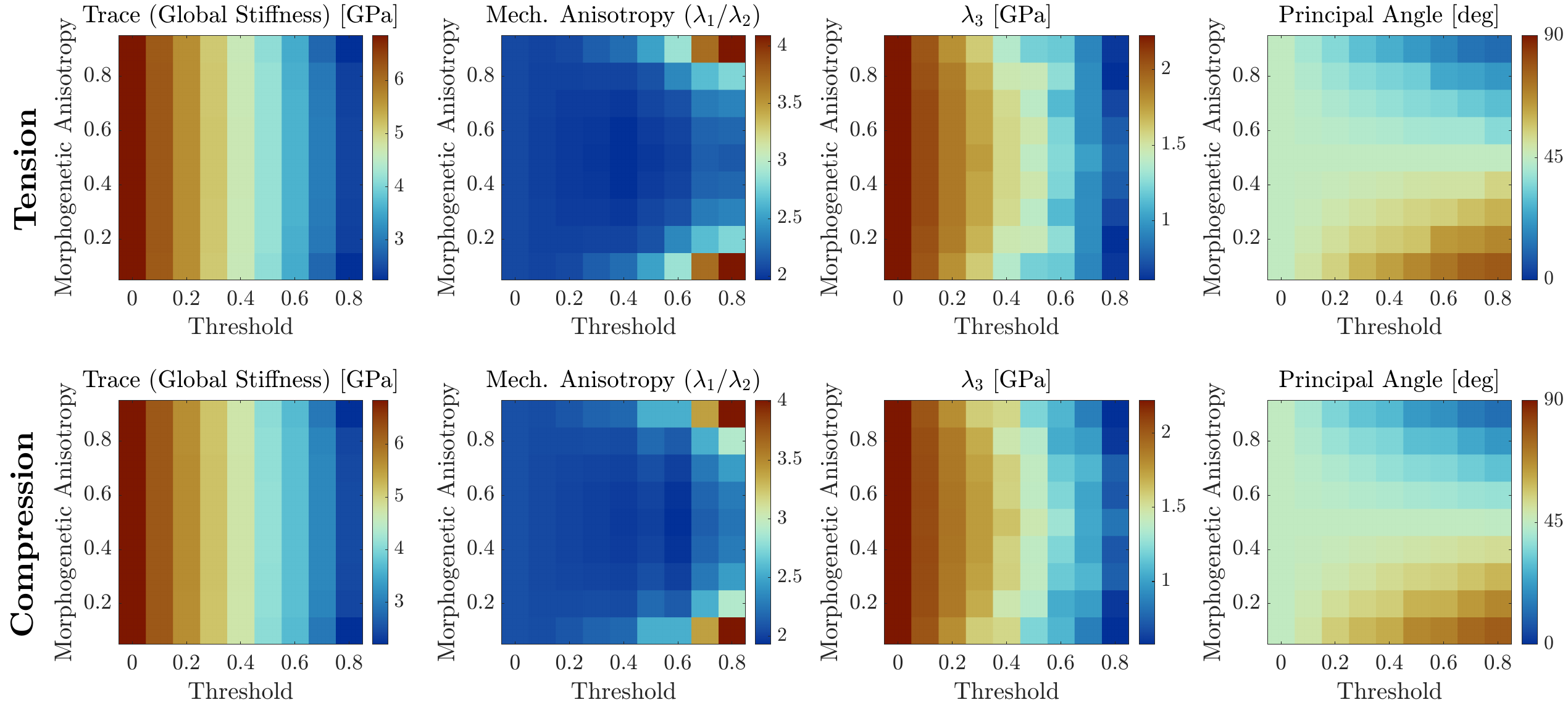}
	\caption{\color{blue} Achievable mechanical properties for the Tension (top) and Compression (bottom) databases. The maps display the stiffness trace, mechanical anisotropy, shear modulus $\lambda_3$, and principal angle as a function of the morphogenetic parameters (threshold and anisotropy).}
	\label{fig:achievable_props}
\end{figure}

To verify this, we generated a database of microstructures and computed their effective elastic properties under both tension and compression. Figure~\ref{fig:achievable_props} shows the achievable mechanical properties for both databases. \textcolor{revtwo}{Specifically, the mapping reveals the following invariant bounds across the explored parameter space: the stiffness trace spans $\lambda_1 + \lambda_2 + \lambda_3 \in [2.31,\;7.95]$~GPa, the principal stiffness anisotropy ranges within $\lambda_1/\lambda_2 \in [1.93,\;5.72]$, and the shear eigenvalue covers $\lambda_3 \in [0.47,\;2.24]$~GPa.}

\begin{figure}[H]
	\centering
	\includegraphics[width=1.0\textwidth]{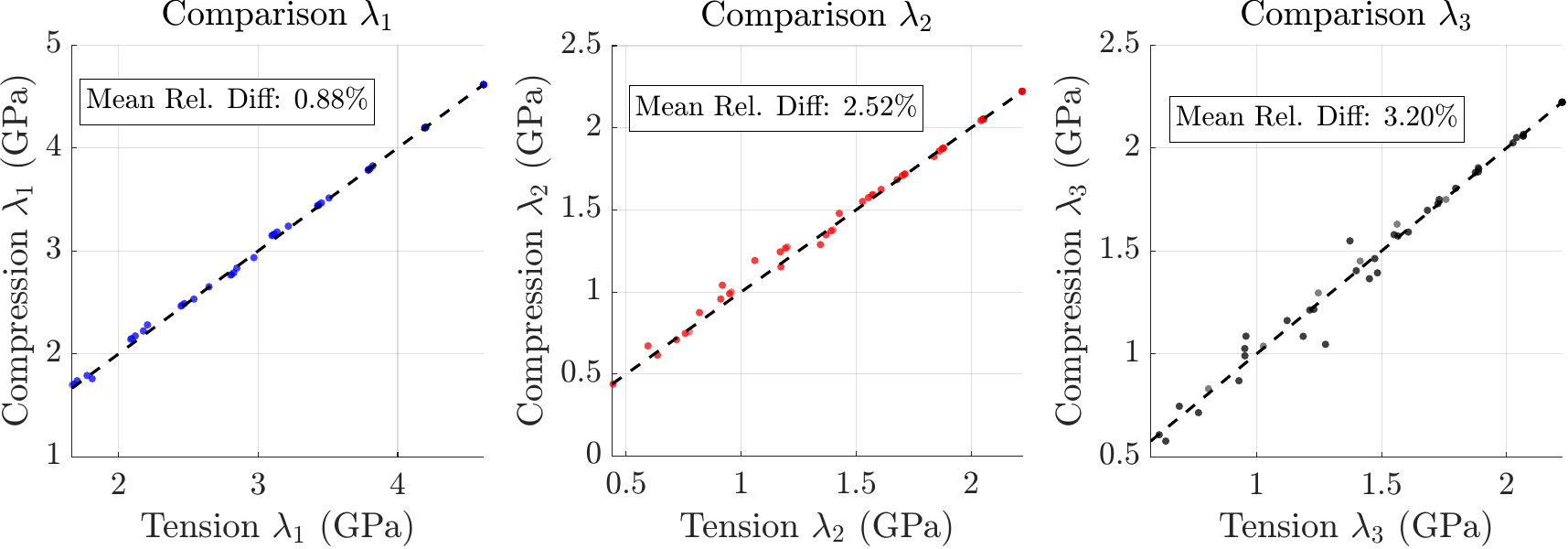}
	\caption{\color{blue} Comparison of effective stiffness eigenvalues ($\lambda_1$, $\lambda_2$, $\lambda_3$) and principal angles computed for a set of generated microstructures under tension (x-axis) and compression (y-axis). The mean relative difference confirms the validity of using a single database.}
	\label{fig:tension_compression_db}
\end{figure}

Figure~\ref{fig:tension_compression_db} provides a direct comparison of the eigenvalues extracted from both databases. The data points align along the $y=x$ line, demonstrating the equivalence of the two databases in the linear elastic regime, with minor deviations attributable to differences in mesh discretization.

The effective stiffness properties are analyzed by diagonalizing the orthotropic tensors. This decomposition yields three eigenvalues $\lambda_1$, $\lambda_2$, and $\lambda_3$, along with a principal angle $\theta$. The angle $\theta$, defined in the first quadrant ($\theta \in [0, 90]^\circ$), corresponds to the orientation of the principal eigenvector associated with the axial modes. 

\textcolor{revtwo}{To rigorously establish the invariant significance of the isolated shear mode, it is essential to trace its definition from the engineering shear behavior. The physical shear stress $\sigma_{12}$ is conventionally related to the engineering shear strain $\gamma_{12}$ by the classical shear modulus $G_{12}$ (or $C_{66}$) via $\sigma_{12} = C_{66} \gamma_{12} = 2 C_{66} \epsilon_{12}$, where $\epsilon_{12}$ is the tensorial shear strain component. However, the Mandel--Kelvin isomorphic vectorization defines the strain state vector as $\underline{\bm{\epsilon}} = [\epsilon_{11}, \epsilon_{22}, \sqrt{2}\epsilon_{12}]^{\mathrm{T}}$. To strictly preserve the strain energy density (a condition for the self-adjointness of the Bond matrix), the constitutive relation on the third axis imposes $\sqrt{2}\sigma_{12} = \underline{\underline{\mathbf{C}}}_{\!\text{MK}}^{(3,3)} ( \sqrt{2}\epsilon_{12} )$. Substituting the conventional relation $\sigma_{12} = 2 C_{66} \epsilon_{12}$ directly yields $\underline{\underline{\mathbf{C}}}_{\!\text{MK}}^{(3,3)} = 2 C_{66}$. Because this third component is completely decoupled from the normal modes when evaluated in the principal axes, it corresponds precisely to the third invariant eigenvalue of the elasticity tensor: $\lambda_3 = 2 C_{66}$. Hence, the factor of $2$ ensures an energy-preserving transition from the engineering shear strain to the pure invariant shear mode.}

This linear approximation is confirmed by the direct correspondence between the results obtained in uniaxial tension and compression (Fig.~\ref{fig:erreurs_essais_tension_free} and Fig.~\ref{fig:erreurs_essais_compression_symmetric}).

\newpage

\section*{\textcolor{reveditor}{Supplementary Note 9:} Transformational Objectives and Synthesized Material Properties}

\begin{figure}[H]
	\centering
	\includegraphics[width=0.5\textwidth]{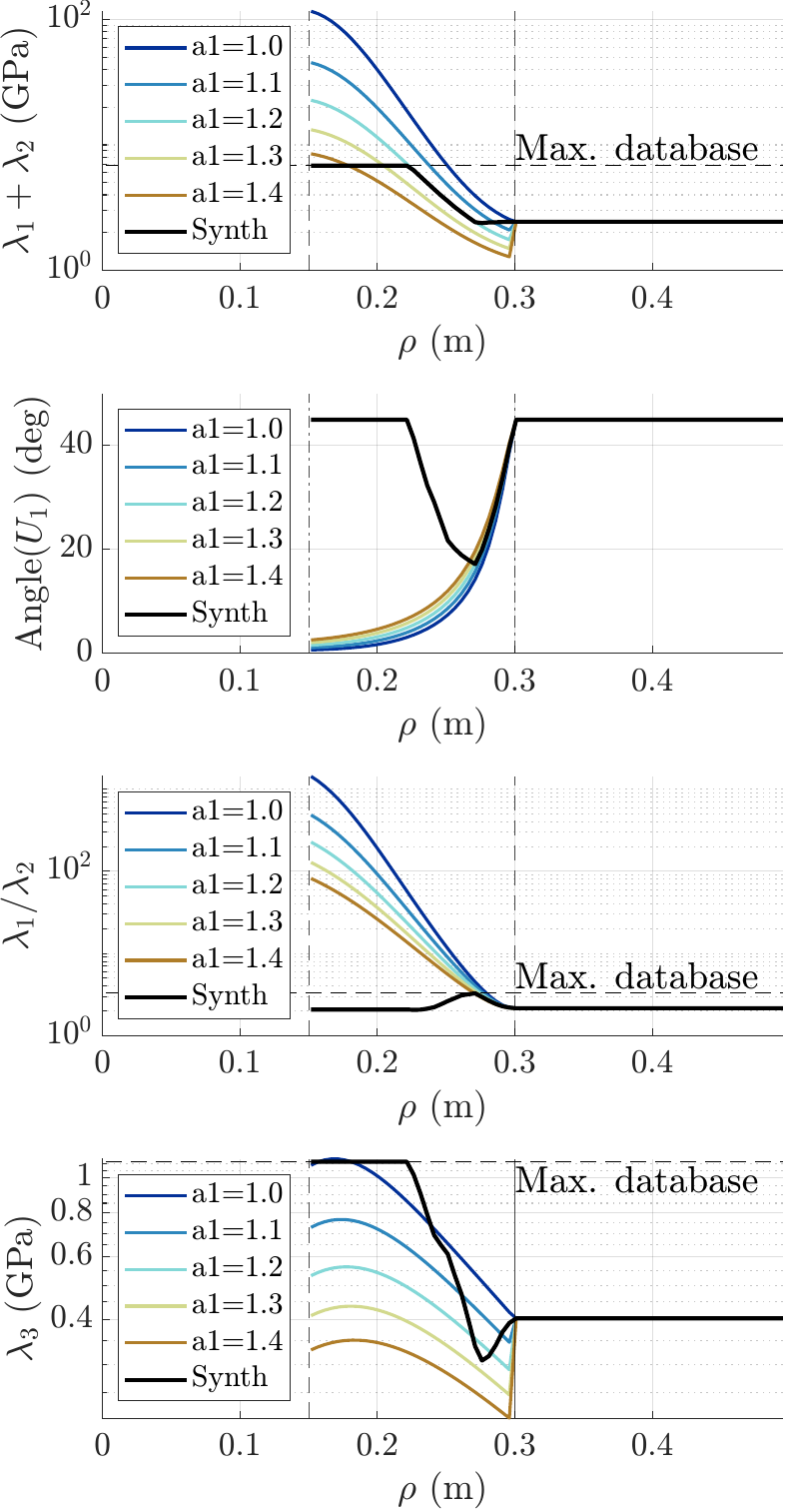}
	\caption{\color{blue} Comparison between the theoretical stiffness objectives derived from the cubic coordinate transformation (colored lines, for $a_1 \in [1.0, 1.4]$) and the properties of the synthesized metamaterial (black line). The comparison is performed along the central radial line (where $\rho$ is the radius). (Top) Sum of axial eigenvalues $\lambda_1 + \lambda_2$. (Middle Top) Principal orientation angle $\theta$. (Middle Bottom) Anisotropy ratio $\lambda_1 / \lambda_2$. (Bottom) Shear eigenvalue $\lambda_3$ ($C_{66}$). The dashed horizontal lines indicate the maximum achievable values within the material database (0\% porosity limit).}
	\label{fig:database_vs_objective}
\end{figure}

The synthesis of the cloaking medium requires matching the target stiffness tensors derived from the coordinate transformation with the available microstructures in our database. A fundamental limitation arises from the theoretical singularity at the inner radius ($R_1$), which demands extreme material properties. To address this, we adopt an adiabatic approach with a transformation parameter $a_1 > 1$ (typically $a_1 \approx 1.2$), which mitigates the singularity. Nevertheless, the achievable stiffness remains physically bounded by the Young's modulus of the constituent material (0\% porosity limit), as shown by the database maxima in Fig.~\ref{fig:database_vs_objective}.

The selection algorithm prioritized two parameters obtained from the diagonalization of the orthotropic tensors: the sum of the axial eigenvalues ($\lambda_1 + \lambda_2$) and the principal orientation angle $\theta$ of the eigenvector $u_1$. The shear component ($\lambda_3$), while not explicitly constrained during optimization, still partially aligns with the objectives derived from the transformational method.

Figure~\ref{fig:database_vs_objective} compares the theoretical targets (for $a_1$ ranging from 1.0 to 1.4) and the synthesized properties. Despite the inherent constraints of the generative model which prevent a perfect fit, especially where targets exceed the database limits, this approximation yielded a robust functional structure. This single synthesized design successfully demonstrated cloaking capabilities across six distinct mechanical loading modes, evidencing the method's robustness and the high fidelity of the fabrication process relative to numerical predictions. Furthermore, given the rotational symmetry of the target transformation and the generated cloak, the radial analysis presented in Fig.~\ref{fig:database_vs_objective} is representative of the error distribution throughout the entire annular domain.

\newpage

\section*{\textcolor{reveditor}{Supplementary Note 10: Experimental and Numerical Setup Details}}

\textcolor{reveditor}{This note presents the finite-element and experimental setups used to validate the mechanical cloaking demonstration presented in the main text. These figures have been moved from the main article to comply with the 10 display-item limit of Nature Communications.}

\begin{figure}[H]
	\centering
	\includegraphics[width=0.35\textwidth]{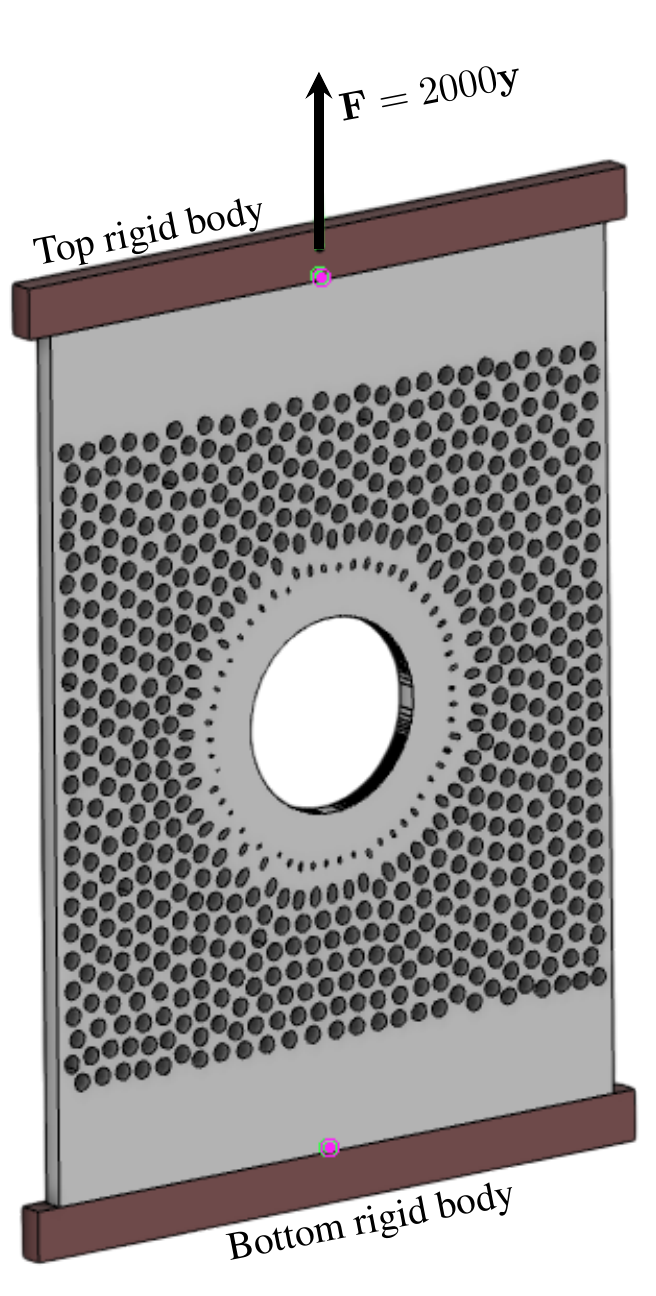}
	\caption{\textcolor{reveditor}{Finite-element setup in \textit{3DEXPERIENCE} SIMULIA. Rigid bodies are tied to the plate ends to emulate the jaws; a vertical tensile force is applied at the mass-center of the top rigid body while the bottom rigid body is fully fixed. The plate is meshed with 10-node quadratic tetrahedra (C3D10), comprising 215{,}103 elements for the cloaked specimen. The plate dimensions are $190 \times 130$~mm$^2$.}}
	\label{fig:3DEXP}
\end{figure}

\begin{figure}[H]
	\centering
	\includegraphics[width=0.7\textwidth]{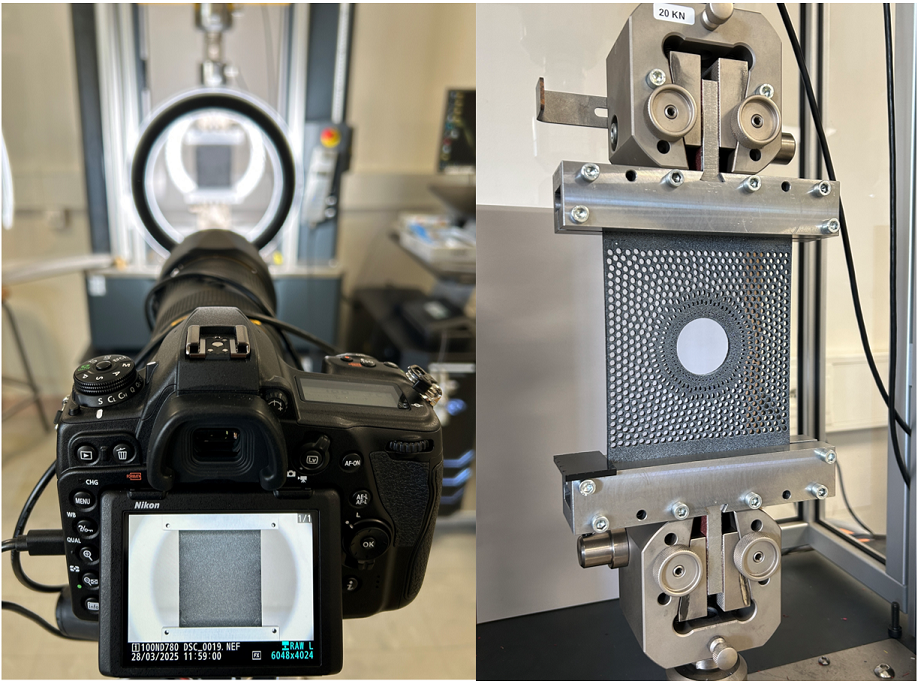}
	\caption{\textcolor{reveditor}{Experimental setup. Displacement fields during tensile tests are measured using a digital image correlation (DIC) method. Each POM plate ($190 \times 130 \times 3$~mm$^3$) is screwed onto custom steel grip bars at five attachment points per plate. A speckle pattern is applied for DIC tracking. }}
	\label{fig:Banc}
\end{figure}

\begin{figure}[H]
	\centering
	\includegraphics[width=0.55\textwidth]{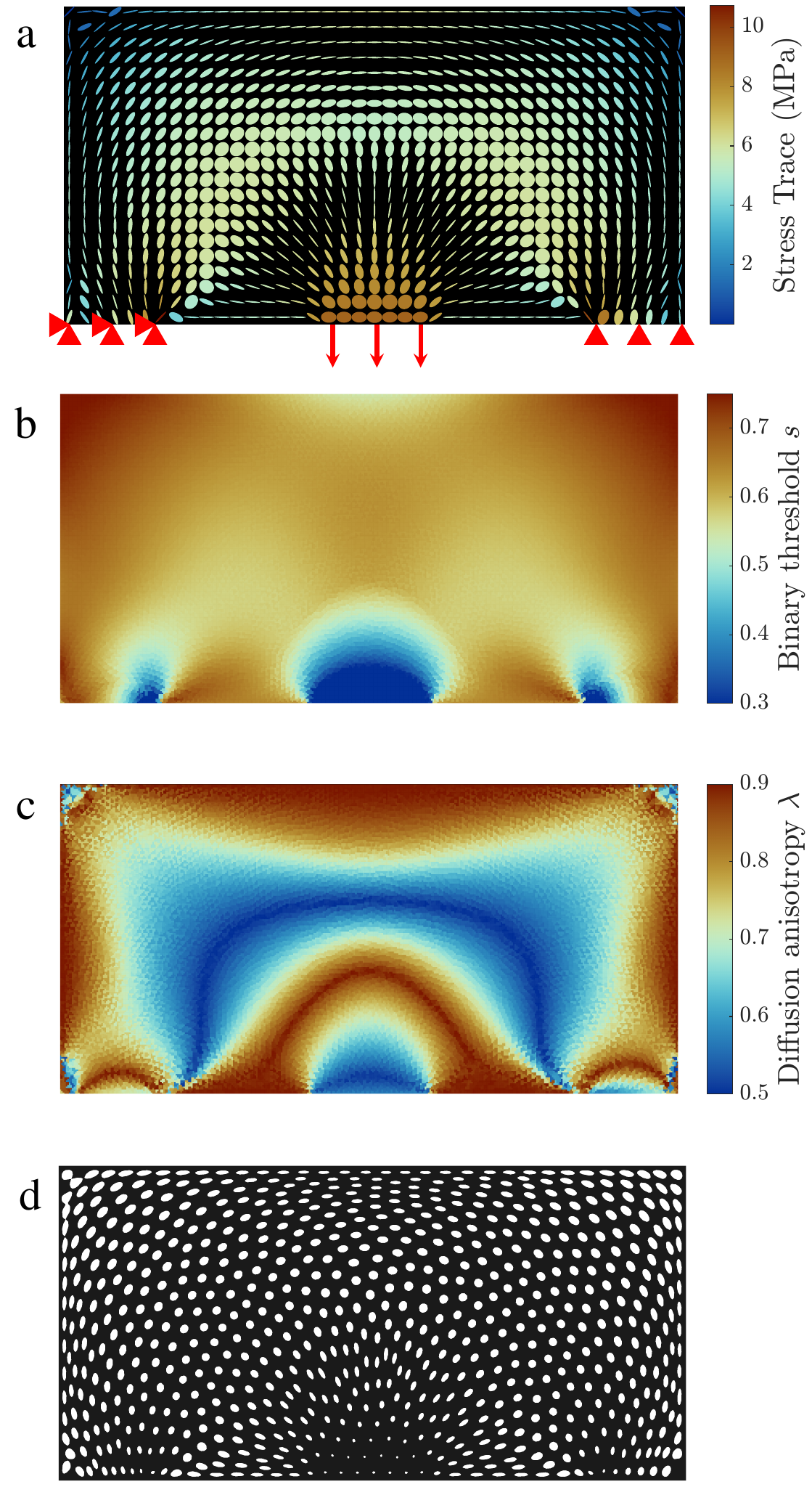}
	\caption{\textcolor{reveditor}{Stress-driven generation workflow. (a) Initial homogeneous finite-element simulation of the target domain (distributed three-point bending case) to extract the principal stress trajectories and magnitudes. (b)~Mapping of the stress tensor trace to the segmentation threshold $s$, defining the local density layout: high stress (red) leads to dense material, low stress (blue) to porous regions. (c) Mapping of the stress deviator to the diffusion anisotropy parameter $\lambda$, aligning the reaction-diffusion growth process with the principal load paths. (d) Resulting generated structure (intermediate resolution $402 \times 802$~px, physical domain $200 \times 100$~mm$^2$) exhibiting a self-organized, functionally graded microstructure with oriented inclusions matched to the mechanical demand.}}
	\label{fig:Field_Guiding}
\end{figure}

\section*{\textcolor{reveditor}{Supplementary References}}

\end{document}